\documentclass[twocolumn,epjc3]{svjour3}          

\RequirePackage{graphicx}
\RequirePackage{epsfig,amsmath}
\RequirePackage{rotate}
\RequirePackage{rotating}
\RequirePackage{multirow}
\RequirePackage{longtable}
\RequirePackage[T1]{fontenc}
\RequirePackage{array}
\RequirePackage{flushend}
\RequirePackage[numbers,sort&compress]{natbib}

\smartqed  

\usepackage{widetext}
\usepackage{epsfig,amsmath}
\usepackage{undertilde}
\journalname{Eur. Phys. J. C}

 \RequirePackage{mathptmx}      

\def\beq{\begin{equation}}
\def\eeq{\end{equation}}
\def\eq{\end{equation}}
\def\to{\rightarrow}

\def\bsg{\ifmmode B\to X_s\gamma\else $B\to X_s\gamma$\fi}
\def\bsll{\ifmmode B\to X_s\ell^+\ell^-\else $B\to X_s\ell^+\ell^-$\fi}
\def\bstt{\ifmmode B\to X_s\tau^+\tau^-\else $B\to X_s\tau^+\tau^-$\fi}
\def\shat{\ifmmode \hat{s}\else $\hat{s}$\fi}

\def\EmissT{\not \! \!  E_{T}}

\def\Emiss{\not  \! \! E}

\def\nolep{0\ell+jets+\EmissT}
\def\onelep{1\ell+jets+\EmissT}
\def\twolep{2\ell+jets+\EmissT}
\def\threelep{3\ell+0jets+\EmissT}

\def\ETsum{E_T^{\rm sum}}

\def\s2b{s_{2\beta}}

\def\u1{G_{SM}\otimes U(1)^{\prime}}

\newcommand{\newc}{\newcommand}

\newc{\lcal}{\int {\cal L}dt}

\newc{\lsp}{{\widetilde{\chi}^0_1}}
\newc{\niki}{{\widetilde{\chi}^0_2}}
\newc{\nuc}{{\widetilde{\chi}^0_3}}
\newc{\ndort}{{\widetilde{\chi}^0_4}}
\newc{\nbes}{{\widetilde{\chi}^0_5}}
\newc{\nalti}{{\widetilde{\chi}^0_6}}
\newc{\mnbir}{{M_{\widetilde{\chi}^0_1}}}
\newc{\mniki}{{M_{\widetilde{\chi}^0_2}}}
\newc{\mnuc}{{M_{\widetilde{\chi}^0_3}}}
\newc{\mndort}{{M_{\widetilde{\chi}^0_4}}}
\newc{\mnbes}{{M_{\widetilde{\chi}^0_5}}}
\newc{\mnalti}{{M_{\widetilde{\chi}^0_6}}}
\newc{\stauR}{{\widetilde{\tau}_R}}
\newc{\stau}{{\widetilde{\tau}_1}}
\newc{\staut}{{\widetilde{\tau}_2}}
\newc{\mstop}{m_{\widetilde{t}}}
\newc{\mHpm}{m_{H^\pm}}
\newc{\simgt}{\lower.7ex\hbox{$\;\stackrel{\textstyle>}{\sim}\;$}}
\newc{\simlt}{\lower.7ex\hbox{$\;\stackrel{\textstyle<}{\sim}\;$}}
\newc{\ie}{{\it i.e.}}
\newc{\etal}{{\it et al.}}
\newc{\eg}{{\it e.g.}}
\newc{\kev}{\hbox{\rm\,keV}}
\newc{\mev}{\hbox{\rm\,MeV}}
\newc{\gev}{\hbox{\rm\,GeV}}
\newc{\tev}{\hbox{\rm\,TeV}}
\newc{\xpb}{\hbox{\rm\, pb}}
\newc{\xfb}{\hbox{\rm\, fb}}

\newc{\mtop}{m_t}
\newc{\mbot}{m_b}
\newc{\mz}{m_Z}
\newc{\mw}{M_W}
\newc{\alphasmz}{\alpha_s(m_Z^2)}
\newc{\swsq}{\sin^2\theta_W}
\newc{\tw}{\tan\theta_W}
\newc{\cw}{\cos\theta_W}
\newc{\sw}{\sin\theta_W}
\newc{\BR}{\hbox{\rm BR}}
\newc{\zbb}{Z\to b\bar}
\newc{\Gb}{\Gamma (Z\to b\bar b)}
\newc{\Gh}{\Gamma (Z\to \hbox{\rm hadrons})}
\newc{\rbsm}{R_b^\hbox{\rm sm}}
\newc{\rbsusy}{R_b^\hbox{\rm susy}}
\newc{\drb}{\delta R_b}

\newc{\sgn}{\mbox{sgn}}
\newc{\tbeta}{\tan\beta}
\newc{\uL}{{\widetilde{u}_L}}
\newc{\uR}{{\widetilde{u}_R}}
\newc{\cL}{{\widetilde{c}_L}}
\newc{\cR}{{\widetilde{c}_R}}
\newc{\tL}{{\widetilde{t}_L}}
\newc{\tR}{{\widetilde{t}_R}}
\newc{\dL}{{\widetilde{d}_L}}
\newc{\dR}{{\widetilde{d}_R}}
\newc{\sL}{{\widetilde{s}_L}}
\newc{\sR}{{\widetilde{s}_R}}
\newc{\bL}{{\widetilde{b}_L}}
\newc{\bR}{{\widetilde{b}_R}}
\newc{\eL}{{\widetilde{e}_L}}
\newc{\eR}{{\widetilde{e}_R}}
\newc{\mhp}{m_{H^\pm}}
\newc{\mhalf}{m_{1/2}}
\newc{\emt}{{e/\mu /\tau}}

\newc{\lR}{\widetilde{\ell}_R}
\newc{\lL}{\widetilde{\ell}_L}
\newc{\nL}{\widetilde{\nu}_{\ell_L}}
\newc{\nR}{\widetilde{\nu}_{\ell_R}}
\newc{\neL}{\widetilde{\nu}_{e_L}}
\newc{\nmL}{\widetilde{\nu}_{\mu_L}}
\newc{\nlL}{\widetilde{\nu}_{\tau_L}}
\newc{\neR}{\widetilde{\nu}_{e_R}}
\newc{\nmR}{\widetilde{\nu}_{\mu_R}}
\newc{\nlR}{\widetilde{\nu}_{\tau_R}}
\newc{\naa}{\widetilde{\chi}^0_1}
\newc{\nbb}{\widetilde{\chi}^0_2}
\newc{\ncc}{\widetilde{\chi}^0_3}
\newc{\ndd}{\widetilde{\chi}^0_4}
\newc{\nee}{\widetilde{\chi}^0_5}
\newc{\nff}{\widetilde{\chi}^0_6}
\newc{\caa}{\widetilde{\chi}^{\pm}_1}
\newc{\cbb}{\widetilde{\chi}^{\pm}_2}
\newc{\phit}{\phi_t}
\newc{\phib}{\phi_b}
\newc{\phiew}{\phi_{ew}}
\newc{\htz}{h^0_t}
\newc{\hbz}{h^0_b}
\newc{\hewz}{h^0_{ew}}
\newc{\hsmz}{h^0_{sm}}
\newc{\huz}{h^0_u}
\newc{\hsusyz}{h^0_{susy}}
\newc{\lmop}{\rm LM1^\prime}
\newc{\lmtp}{\rm LM2^\prime}
\newc{\lmsp}{\rm LM6^\prime}
\newc{\smin}{\hat{s}_{\rm min}^{1/2}}

\def\slashchar#1{\setbox0=\hbox{$#1$}           
   \dimen0=\wd0                                 
   \setbox1=\hbox{/} \dimen1=\wd1               
   \ifdim\dimen0>\dimen1                        
      \rlap{\hbox to \dimen0{\hfil/\hfil}}      
      #1                                        
   \else                                        
      \rlap{\hbox to \dimen1{\hfil$#1$\hfil}}   
      /                                         
   \fi}                                         %
\begin{document}

\title{Neutralino and Chargino Production in $U(1)'$ at the LHC}


\author{ Mariana Frank\thanksref{e1,a1} \and
Levent Selbuz\thanksref{e2,a1,a2} \and Ismail Turan\thanksref{e3,a3}
}


\thankstext{e1}{e-mail: mariana.frank@concordia.ca}
\thankstext{e2}{e-mail: levent.selbuz@eng.ankara.edu.tr}
\thankstext{e3}{e-mail: ituran@metu.edu.tr}

%
\institute{Department of Physics, Concordia University,
7141 Sherbrooke St. West, Montreal, Quebec, Canada H4B 1R6,\label{a1}
\and
Department of Engineering Physics, Ankara
University, TR06100 Ankara, Turkey,\label{a2}
\and
Department of Physics, Middle East Technical University, Ankara TR06800,
Turkey.\label{a3}
}

\date{Received: date / Accepted: date}

\maketitle


\begin{abstract}
We examine the production and decay
modes of neutralinos and charginos in a softly-broken supersymmetric
model with an extra Abelian symmetry $U(1)^\prime$. We perform the study in a
$U(1)^\prime$ model with a secluded sector, where the tension
between the electroweak scale and developing a large enough mass for
$Z^\prime$ is resolved by incorporating three additional $SU(2)$
singlet fields into the model. Although the chargino sector is the
same as in the MSSM,  the neutralino sector of the model is very
rich:  five new fermion fields are added to the neutral sector bring the total neutralino states
to nine.  We implement the model into
standard packages and perform a detailed and systematic analysis of production and decay modes at the LHC, for three different
scenarios,  consistent with the Higgs data and relic density constraints. We concentrate on final signals (1)
$\onelep$, (2) $\twolep$ and (3) $\threelep$, and comment on the case with  $\nolep$.
We discuss backgrounds and indicate how these signals can be observed, and how the model can be distinguished from other supersymmetric model scenarios.
\end{abstract}
\keywords{Supersymmetry,
Neutralino, Chargino, LSP, LHC.}
 \PACS{12.60.Cn,12.60.Jv,14.80.Ly.}

\section{Introduction and Motivation}\label{sec:intro}

After the recent discovery of the new resonance most likely to be the standard model (SM) Higgs
boson at ATLAS \cite{:2012gk} and CMS \cite{:2012gu}, the top priority
for LHC shifts to the search for physics effects beyond the SM, in particular for
supersymmetry as the leading candidate. The Minimal Supersymmetric Standard
Model (MSSM),
motivated by the resolution of such long standing problems in
SM as the gauge hierarchy problem, the existence of dark matter, and
the  gauge unification,
 is arguably the most popular `new physics' scenario as  the
 perturbative extension of the SM beyond electroweak scales.
 However, recent LHC results \cite{:2012ct}
 rule out some of the parameter regions of the constrained version of MSSM and, if this particular version of SUSY is realized in nature, point towards a heavy spectrum of supersymmetric partners.

 Minimal extensions of the SM gauge symmetry by additional $U(1)^\prime$ Abelian groups
are well motivated,   not so much based on resolving some of problems in MSSM, but by the fact that such an
extension is justified in superstring
theories \cite{Cvetic:1995rj},  grand unified theories
\cite{Hewett:1988xc} and  in dynamically   broken electroweak theories
\cite{Hill:2002ap}. The additional gauge group introduces one  extra neutral gauge boson $Z^\prime$. The simplest version of $U(1)^\prime$ extended supersymmetric models also involve an additional singlet $S$, charged under $U(1)^\prime$, whose vacuum expectation value (VEV) is responsible for the breaking of $U(1)^\prime$. This VEV simultaneous generates dynamically an effective $\mu$ term, an elegant resolution of the
so-called $\mu$ problem \cite{muprob}, and is responsible for the mass of the $Z^\prime$ boson.  Some versions of these extended symmetries also allow
right-handed neutrinos into the spectrum.  Small neutrino masses consistent with neutrino
oscillation phenomenology are usually explained by the see-saw
mechanism \cite{Minkowski:1977sc}. In the Type II see-saw mechanism, large
Majorana masses for right-handed neutrinos are responsible for inducing small Majorana
masses for left-handed neutrinos.  The choice of $U(1)^\prime$ symmetry
would determine the magnitude and type of neutrino masses \cite{Kang:2004ix}:  $U(1)^\prime$ extended forms of the MSSM can
contain Dirac-type neutrino masses as in  \cite{biz3}, and viable models exist for Majorana masses as well \cite{biz2}. The $U(1)^\prime$ model shares some of the the advantages of the next-to-minimal supersymmetric standard models (NMSSM). In the MSSM at tree-level, the Higgs mass is bound by $m_h \le M_Z$. To alleviate this problem,  large stop masses and large trilinear $A_t$-terms are added to the MSSM \cite{Heinemeyer:2011aa}. In $U(1)'$, the addition of one singlet field provides new tree-level contributions to the $F$-term, which stabilize the Higgs mass naturally at a larger value \cite{Ross:2012nr}, thus accommodating a lightest Higgs boson mass at 125 GeV.

In the minimal version of extended $U(1)^\prime$ symmetry models, loops generate a mixing term for $Z-Z^\prime$ bosons, which in turn is constrained by the electroweak precision data to be ${\cal O}(10^{-3})$, or smaller, while collider constraints on $Z^\prime$ mass require it to be heavy. In the minimal $U(1)^\prime$
model, explored in Ref.~\cite{Ali:2009md},  the
difficulty to induce a small $\mu_{eff}$ while satisfying the
$Z^\prime$ mass bound, which is around 1 TeV, stems from the fact that both
 are proportional to the VEV of the additional scalar field $S$. The resolution is provided in a non-minimal version of the $U(1)^\prime$ extended MSSM,  in which  several singlet 
fields ($S_i$) are introduced to resolve the conflict between maintaining the electroweak scale and developing a
large enough mass for $Z^\prime$. One needs
three additional scalars to ameliorate the picture, and the
VEVs of the new scalars must be kept large \cite{Erler:2002pr,Chiang:2008ud}.  We refer to this version of
the model  as secluded $U(1)^\prime$, an abbreviated
notation for  the gauge symmetry underlying the model,
 $SU(3)_c \otimes SU(2)_L \otimes U(1)_Y \otimes
U(1)^\prime$, with a non-minimal $U(1)^\prime$.  A comparative study of LHC signals
of sneutrino production and decays in the MSSM and in a
supersymmetric model with a secluded $U(1)^\prime$ breaking sector
has been performed in \cite{Demir:2010is}.

Direct or indirect detection of the superpartners of the Standard
Model particles, considered the definitive signal for supersymmetry, is an
important part of the experimental program of the LHC. Two distinct phenomenological approaches to SUSY searches are possible.
One approach is based on the latest available experimental
information. This method has the advantage of incorporating all the
relevant experimental constraints, but the disadvantage of becoming
quickly obsolete, as more data becomes available; also experimental
data forecasts rarely impose direct and precise constraints, as
many free parameters are involved. The other approach is to look
into interesting benchmark scenarios in models, which illustrate
model-specific possibilities. These benchmarks may incorporate some,
but perhaps not all, present experimental constraints, and serve as
indicators of possible experimental signatures. For instance, the
cosmological relic density constraint  for models where the lightest
neutralino is the Lightest Supersymmetric Particle (LSP) and the neutral Higgs mass are
definite constraints; so are consistency with low-energy
phenomenology, such as flavor-changing and CP-violating processes.
We follow the latter approach here.

The LHC has already devoted a great deal of time and effort to
searches for supersymmetric partners. Gluinos and scalar quarks are
expected to be produced copiously at a hadron collider, though no
signals are seen \cite{LHCgsq}. However, these states are expected
to be heavy, and, except for the LSP in the $R-$parity conserving
supersymmetry, all superpartners are expected to decay
instantaneously into SM particles plus the LSP, detected as missing
energy.

Neutralinos and charginos, expected
to be lighter, can play an important role as they occur in various
steps in the cascade decays of certain supersymmetric particles
(squarks, gluinos, etc.), and thus they would be be abundantly produced at the
LHC.   Besides direct signals at the colliders, charginos and
neutralinos can give indirect indications of their existence. Both
can  have implications  on Higgs physics.  For instance, it is
possible that the Higgs can decay in a non-standard fashion, with invisible width due to decays into
neutralinos \cite{Godbole:2003it,Frank:2013yta}, while the charginos could be responsible for the
enhancement of the Higgs decays into $\gamma \gamma$
\cite{Hchargino}. The production of neutralinos at hadron colliders is an important
part of the program of SUSY searches. One special reason is related
to the possibility that the lightest neutralino state
($\tilde\chi^0_{1}$) is in fact the
LSP.  Searches for charginos and neutralinos have not yielded any results
so far. However, all searches come with conditions attached, due to
the many alternative models, different sources of SUSY breaking,
classes of compactification. Particularly, the searches rely on having gluinos and squarks below the TeV scale,  make specific assumptions on the nature of the LSP,  and most analyses focus on MSSM.

We summarize the results of
some of the recent searches.  At ATLAS, chargino masses between 110
and 340 GeV are excluded in direct production of wino-like pairs,
decaying into LSP via on-shell sleptons, for a 10 GeV neutralino, at
95\% C.L.  For models with decays into intermediate degenerate sleptons,
the lightest chargino ${\tilde \chi}^+_1$ and second lightest
neutralino ${\tilde \chi}^0_2$  are ruled out up to masses of 500
GeV \cite{atlascharginos}. CMS analyzed final states with three leptons
in conjunction with two $jets$  to rule out chargino and neutralino
masses between 200 and 500 GeV,  for models where BR$ ({\tilde
\chi}^{0(-)} \to Z (W)$ leptons) is large \cite{CMScharginos}.

 Despite all the negative searches, one might argue that, even if no direct signals of supersymmetry have been observed, the presence of dark matter in the universe is already an indirect signal for supersymmetry.   In most variants of the MSSM consistent with relic density
calculations, the LSP is the lightest neutralino.  the Thus studies of possible supersymmetric particles at colliders are worthwhile pursuits.

 The production of neutralinos is of special interest in the secluded sector $U(1)^\prime$  model, as the model contains five more neutralinos than MSSM. The additional singlet  fields introduced to generate the $Z-Z^\prime$ mass splitting  are difficult to detect, and expected to be heavy. However, their fermion partners, the neutralinos, could be light and enhance the direct and cascade production of supersymmetric particles at colliders. This would then be the best test of the secluded sector. We note that while the additional scalar multiplets are necessary ingredients in the $U(1)^\prime$  model, the behavior of the additional superpartner fields is generic,  typical of any supersymmetric model with additional singlet scalar fields.  From this point of view, an analysis of neutralinos in secluded $U(1)^\prime$ models is more general, and illustrative of the effects of the fermionic partners of singlet fields. In addition, the LSP or the next-to LSP (NLSP) in these models can be the singlino, yielding different decay patterns, as all supersymmetric particles decay eventually into $X+LSP$, with $X$ a mixture of jets, leptons and possibly additional $\Emiss$.

 With this motivation,  we perform a comprehensive study of LHC
signals of neutralino (and chargino) production and decays in a
supersymmetric model with a secluded $U(1)^{\prime}$ breaking sector, concentrating on highlighting the contributions of the additional singlino-like neutralino states.  These appear  can alter the signatures of the secluded $U(1)'$ model as compared to the MSSM.   We analyze the signals, classified according to the number of leptons in the final states, and we also include estimates of possible SM backgrounds
in  three different scenarios. Older analyses are available for MSSM \cite{Baer:1994nr}, though the production, decay and identification of charginos and neutralinos have received some attention very recently, given the failure to find squarks and gluinos at the LHC \cite{Bharucha:2013epa}.
While in a previous work \cite{Demir:2009kc}, we showed that
in a minimal $U(1)^{\prime}$ model (with one extra singlet boson), choosing the right-handed sneutrino as the LSP could be
consistent with the excess positron observed in satellite
experiments, for the purpose of this work, in the secluded sector
$U(1)^{\prime}$, we take the lightest neutralino consistently to be the
lightest supersymmetric particle (LSP) and therefore a dark matter (DM) candidate.

The outline of this paper is as follows. We briefly introduce the
model in Section~\ref{sec:model}, with particular emphasis on the neutralino and chargino sector, then we choose three benchmark scenarios and for each, give the parameters and
physical masses of supersymmetric particles in the  $U(1)^{\prime}$ model in Section~\ref{sec:num}. For each
case, we insure  that the dark matter candidate of the
model yields relic densities consistent with the WMAP range of cold
dark matter density \cite{wmap}. We then perform a
comprehensive analysis of the production, decays and detectability
of neutralinos and chargino within these benchmark supersymmetric
parameter points. During this analysis we focus on three types of detector
signatures: (1) $\onelep$, (2) $\twolep$ and (2)
$\threelep$, and we present the results of our simulation analysis for
the LHC. In Section~\ref{sec:conc} we summarize and conclude the analysis. We list diagrams for some characteristic decay patterns in the three scenarios in  \ref{app}.

\section{The secluded $U(1)^{\prime}$ Model \label{sec:model}}

We summarize here the salient features of the secluded $U(1)^\prime$ model, with particular emphasis on the chargino and neutralino sector.

The superpotential of the model contains Yukawa couplings for quarks and leptons, and the couplings for the exotic fields and is  given by
\begin{eqnarray}\label{eq:superpot}
\widehat{W}&=&h_u\widehat{Q}\cdot \widehat{H}_u \widehat{U}+
h_d\widehat{Q}\cdot \widehat{H}_d \widehat{D} + h_e\widehat{L}\cdot
\widehat{H}_d \widehat{E} \nonumber \\
&+& h_s \widehat{S}\widehat{H}_u \cdot
\widehat{H}_d +  \frac{1}{M_R}  \widehat{S}_1 \widehat{L}\cdot
\widehat{H}_u {\bf h_{\nu}} \widehat{N}+
\bar{h}_s \widehat{S}_1 \widehat{S}_2 \widehat{S}_3 \nonumber \\
&+& \sum_{i=1}^{n_{\cal{Q}}} {h}_Q^i \widehat{S} \widehat{\cal{Q}}_i
\widehat{\cal{\overline{Q}}}_i + \sum_{j=1}^{n_{\cal{L}}} {h}_L^j
\widehat{S} \widehat{\cal{L}}_j \widehat{\cal{\overline{L}}}_j,
\end{eqnarray}
where the fields ${\cal Q},~{\cal L}$ are the exotic fermions, $M_R$ is a large mass scale and
$h_{\nu}$ is the Yukawa coupling responsible for generating neutrino
masses. The $U(1)^\prime$ charge assignments
which generate the term $\mu_{eff}$, of the form $\displaystyle \lambda_s\frac{\langle  S \rangle}{\sqrt{2}}H_u H_d$,  induce mixed anomalies between the $U(1)^\prime$ and the $SU(3)_C \times SU(2)_L \times U(1)_Y$ groups. The cancellation of these anomalies requires introduction of  exotic fermions, vector-like with respect to the MSSM, but chiral under the $U(1)^\prime$ group. These fields introduce additional $D$-terms in the Lagrangian. For anomaly cancellation we require $n_{\cal{Q}}=3$, $Q_{\cal Q}=-1/3$ for the color triplets, and $n_{\cal{L}}=5$, $Q_{\cal L}=-\sqrt{2/5}$ for the singlets, where $Q_i$ is the electric charge of particle $i$ \cite{Demir:2010is}. Thus the  supersymmetric partners of the exotic fermions do not mix with charginos or neutralinos. 

In addition, the Lagrangian contains  soft-breaking terms for the secluded sector
\begin{eqnarray}
 V_{soft}&=& (m^2_{SS_1}S^\dag S_1+m^2_{SS_2}S^\dag S_2+m^2_{S_1S_2}S_1^\dag
S_2+h.c.)\nonumber \\
&+&m^2_{H_u}|H_u|^2+m^2_{H_d}|H_d|^2+m^2_S|S|^2\nonumber \\
&+& \sum_{i=1}^3m^2_{S_i}
|S_i|^2-(A_sh_sSH_uH_d+A_{\bar s}\bar h_sS_1S_2S_3+h.c.).
\end{eqnarray}
The symmetry-breaking sector of the model is very rich. There are a
number of $CP$-even and $CP$-odd Higgs fields. Finding an acceptable
minimum of the Higgs potential is not a trivial task,  even at the
tree level. While we addressed the details of the scalar sector
elsewhere \cite{Frank:2013yta}, we include here some general comments. Once a
minimum is found, the mass of the lightest Higgs boson can be fine-tuned to 125 GeV 
by small variations in the parameter $\bar{h}_s,
h_s, A_{\bar s}, A_{s}$ and the singlet VEVs $v_{s_1}, v_{s_2},
v_{s_3}$. Setting masses for the additional scalars in the TeV range
insures that the mixing with the lightest Higgs boson is small, and
thus it does not spoil  the couplings with the $Z$ boson, or
adversely affect the $4 \ell$ signal observed at the LHC. Additional
Higgs states, in particular the lightest pseudoscalar, will have to
satisfy constraints from $B_s \to \mu^+ \mu^-$ branching ratio
\cite{:2012ct} and may have to be heavy.

The $U(1)^{\prime}$ charges of the fields satisfy a number of
conditions arising from the requirement of
cancellation of gauge and gravitational anomalies. For instance, the
charges for Higgs fields in the model are chosen so that
$\displaystyle
Q^\prime_{S}=-Q^\prime_{S_1}=-Q^\prime_{S_2}=\frac{1}{2}Q^\prime_{S_3},
~~~Q^\prime_{H_u}+Q^\prime_{H_d}+Q^\prime_{S}=0$.
The $U(1)^\prime$ charge of the quark doublet $\widehat{Q}$ is kept
as a free parameter after the  normalization  $Q^\prime_{H_u}=-2$,
$Q^\prime_{H_d}=1$, $Q^\prime_{S}=1$, $Q^\prime_{S_1}=-1$,
$Q^\prime_{S_2}=-1$, $Q^\prime_{S_3}=2$. A detailed analysis of
the secluded sector $U(1)^\prime$ model, including the complete list of conditions for anomalies cancellation  in the model, the Lagrangian as well as the complete charge assignments of the SM and exotic quarks and leptons in the model
 can be found in \cite{Demir:2010is}. We forgo the complete discussion here and concentrate on the chargino and neutralino sector, where we highlight differences with the MSSM.

\subsection{Charginos and Neutralinos}\label{app:C}
In $U(1)^{\prime}$ models chargino sector is unaltered. However,
chargino mass eigenstates become dependent upon $U(1)^{\prime}$
breaking scale through the $\mu_{eff}$ parameter in their mass matrix:
\begin{eqnarray}
M_{\chi^\pm}= \left(\begin{array}{c c}
M_2 & M_W \sqrt{2} \sin{\beta} \\
M_W \sqrt{2} \cos{\beta} & \mu_{eff} \\
\end{array} \right)
\end{eqnarray}
 which can be diagonalized by biunitary transformation
  \begin{eqnarray}
U^{\star} M_{\chi^\pm} V^{-1} = {\rm Diag}(\tilde{M}_{\chi_1^+} ,
\tilde{M}_{\chi_2^+}),
 \end{eqnarray}
  where $U$ and $V$ are unitary mixing
matrices.

More importantly for this study, the $U(1)^{\prime}$ model  has five additional fermion fields in the
neutral sector: the $U(1)^{\prime}$ gauge fermion
$\widetilde{Y}^{\prime}$ and four singlinos $\widetilde{S}$,
$\widetilde{S_1}$, $\widetilde{S_2}$, $\widetilde{S_3}$, in total,
nine neutralino states $\widetilde{\chi}_i^0$ ($i=1,\dots,9$)
\cite{Erler:2002pr}:
\begin{eqnarray}
\label{neutralino-def1} \widetilde{\chi}_i^0 = \sum_{a} {\cal N}^0_{i a}
\widetilde{G}_a\,,
\end{eqnarray}
where the mixing matrix ${\cal N}^0_{i a}$ connects the gauge-basis neutral
fermion states to the physical-basis
neutralinos $\widetilde{\chi}_i^0$. The neutralino masses
$M_{\widetilde{\chi}_i^0}$  are
obtained through diagonalization

\noindent
${\cal N}^0 {\cal{M}} {\cal N}^{0\ T}
= \mbox{Diag}$ $\Big\{M_{\widetilde{\chi}_1^0},$ $\dots,$
$M_{\widetilde{\chi}_9^0}\Big\}$. The $9 \times 9$ neutral fermion mass matrix is

\begin{widetext}
\begin{eqnarray}\label{mneut}
{\cal M}=\left(
 \begin{array}{ccccccccc}
 M_{\tilde Y}&0&-M_{\tilde Y \tilde H_d}&M_{\tilde Y \tilde H_u}&0&M_{\tilde Y
\tilde Y'}&0&0&0 \\[1.ex]
  0&M_{\tilde W}&M_{\tilde W \tilde H_d}&-M_{\tilde W \tilde
H_u}&0&0&0&0&0\\[1.ex]
  -M_{\tilde Y \tilde H_d}&M_{\tilde W \tilde
H_d}&0&-\mu&-\mu_{H_u}&\mu'_{H_d}&0&0&0\\[1.ex]
  M_{\tilde Y \tilde H_u}&-M_{\tilde W \tilde
H_u}&-\mu&0&-\mu_{H_d}&\mu'_{H_u}&0&0&0\\[1.ex]
  0&0&-\mu_{H_u}&-\mu_{H_d}&0&\mu'_S&0&0&0\\[1.ex]
  M_{\tilde Y \tilde Y'}&0&\mu'_{H_d}&\mu'_{H_u}&\mu'_S&M_{\tilde
Y'}&\mu'_{S_1}&\mu'_{S_2}&\mu'_{S_3}\\[1.ex]
  0&0&0&0&0&\mu'_{S_1}&0&-\frac{\bar{h}_s v_{s_3}}{\sqrt{2}}&-\frac{\bar{h}_s
v_{s_2}}{\sqrt{2}}\\[1.ex]
  0&0&0&0&0&\mu'_{S_2}&-\frac{\bar{h}_s v_{s_3}}{\sqrt{2}}&0&-\frac{\bar{h}_s
v_{s_1}}{\sqrt{2}}\\[1.ex]
  0&0&0&0&0&\mu'_{S_3}&-\frac{\bar{h}_s v_{s_2}}{\sqrt{2}}&-\frac{\bar{h}_s
v_{s_1}}{\sqrt{2}}&0\\[1.ex]
 \end{array}
 \right).
\end{eqnarray}
\end{widetext}
The gaugino masses and mixing mass parameter between the $U(1)_Y$ and
$U(1)^{\prime}$ gauginos are generated by the soft symmetry breaking terms.
The remaining entries in (\ref{mneut}) are generated by the MSSM
soft breaking masses in the Higgs sector. The mass mixing terms are
\begin{eqnarray}
M_{\widetilde{Y}\, \widetilde{H}_d} &=& M_Z \sin\theta_W
\cos\beta\,,\quad
M_{\widetilde{Y}\, \widetilde{H}_u} = M_Z \sin\theta_W \sin\beta\,,\nonumber\\
M_{\widetilde{W}\, \widetilde{H}_d} &=& M_Z \cos\theta_W
\cos\beta\,, \quad
M_{\widetilde{W}\, \widetilde{H}_u} = M_Z \cos\theta_W \sin\beta\, ,
\end{eqnarray}
and  the effective $\mu$ couplings in each sector
\begin{eqnarray}
\mu_{H_d} &=& h_s \frac{v_d}{\sqrt{2}}\,,\qquad \mu_{H_u} = h_s
\frac{v_u}{\sqrt{2}}\,,\quad \mu^{\prime}_{H_d}=g_{Y^{\prime}}
Q_{H_d}^{\prime} v_d,\nonumber\\
\mu^{\prime}_{H_u} &=& g_{Y^{\prime}} Q_{H_u}^{\prime} v_u\,\quad
\mu^{\prime}_{S} = g_{Y^{\prime}} Q_{S}^{\prime} v_s\,,\quad
\mu^{\prime}_{S_i} = g_{Y^{\prime}} Q_{S_i}^{\prime} v_{s_i}\,,
\end{eqnarray}
with $g_{Y'}$ the coupling constant of $U(1)^{\prime}$. For
the numerical analysis we choose the usual  value   at GUT scale
$g_{Y'}=\sqrt{\frac{5}{3}}g\tan\theta_W$. The production and decay of neutralinos in the $U(1)^\prime$ model without a secluded sector has been studied previously in \cite{choi06}.

As mentioned previously, compared with MSSM, the $U(1)^\prime$ neutralino sector is extended by an additional gaugino and four additional higgsinos (while the chargino sector is unaltered). The complexity of their production and decay is increased, and specific features depend on the parameters chosen. Clear general signatures emerge if some simplifying assumptions are made, such as for instance assuming the mixing between the additional fields and the MSSM fields is weak. This is because in MSSM, the $4 \times 4$ neutralino mass matrix can be diagonalized analytically, whereas here it cannot be done exactly; though under weak mixing assumption it can be done perturbatively. On general grounds, we expect that the most important of the fermionic components is the singlino $\tilde S$, as this mixes with the doublet higgsino components ${\tilde H}_u$ and ${\tilde H}_d$, whereas the singlet fermions ${\tilde S}_1$, ${\tilde S}_2$ and ${\tilde S}_3$ couple to each other. Thus, the production and decays of neutralinos would be mostly influenced by either the mixed state of ${\tilde H}_u$ and ${\tilde H}_d$ with $\tilde S$; or by decays into pure singlino states ${\tilde S}_1, {\tilde S}_2$ and ${\tilde S}_3$.

As in MSSM, the main production mechanism for neutralinos proceeds through the $Z$ boson, and the decays through $Z$ boson ($W^\pm$ for charginos) are likely to dominate, if kinematically accessible. If the additional neutralino states are heavy, they would be rarely produced and unlikely observed, thus when considering benchmark points for the parameter space we will take at least some to be light.

The singlino components modify the production cross sections and decay branching ratios.   For instance, normally in MSSM  annihilation of the lightest neutralinos through a $Z$ resonance  is expected to be small for small $\tan \beta$ \cite{Barger:2004bz}, which is not the case for singlinos.
If the singlino is light, the main $pp$ production will be through $Z$ boson and into $\tilde S \tilde S$, and it would be sufficiently enhanced to compete with other channels. In general, neutralino production is determined by a) the mixing among neutralino states, and b) the mass and kinetic mixing parameters of the $U(1)$ and $U(1)^\prime$ gauge groups.

 For the decays of neutralinos,  the two-body decays ${\tilde \chi}_i^0 \to {\tilde \chi}_j^0 Z$ ($Z^\prime$, if accessible) are important. The other two-body decay of neutralinos which is important, if kinematically allowed, is through sleptons $\tilde l_{L,R}$ (we assume squarks ${\tilde q}_k$ are much heavier); while the decays through a CP-even or CP-odd Higgs boson, even if allowed, are subdominant. In principle, neutralino radiative decays are important when the gap between two neutralino masses becomes very small  \cite{choi06}, and $\chi^0_k \to \chi^0_j \gamma$ are phase suppressed, but less so than the competing standard decays, because of the zero mass of the photon. This is true even for three-particle decays into a lighter lepton and $l^+l^-$ pair. While the small mass gap case occurs for a pair of neutralinos in each of the benchmark scenarios chosen, the radiative decay does not play an important role because  the production of that particular neutralino pair is subdominant.

In the following section, we discuss the specific differences in the decay signatures between $U(1)^\prime$ and the MSSM for each benchmark set.

\section{Charginos and Neutralinos in $U(1)^{\prime}$ at the LHC  \label{sec:num} }

\subsection{$U(1)^\prime$ Benchmark Points and Relic Density}

Charginos and neutralinos, once produced, will
decay following a pattern dictated by the benchmark
 parameters of the model.  These scenarios would give definite predictions for the production and abundance of the lightest neutralino, assumed here to be the LSP. We proceed by evaluating the relic density of the lightest neutralino in the model, and subject it to the constraints from WMAP of cold  dark matter.

 For this task we specify  three benchmark
scenarios for the secluded $U(1)^\prime$, denoted as Scenario A,  Scenario B and Scenario C,  by fixing the
additional parameters to agree with phenomenological constraints on
masses \cite{Battaglia:2003ab}.

 Finding an acceptable minimum of the Higgs potential is highly nontrivial even at the tree
level. Requiring the tadpole conditions and positive-definiteness of the squared
masses of the Higgs bosons, the global minimum is shifted from v$\ne 246$ GeV,
due to the presence of the Higgs singlets in the Higgs potential. The procedure is roughly the following: first soft SUSY breaking masses and trilinear couplings are taken at arbitrary values. After a minimum is found, all dimensionful parameters are rescaled so that the minimum occurs at v$=246$ GeV. This procedure determines the Higgs VEVs through tadpole conditions, as well as $\tan \beta$ \cite{Erler:2002pr}. Alternatively, one can start by fixing the Higgs VEVs, and then looking for minimum \cite{Chiang:2008ud}. The desired minumum does not always exist.  We rely on previously established benchmark scenarios  \cite{Erler:2002pr,Chiang:2008ud}, which satisfy all the theoretical and experimental requirements, and in particular  generate correct $Z^\prime-Z$ mass hierarchy and  a normal sparticle spectra (squark and slepton similar to that in MSSM, with additional particles in the chargino/neutralino spectrum), acceptable effective $\mu$ parameter, and avoid unwanted global symmetries.

 The benchmark points were required to obey three important conditions:
\begin{itemize}
\item The scenarios chosen had to insure the stability of the vacuum, as in \cite{Erler:2002pr,Chiang:2008ud};
\item The points had to satisfy relic density constraints for the LSP, the lightest neutralino;  and
\item Of the parameter points satisfying the above two conditions,  benchmarks were chosen to enhance some signals of the model in neutralino and chargino decays.
\end{itemize}
 Of possible choices, we selected scenarios where the singlinos are light, to highlight characteristics of the $U(1)^\prime$ model, as discussed in the previous subsection (\ref{app:C}). The three benchmark scenarios  are
given in Table~\ref{tab:inputs}. We show VEVs, Yukawa couplings, trilinear couplings, mass ratios and mixings for the gauginos and bare scalar fermion masses.   The parameters for Scenario A are based on \cite{Erler:2002pr}, while for Scenario B and C they are loosely based on Case II in \cite{Chiang:2008ud}.
We varied the parameters slightly to insure that in addition to the constraints above, in each scenario we obtain a light CP-even SM-like Higgs boson with (tree-level) mass $m_{h^0}\approx 125$ GeV.  Note also that the low value of $\tan \beta \approx 1$ is favored by  constraints from $B_s \to \mu^+ \mu^-$ branching ratio \cite{:2012ct}. This branching ratio, proportional to $\displaystyle \frac{(\tan \beta)^6}{m_{A^0}^4}$ in MSSM, does not show significant deviations from the SM prediction, so it favors regions of low $\tan \beta$'s and heavier pseudoscalar ${A^0}$.  In  Scenario A,  the two lightest pseudoscalar bosons are very light, but these are both singlets \cite{Erler:2002pr} which do not couple directly to quarks and leptons \cite{Ali:2009md}, and thus the bound for the MSSM-like  pseudoscalars does not apply. A complete resolution of this problem is beyond the scope of this work and is dealt with  in \cite{Frank:2013yta}. For the purpose of this analysis however, we note that a definite conclusion at this point may be premature, as a comprehensive recent analysis still allows for sizable contributions from BSM \cite{Buras:2013uqa}.

For each benchmark scenario, the
mass spectra for the supersymmetric partners obtained are given in Table~\ref{tab:spec}. The mass of the additional $Z^\prime$ boson is 
\begin{equation}
M_{Z^\prime}^2=g_{Y^\prime}^2 \left (Q_{H_d}^{\prime\, 2}v_d^2+ Q_{H_u}^{\prime\, 2}v_u^2+ Q_{S}^{\prime\, 2}v_s^2+ \sum_{i=1}^3Q_{S_{i}}^{\prime\, 2}v_{s_{i}}^2\right ),
\end{equation} 
and is equal to $M_{Z^\prime}=2015.8$ GeV for Scenario A, $M_{Z^\prime}=1414.7$ GeV for Scenario B and $M_{Z^\prime}=1412.4$ GeV for Scenario C. 
As seen from Table~\ref{tab:inputs}, the VEVs of the additional
scalars ($S_1,S_2$ and $S_3$) $v_{s_i}, i=1,2,3$ are mostly taken
above the TeV scale so that the $Z^\prime$ mass bound is satisfied
no matter what the VEV of the scalar field $S$ is chosen.
For convenience, the parameters $\mu_{eff}$ and $h_s$ are taken as
free parameters and the VEV of $S$ is determined accordingly using
the relation
\begin{eqnarray}
\hspace{0.5cm} \mu_{eff} =\frac{h_{s} \langle S \rangle}{\sqrt{2}}.
\hspace{0.5cm} \label{eq:rho}
\end{eqnarray}

\begin{table*}[htbp]
\setlength{\voffset}{-0.5in}
 \begin{center}
\caption{\label{tab:inputs}\sl\small The benchmark points for the
$U(1)^{\prime}$ model: Scenario A, Scenario B and Scenario C.}
\setlength{\extrarowheight}{-5.8pt} \small
\begin{tabular*}{0.99\textwidth}{@{\extracolsep{\fill}} ccccccc}
\hline\hline
 $\rm Parameters$ & $\rm Scenario~ A$  &$\rm Scenario
~B$ &$\rm Scenario ~C$&
 \\ \cline{1-6}\cline{1-7}
 $\tan\beta$&$1.01$&$1.175$&$
 1.175$\\
 $Q_{Q}^{\prime}$&-2&0&0\\
 $\mu(\mu_{eff})$&139.05&282.8&265\\
 $h_{\nu}$&1&1&1\\
 $h_s$&0.75&0.8&0.8\\
 $\bar{h}_s$&0.073&0.1&0.1\\
 $A_s$&195.5&522&490\\
 $A_{\bar {s}}$&195.5&522&490\\
 $v_{s_1}$&1782.4&100&100\\
 $v_{s_2}$&1782.4&3000&3000\\
 $v_{s_3}$&1778.1&100&100\\
 $R_{Y^\prime}$&$12$&0.8&5\\
 $R_{Y Y^\prime}$&$10$&8&4.8\\
 $M_{\tilde \nu{_e{_R}}}$&$600$&1700&1700\\
 $M_{\tilde \nu{_\mu{_R}}}$&$650$&1750&1750\\
 $M_{\tilde \nu{_\tau{_R}}}$&$700$&1800&1800\\
 $M_1$&-100&100&-400\\
 $M_2$&-800&700&212\\
 $M_3$&1000&1000&1000\\
 $M_{L_1}$&250&600&573\\
 $M_{E_1}$&260&300&300\\
 $M_{Q_1}$&950&1000&1000\\
 $M_{U_1}$&900&1900&1900\\
 $M_{D_1}$&890&1200&1200\\
 $M_{L_2}$&250&600&573\\
 $M_{E_2}$&260&300&300\\
 $M_{Q_2}$&950&1000&1000\\
 $M_{U_2}$&900&1900&1900\\
 $M_{D_2}$&890&1200&1200\\
 $M_{L_3}$&240&575&573\\
 $M_{E_3}$&250&275&275\\
 $M_{Q_3}$&850&1400&1400\\
 $M_{U_3}$&800&2100&2100\\
 $M_{D_3}$&880&1500&1500\\
 $M^2_{SS_{1}}$&$-382.3$&$(306)^2$&$(306)^2$\\
 $M^2_{SS_{2}}$&$-382.3$&$(56)^2$&$(56)^2$\\
 $M^2_{S_{1}S_{2}}$&0&$0$&$0$\\
 $A_t$&-697.75&-697.75&-697.75\\
 $A_b$&-959.66&-959.66&-959.66\\
 $A_\tau$&-138.7&-138.7&-138.7\\
\hline\hline
\end{tabular*}
\end{center}
 \end{table*}
 
\begin{table*}[htbp]
\begin{center}
\caption{\label{tab:spec}
The  mass spectra for the supersymmetric sector and the relic
density $\Omega_{\rm DM}$ values of the
benchmark points given in Table~\ref{tab:inputs} for the
secluded $U(1)^\prime$. The tree-level values of the masses for the light CP-even Higgs bosons are included.}
\setlength{\extrarowheight}{-5.8pt} \small
\begin{tabular*}{0.99\textwidth}{@{\extracolsep{\fill}} ccccccc}
\hline\hline
 $\rm Masses$ & $\rm Scenario~ A$  &$\rm Scenario~
B$ &$\rm Scenario ~C$&
 \\ \cline{1-6}\cline{1-7}
 $m_{\tilde\chi^0_1}$&72.1&50.9&56.9\\
 $m_{\tilde\chi^0_2}$&78.5&71.5&154.6\\
$m_{\tilde\chi^0_3}$&94.2&211.4&154.9\\
 $m_{\tilde\chi^0_4}$&151.7&212.5&211.4\\
 $m_{\tilde\chi^0_5}$&188.9&278.8&212.7\\
 $m_{\tilde\chi^0_6}$&217.5&339.6&318.7\\
 $m_{\tilde\chi^0_7}$&806.7&714.7&324.5\\
 $m_{\tilde\chi^0_8}$&1771.9&1577.4&1435.7\\
 $m_{\tilde\chi^0_9}$&2901.3&1673.9&3654.1\\
 $m_{\tilde\chi^\pm_1}$&145.8&268.1&154.6\\
 $m_{\tilde\chi^\pm_2}$&806.7&714.7&322.5\\
 $m_{\tilde e_L}$&259.1&217.3&120.3\\
 $m_{\tilde e_R}$&249.5&1155.7&1157.5\\
 $m_{\tilde \mu_L}$&259.1&217.3&120.3\\
 $m_{\tilde \mu_R}$&249.5&1155.7&1157.5\\
 $m_{\tilde \tau_1}$&239.0&133.7&120.3\\
 $m_{\tilde \tau_2}$&249.5&1149.4&1151.3\\
 $m_{\tilde\nu_e}$&258.9&215.0&116.0\\
 $m_{\tilde\nu_{\mu}}$&258.9&215.0&116.0\\
 $m_{\tilde\nu_{\tau}}$&249.3&129.8&116.0\\
 $m_{\tilde\nu_{e_R}}$&597.9&643.3&636.6\\
 $m_{\tilde\nu_{\mu_R}}$&648.1&765.7&760.1\\
 $m_{\tilde\nu_{\tau_R}}$&698.2&874.0&869.0\\
 \hline \hline
$m_{h^0}$&125.9&125.6&126.5\\
 \hline \hline
$\Omega_{\rm DM}h^2$ &0.102 & 0.114 &0.106&\\
 \hline\hline
\end{tabular*}
\end{center}
 \end{table*}

The parameters in the supersymmetric sectors for each scenario has been chosen as follows. In Scenario A, as seen in  Table~\ref{tab:spec},
both left
and right scalar leptons are light and close in mass, but the NLSP ${\tilde \chi}_2^0$, and the lightest chargino ${\tilde \chi}^\pm_1$, are lighter than the sleptons. This favors decays into LSP and $W^\pm$ or $h$ \footnote{The Higgs sector parameters can be fine-tuned and do not affect the specific calculations in this paper.}
and a reduced yield of leptons compared to the case where the
 two body decays of neutralinos into either mass-shell scalar
leptons is open.  Scenario A has six light neutralinos (below 500 GeV), to
highlight the spectrum and signal outcomes from additional neutralinos at the LHC. The fourth
neutralino and lightest chargino are close in mass. In this
scenario, dominant decays will be into chargino-neutralino pairs.  In
Scenario B, the right  scalar leptons are heavy, but the lightest chargino ${\tilde \chi}^\pm_1$ is heavy enough to decay through the left-handed slepton, while the NLSP ${\tilde \chi}^0_2$ decays through three-body decays to ${\tilde \chi}^0_1 l^+ l^-$.
 In this scenario, the
dominant decays will be into neutralino pairs, though there would be one important chargino-neutralino associated production channel. In Scenario C, both the NLSP ${\tilde \chi}^\pm_2$ and the lightest chargino ${\tilde \chi}^\pm_1$ are heavier than the sleptons and the sneutrinos, allowing for two body decays ${\tilde \chi}^0_2 \to  {\tilde l}^\pm l^\mp$ and ${\tilde \chi}^\pm_1 \to {\tilde l}^\pm \nu, {\tilde \nu}  l^\pm$,  and yielding a significant number of leptons in the final state. This scenario
has been designed to maximize the  $\threelep$ signal. The production cross section is dominated by the NLSP plus the lightest chargino.  Another  significant difference between Scenarios A,  B, and Scenario C, is that in A and B, the $U(1)$  bino mass $M_1$ is the lightest, while in C the wino mass $M_2$ is lighter. This insures that the NLSP has a significant wino component, maximizing the decay into $\threelep$.

We give the values for the lightest SM-like Higgs boson masses for all three scenarios.
  We also include the values of the relic density in  Table~\ref{tab:spec}, together with the LSP, the lightest neutralino $\lsp$, with
masses 72.1 GeV, 50.9 GeV and 56.9 GeV,  for Scenario A, Scenario B
and Scenario C,  respectively.
The calculation of the relic density is performed including the model files from
 {\tt CalcHEP}  \cite{calchep}
into  the {\tt MicrOmegas} package \cite{Belanger:2008sj}. All the numbers obtained are within the $1\sigma$ range of the WMAP
result \cite{wmap}  obtained from the Sloan
Digital Sky Survey \cite{Spergel:2006hy}
 \begin{eqnarray}
\Omega_{DM} h^2 = 0.111^{+0.011}_{-0.015}\,.
\end{eqnarray}
The relic density of dark matter $\Omega_{\rm DM} h^2$ is
very sensitive to the parameter $\displaystyle R_{Y^\prime}\equiv
M_{\widetilde{Y}^{\prime}}/M_{\widetilde{Y}}$ from
Table~\ref{tab:inputs}.

The composition of the physical neutralino states $\widetilde{\chi}_i^0,
i=1,2,...,9$ from Table~\ref{tab:spec}, in terms of the bare bino, wino, bino', higgsino and singlino
components of the states in the Lagrangian is given in Table~\ref{tab:neut_comps}, for Scenarios A,  B and  C. This table shows clearly differences between the three scenarios in neutralino compositions. For instance, in Scenario A, the LSP is mostly bino $\tilde B$, while the NLSP is 73\% singlino, with an admixture of $\tilde H_u$ and $\tilde H_d$.  In Scenario B the LSP is 62\% singlino $\tilde S$ (with $\tilde H_u, \tilde B$ and $\tilde S_2$ admixtures), and the NLSP is 58\% bino, with a mixtures of $\tilde S$ and $\tilde S_2$ and $\tilde H_d$. In Scenario C the LSP is 80\% singlino $\tilde S$, while the NLSP is mostly wino, with a small admixture of $\tilde H_u$ and $\tilde H_d$. Thus in all the scenarios chosen, one or more of the singlinos are light to highlight differences with the MSSM spectrum.
\begin{table*}[htbp]
\caption{\label{tab:neut_comps}\sl\small The bino, wino, bino', higgsino
and singlino composition of the neutralinos $\widetilde{\chi}_i^0,
i=1,2,...,9$ for Scenario A, Scenario B and Scenario C. }
  \begin{center}
    \setlength{\extrarowheight}{-2.0pt}
 \small
 \begin{tabular*}{0.99\textwidth}{@{\extracolsep{\fill}} cccccccccc}
 \hline\hline
  $\rm Scenario~ A$&$\tilde\chi^0_1$&$\tilde\chi^0_2$
&$\tilde\chi^0_3$&$\tilde\chi^0_4$&$\tilde\chi^0_5$&$\tilde\chi^0_6$&
$\tilde\chi^0_7$&$\tilde\chi^0_8$&$\tilde\chi^0_9$\\
  \hline\hline
  $\tilde B$&0.889  &-0.004  &0.0 &-0.151  &0.0 &0.004 &-0.007 &0.324 &0.283 \\
  $\tilde W^3$&0.022  &0.0  &0.0 &0.081  &0.0 &0.0 &0.996 &-0.002 &0.001\\
  $\tilde H_{d}^0$ &0.131 &-0.360  &0.0 &0.692  &0.0 &0.607 &-0.059&-0.035&-0.018 \\
  $ \tilde H_{u}^0$ &-0.156 &-0.365  &0.0 &-0.682  &0.0 &0.605 &0.059 &0.065&0.039  \\
  $ \tilde S$ &0.025  &0.855  &0.0&-0.013  &0.0 &0.514 &0.0 &-0.042&-0.032\\
  $ \tilde B^{\prime}$ &-0.033  &0.0  &0.0 &-0.004  &0.0 &0.0 &-0.001 &-0.604&0.795   \\
  $\tilde S_{1}$ &-0.165  &0.027  &-0.707 &0.065  &0.577 & 0.001& -0.001&0.295&0.217 \\
  $\tilde S_{2}$ &-0.165  &0.027  &0.707 &0.065  &0.577 &0.001 &-0.001 &0.295&0.217   \\
  $\tilde S_{3}$ &0.331  &-0.055  &0.0 &-0.130 &0.577 &-0.003 &0.002 &-0.589&-0.434  \\
  \hline
  $\rm Scenario~ B$&& &&&&&&&\\
  \hline
  $\tilde B$       &0.349&0.764&0.042&0.007&-0.220 &-0.011&-0.021 &-0.336&0.359 \\
  $\tilde W^3$     &-0.017&-0.017&-0.002&0.0&-0.180 &0.005&0.983 &0.002 &0.006 \\
  $\tilde H_{d}^0$ &-0.142&0.246&0.007& 0.0&0.684 &0.658&0.124 &-0.032&0.032 \\
  $ \tilde H_{u}^0$&-0.312&0.0246&-0.011&-0.002&-0.669 &0.651&-0.131&0.072 &-0.077 \\
  $ \tilde S$      &0.790 &-0.458&0.008 &0.002&-0.041 &0.377 &-0.003 &-0.100  &0.099\\
  $ \tilde B^{\prime }$     &-0.013&-0.019&0.006&-0.002 &0.002 &0.002&-0.006 &0.701&0.712  \\
  $\tilde S_{1}$  &-0.016&-0.020&0.706&0.706 &-0.002 &0.0&0.0 &0.018&-0.022  \\
  $\tilde S_{2}$   &0.365&0.377&-0.043&0.028 &-0.014 &0.010&0.012 &0.613&-0.586  \\
  $\tilde S_{3}$   &-0.005&-0.001&-0.704&0.707 &0.003 &0.0&-0.001 &-0.035&0.044  \\
  \hline
  $\rm Scenario~ C$&& &&&&&&&\\
  \hline
$\tilde B$       &0.016&-0.011&-0.639&-0.023&-0.040&0.017&0.005&0.596&0.481 \\
  $\tilde W^3$     &-0.056&-0.801&0.024&-0.027&0.001&0.593&0.010&-0.007&0.0 \\
  $\tilde H_{d}^0$ &-0.217&0.442&-0.049&-0.002&-0.002&0.568&0.654&-0.052&-0.011 \\
  $ \tilde H_{u}^0$&-0.329&-0.378&0.104&-0.002&0.004&-0.556&0.645&0.097&0.032 \\
  $ \tilde S$      &0.903&-0.096&-0.090&0.010&-0.003&-0.048&0.392&-0.087&-0.047\\
  $ \tilde B^{\prime}$     &-0.006&-0.009&0.084&0.007&0.004&-0.024&0.001&-0.568&0.817  \\
  $\tilde S_{1}$  &-0.007&-0.014&-0.042&0.706&0.705&0.013&0.0&0.021&0.009 \\
  $\tilde S_{2}$   &0.157&0.088&0.744&-0.048&0.073&0.107&0.005&0.546&0.308 \\
  $\tilde S_{3}$   &-0.001&0.009&-0.074&-0.704&0.703&-0.018&0.0&-0.042&-0.019  \\
  \hline
    \hline
   \end{tabular*}
\end{center}
 \end{table*}

The production cross sections for the scattering $pp \to
{\tilde\chi_i\tilde\chi_j}$ processes at the LHC with $\sqrt{s}=14$ TeV are shown in
Table~\ref{tab:crsrlc} for three benchmark scenarios of the secluded
$U(1)^\prime$ model. The values were obtained implementing the
secluded $U(1)^\prime$ model into {\tt CalcHEP} \cite{calchep} with
the help of {\tt LanHEP} \cite{Semenov:2008jy}. The parton
distributions have been parametrized by using {\tt
CTEQ6M} of {\tt LHAPDF} \cite{Whalley:2005nh}.  For background calculations, including SM backgrounds and QCD corrections,  we have used {\tt Pythia8.150} \cite{Sjostrand:2007gs}. We outline  the distinctive features of each benchmark scenario.

\begin{table*}[htbp]
  \begin{center}
\caption{\sl \small Total cross sections for production of
$\tilde\chi_{i}^0\tilde\chi_{j}^0$,
$\tilde\chi_{i}^0\tilde\chi_{j}^{\pm}$  and
$\tilde\chi_{i}^{+}\tilde\chi_{j}^{-}$ at the LHC with $\sqrt{s}=14$ TeV  for the three scenarios considered.}
 \label{tab:crsrlc}
 \begin{tabular*}{0.99\textwidth}{@{\extracolsep{\fill}} ccccccc}
\hline\hline
 $\rm Observables $ &  $\rm Scenario~ A$  &$\rm Scenario~ B$ &$\rm Scenario~
 C$&
 \\ \cline{1-7}\cline{3-7}
$\rm \sigma(pp\rightarrow \tilde\chi^0_1\tilde\chi^0_2
)/fb$ &238 & 628& $< 10$ \\
$\rm \sigma(pp\rightarrow \tilde\chi^0_1\tilde\chi^0_6
)/fb$&$< 10$ & 169& $< 10$ \\
$\rm \sigma(pp\rightarrow \tilde\chi^0_2\tilde\chi^0_2
)/fb$ &55 & $< 10$& $< 10$\\
$\rm \sigma(pp\rightarrow \tilde\chi^0_2\tilde\chi^0_4
)/fb$ &153 & $< 10$ & $< 10$\\
$\rm \sigma(pp\rightarrow \tilde\chi^0_3\tilde\chi^0_4
)/fb$ &$< 10$ & 1146& $< 10$ \\
$\rm \sigma(pp\rightarrow \tilde\chi^0_4\tilde\chi^0_6
)/fb$ &225 & $< 10$& $< 10$ \\
$\rm \sigma(pp\rightarrow \tilde\chi^0_5\tilde\chi^0_6
)/fb$ &$< 10$ & 780& $< 10$ \\
$\rm \sigma_{TOT}(pp\rightarrow \tilde\chi^0_i\tilde\chi^0_j
)/fb$ & $<743>$ &$<4827>$ & $< 10$ \\
\cline{1-7} \hline
   $\rm \sigma(pp\rightarrow \tilde\chi^0_2\tilde\chi^{\pm}_1
)/fb$ &279 & $< 10$&2170  \\
$\rm \sigma(pp\rightarrow \tilde\chi^0_4\tilde\chi^{\pm}_1
)/fb$&1037 & $< 10$& $< 10$ \\
$\rm \sigma(pp\rightarrow \tilde\chi^0_5\tilde\chi^{\pm}_1
)/fb$&$< 10$ & 113& $< 10$ \\
$\rm \sigma(pp\rightarrow \tilde\chi^0_6\tilde\chi^{\pm}_1
)/fb$ &369 & 62& $< 10$ \\
$\rm \sigma_{TOT}(pp\rightarrow \tilde\chi_{1}^{\pm}\tilde\chi_{i}^0
)/fb$ &$<1739>$ &$<235>$&<2368> \\
\hline
 $\rm \sigma(pp\rightarrow \tilde\chi^{+}_1\tilde\chi^{-}_1
)/fb$ &693 & 1120&$< 10$ \\
$\rm \sigma_{TOT}(pp\rightarrow \tilde\chi_{i}^{+}\tilde\chi_{j}^{-}
)/fb$ &$<694>$ &$<1166>$&$< 10$ \\
\hline\hline
\end{tabular*}
\end{center}
\end{table*}
 
The total cross
sections in Scenario A are of the order 1 $pb$ for $pp \to
{\tilde \chi}_4^0 {\tilde \chi}_1^{\pm}$ and large  for, in order,
$pp \to {\tilde \chi}_1^{\pm} {\tilde \chi}_1^{\mp},   {\tilde \chi}_6^0 {\tilde \chi}_1^{\pm},~ {\tilde
\chi}_2^0{\tilde \chi}_1^{\pm},~ {\tilde \chi}_1^0 {\tilde
\chi}_2^0,~ {\tilde \chi}_4^0{\tilde \chi}_6^0$ and $pp \to {\tilde
\chi}_2^0 {\tilde \chi}_4^0$ (hundreds of $fb$).  The dominant chargino-neutralino decay into ${\tilde \chi}_1^{\pm} {\tilde \chi}_4^0 $  is MSSM-type  into ${\tilde W}^\pm$, and a maximal mixture of ${\tilde H}_u$ and ${\tilde H}_d$. The ${\tilde \chi}_4^0$ decays further through a pseudoscalar Higgs, on- or off-mass shell. The next significant chargino-neutralino decay is into  ${\tilde \chi}_6^0 {\tilde \chi}_1^{\pm}$, where ${\tilde \chi}_6^0 $ is an almost even mixture of  ${\tilde H}_u$,  ${\tilde H}_d$ and $\tilde S$, with the singlino admixture reducing the production cross section by a factor of 3. The neutralino ${\tilde \chi}_6^0$ can decay through $Z$ and/or $H$ bosons.
The cross section for  ${\tilde \chi}_2^0{\tilde \chi}_1^\pm $ is reduced even further, as the ${\tilde \chi}_2^0$ state is mostly $\tilde S$ with a small admixture of  ${\tilde H}_u$ and ${\tilde H}_d$. Here the neutralino ${\tilde \chi}_2^0$ can decay further through scalar leptons, $Z$ and $H$ bosons (when kinematically allowed).

In Scenario B the
dominant decays are into neutralino pairs, again in order: $pp \to {\tilde
\chi}_3^0 {\tilde \chi}_4^{0},~ {\tilde \chi}_5^0{\tilde
\chi}_6^{0},~ {\tilde \chi}_1^0 {\tilde \chi}_2^0,~ {\tilde
\chi}_1^0{\tilde \chi}_6^0$, while the decays into charginos are
dominated by   $pp \to {\tilde \chi}_1^\pm {\tilde \chi}_1^\mp$.  The dominant production here is into  non-MSSM channel ${\tilde
\chi}_3^0 {\tilde \chi}_4^{0}$ with ${\tilde \chi}_3^0$ and ${\tilde \chi}_4^0$ neutralino both (orthogonal) maximal combinations of singlinos, ${\tilde S}_1$ and ${\tilde S}_2$. The following dominant two neutralino channels  are into ${\tilde \chi}_5^0$  (which is a mixture of ${\tilde H}_u$,  ${\tilde H}_d$), and ${\tilde \chi}_6^0$ (a mixture of ${\tilde H}_u$,  ${\tilde H}_d$, with 10\% singlino $\tilde S$ component).

In both Scenarios A and B, the ${\tilde \chi}_1^{\pm} {\tilde \chi}_1^{\mp} $ chargino pair-production is significant, and in Scenario B it competes with the largest neutralino pair production.

In
Scenario C, the dominant decay is $pp \to {\tilde \chi}_2^0 {\tilde
\chi}_1^{\pm}$ while all others are negligible. The ${\tilde \chi}_2^0$ is mostly gaugino $\tilde W^3$, with a significant ${\tilde H}_u$ and  ${\tilde H}_d$ admixture, and it can further decay through sleptons, while the chargino ${\tilde \chi}_1^\pm$ is wino-like, and can decay through $W, ~{\tilde l}_L$ or ${\tilde \nu}_L$.

To sum up, cross
sections in Scenario A are dominated by chargino-neutralino
production, in Scenario B  by
 neutralino pair production, while for both scenarios the cross section for lightest chargino pair production is large. Scenario C is dominated by a single
chargino-neutralino decay ${\tilde \chi}_2^0{\tilde \chi}_1^{\pm}$,
chosen to enhance the three-lepton signal.

The decay channels of heavy neutralinos depend on their masses and
the masses and couplings of other sparticles and Higgs bosons. A
sufficiently heavy neutralino can decay via tree-level two-body
channels containing a Z (W), or a Higgs boson,  a lighter
neutralino, (chargino)  yielding a sfermion-fermion pair. The main decay modes for the charginos and neutralinos in each scenario are given schematically in the Feynman diagrams of  \ref {app}.

Finally, we comment on the exotics predicted by the model. Although they do not directly affect the spectrum of charginos and neutralinos, in all three scenarios the exotic quarks and leptons, ${\cal Q}_i, {\cal L}_i$ (vector-like under MSSM, chiral under $U(1)^\prime$),  required to cancel anomalies,  are predicted to be light, and are a feature of this model. Their masses are
 \begin{equation}
 m_{\cal Q}=\frac {h_{Q}v_s}{\sqrt{2}}, \qquad m_{\cal L}=\frac {h_{ L}v_s}{\sqrt{2}}, \nonumber
 \end{equation}
 and thus in scenarios A, B and C, they can be as light as 100-300 GeV. These exotic quarks and leptons do not have a Yukawa coupling to the doublet Higgs and, as they are vector-like under MSSM,  they do not enhance the observed Higgs production cross section, assuming that the lightest Higgs boson is SM-like. They would affect production of the heavier neutral Higgs boson, and might contribute to CP mixing between the heaviest scalar and the pseudoscalar Higgs bosons \cite{Ham:2007kc}. But definite constraints on their masses and couplings would come only from direct searches. So far limits have assumed that the exotic quarks ${\cal Q}_i$ and their superpartners ${\tilde {\cal Q}}_i$ can be pair-produced at the LHC by QCD-processes, and then decay into ${\cal Q} \to tW,\, bZ$ and ${\cal Q} \to bH_0$, if driven by mixing with a third generation quark of the same charge. The current limit on the mass of such a quark is $m_{\cal Q}>590$ GeV \cite{ATLAS-CONF}, in apparent conflict with the masses in the model discussed here. However, if we justify the $U(1)^\prime$ model as being obtained from the breaking of $E_6$, a mixing with ordinary SM fermions is forbidden in supersymmetric $E_6$ if $R$-parity is conserved \cite{Langacker:2008yv}, which we  assume here\footnote{The branching ratios into  $ tW,\, bZ$ and $ bH_0$ for this analysis are assumed to be 42\%, 31\% and 27\%, respectively, for $m_{\cal Q}=500$ GeV, and the mass restrictions depend crucially on the assumed branching ratios. For reduced ratios, as in our parameter space, the limits disappear.}. The cross sections for the scalar partners are one order of magnitude smaller, and smaller also for the exotic leptons. (Note also that the masses of the superpartners of the exotic fermions is determined by the soft  masses $m^2_{\tilde D}$ and $m^2_{\tilde D^c}$, which are not constrained to be small.) Thus these limits  from ATLAS do not apply here. 
 
 An alternative would be that such exotic quarks could be stable at the renormalizable level due to the $U(1)^\prime$ symmetry, or another accidental symmetry  \cite{Kang:2007ib}. They would decay through higher-dimensional operators, on a time scale short enough to avoid cosmological problems \cite{Kawasaki:2004qu}, involving singlets under SM with VEVs which would induce extremely small mixings with ordinary quarks. Mass limits on such stable charged particles exist, but only for lepton-like particles  with $|Q_D|=e/3$ produced in Drell-Yan processes. Their masses  are constrained to be $m_{D}>200$ GeV \cite{Chatrchyan:2013oca}. Specific examples of such exotic quarks and squarks from $E_6$ appear in \cite{Langacker:2008yv}. See also \cite{Ham:2009bu} for an alternative analysis of the effects of exotic quarks.

\subsection{Chargino and Neutralino  Signals at the LHC }

\begin{figure*}[htbp]
\begin{center}$
    \begin{array}{c}
      \includegraphics[width=4.8in,height=0.8in]{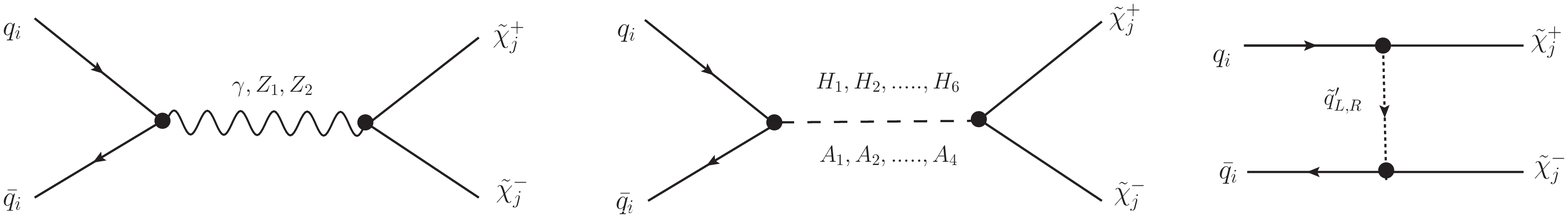}\\
    \includegraphics[width=4.8in,height=0.8in]{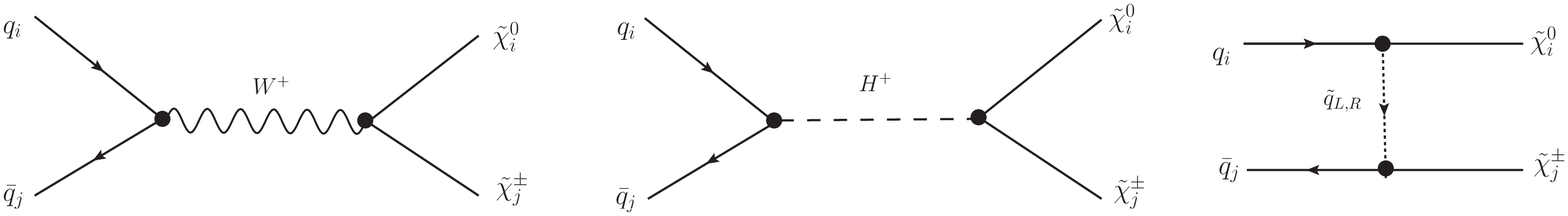}\\
    \includegraphics[width=4.8in,height=0.8in]{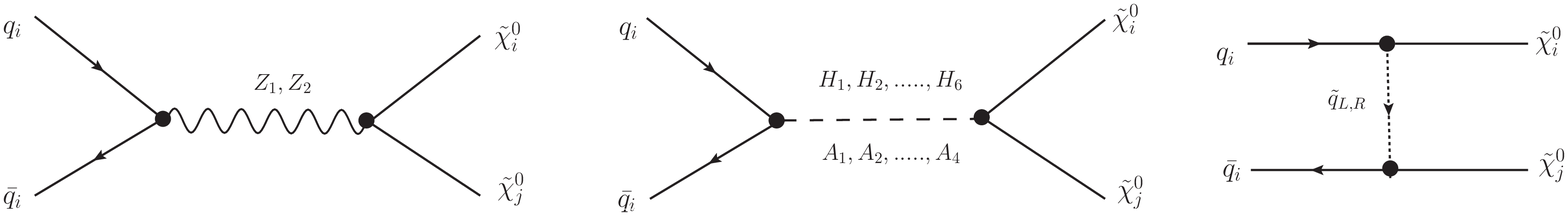} \\
    \end{array}$
\end{center}
\vskip -0.1in
      \caption{The Feynman diagrams for the production of the
chargino and neutralino in the secluded $U(1)^\prime$ model. The top
row shows chargino pair production only, the middle row the associated
chargino-neutralino, and the bottom row shows the neutralino pair
production. } \label{fig:Feyndiagrams}
\end{figure*}

After defining the benchmark points for $U(1)^\prime$, describing the basic features of production and decay processes, and calculating the the relic
density, we proceed to analyze the neutralino and chargino
signals at LHC.
Fig. \ref{fig:Feyndiagrams}   shows the Feynman diagrams contributing to chargino and neutralino production in the secluded $U(1)'$ model. We leave the diagrams for the characteristic decay patterns in the three scenarios for \ref{app}.

To determine and analyze all possible signals for the three scenarios we
need to look at the decay topology of these particles, and classify
signals according to the final number of leptons present in the signal
events.
We impose the following basic cuts  to suppress the SM background,
where relevant. We call the set {\it cut-1}:
\begin{itemize}
\item(i) Each isolated charged lepton (electron or muon) has a minimum transverse momentum $p_T(\ell)>15$ GeV;
\item(ii) The missing transverse energy must be larger than $\EmissT> 100$ GeV;
\item (iii) If two leptons or more are produced, they are constrained to be in the central region by the condition on pseudorapidity $|\eta|<2$ (and the same condition holds for the lepton in the single lepton channel);
\item (iv) The cone size between two charged leptons $\Delta R_{\ell \ell} \ge 0.4$, where $\Delta R_{\ell \ell}$ is defined in the pseudorapidity-azimuthtal angle plane as $\Delta R_{\ell \ell}=(\Delta \eta^2+ \Delta \phi^2)^{1/2}$.
\end{itemize}

\begin{table*}[htbp]
\caption{\sl\small The SM background cross sections in $fb$
after {\it cut-1} and {\it cut-2a(b)} for Scenarios A, B (C), defined in the text. The ``$-$'' along the columns indicates that, for the given process,  there is no background for that channel.}
 \label{tab:cuts-bckgr}
\begin{tabular*}{0.97\textwidth}{@{\extracolsep{\fill}} ccccc}
\hline\hline
$\rm Background\,[fb]$& Cut &$1\ell 2j \EmissT$&$2\ell 2j \EmissT$&$3\ell 0j \EmissT$\\
\hline\hline
$\rm W\, jet$ &{\it cut-1}&$21288.0 $&$-$ &$-$\\
&{\it cut-2}&$ 134.0$&$-$  &$-$\\
$\rm Z\, jet$ &{\it cut-1}&$84.0  $&$28.0$ &$-$\\
&{\it cut-2}&$1.75 $& $1.75$&$-$\\
$\rm ZZ$ &{\it cut-1}&$8.4 \times 10^{-3}$&4.3&$-$\\
&{\it cut-2}&$9.3 \times 10^{-5}$&$1.7 \times 10^{-4}$ &$-$\\
 $\rm WW$&{\it cut-1}&156.9&11.2&$-$\\
&{\it  cut-2}&1.4&0.3&$-$\\
  $\rm WZ$&{\it  cut-1}&98.3&$2.1$&4.5\\
&{\it cut-2}&0.4&$<1.5\times 10^{-3}$&0.4\\
 $\rm t\bar t$&{\it  cut-1}&2502.9&205.5&$-$\\
&{\it cut-2}&6.7&$5.77$&$-$\\
\hline
$\rm Total:$&{\it cut-1}&24130.1&$251.1$&4.5\\
&{\it cut-2}&144.2&$7.82$&0.4\\
\hline\hline
\end{tabular*}
 \end{table*}

The  contributions to
the background in the signal regions come from SM processes. In
the $\onelep$ case, the background arises from  $pp\rightarrow\bar tt, \,W+$jets, $Z+$jets and di-bosons ($WW+WZ+ZZ)$.
For the $\twolep$ mode, $pp\rightarrow\bar tt, \, WW,\, WZ, \,ZZ, \, Z$+jets yield the dominant background. And the process $pp\rightarrow WZ$ is
the background for the $\threelep$ decay mode.
 We found that further cuts are  needed to reduce the SM background more. As the topology of signal and background events is somewhat similar after
first level selection cuts, the difference in the angular distributions and circularity is
not significant either, it is thus not very useful to apply cuts on these variables too.
Also the difference in the lepton $p_T^\ell$ distributions is not very pronounced as signal
leptons are produced in cascade decays, thus loosing information about
the original process which created them. Thus we modified only the missing transverse energy $\EmissT$ cut as compared to the first set.  The reason for this cut is to select events with two LSP, acting as a large source of $\EmissT$, over events where the $\EmissT$ comes from the neutrinos, which can be produced with high energy from the decay of a $W$ or $Z$ boson.  We call this {\it cut-2}, and distinguish between scenarios:
\begin{itemize}
\item $\EmissT> 500$ GeV for the Scenarios A and B, {\it cut-2a}, and
\item $\EmissT> 200$ GeV for the scenario C, {\it cut-2b}.
\end{itemize}
In addition, one could require high jet multiplicity cuts, as the production cross section for $W+$jets decreases as the number of jets increases. But we found that these cuts could reduce the signal as well, and that the above cut sufficient to eliminate most the unwanted background. In Table \ref{tab:cuts-bckgr} we list the SM background contributions (given along rows) to the cross sections of the signals (given in between the $4^{\rm th}$ to $6^{\rm th}$ columns of Table \ref{tab:cuts-sign}) after the {\it cut-1} set is imposed, including as well as the numbers after a second more restrictive set, called {\it cut-2}, is considered. The background, particularly for the $1\ell 2j \EmissT$ is quite large, but is reduced when increasing the number of leptons in the final state, and requiring $0jets$ reduces the background drastically. The effect of {\it cut-2} is seen most clearly on reductions in W+jets and $t \bar t $ backgrounds. The symbol ``-'' means that this  particular decay is not a background for the signal studied.

As can be seen from the numbers in Table \ref{tab:cuts-bckgr}, after imposing  {\it cut-2a} for Scenarios A and B, and {\it cut-2b} for Scenario C, as compared to {\it cut-1}, the signal cross sections is reduced on average to around 6 parts in a thousand for the monolepton signal, around 3 percent for the dilepton signal, and around 9 percent for the trilepton signal.

We use the following formula for the significance of the signals (signal-to-background):
\begin{eqnarray}
\beta_{\alpha}^{ij}(r)=\frac{N_{\alpha}^{ij}}{\sqrt{N_{SM}^{bg}+r\sum_{k,l\neq
i,j}N_{\alpha}^{kl}}}
\end{eqnarray}
where $N$ are the number of events and $\alpha=A,B,C$ represent Scenarios A, B, and C, respectively. The indices $i,j$ run over the chargino, neutralino states contributing to the signals. The parameter $r$ can take two values, 0 or 1. The case with $r=0$ corresponds to the significance with no sizable contribution from the $U(1)'$ model to the background. Whenever there is a need to consider any contamination from the other $U(1)^\prime$ channels, $r=1$ is taken.

In Table \ref{tab:cuts-sign} we list the cross sections obtained
after we perform both {\it cut-1} and {\it cut-2a} for the signals
in Scenarios A and B, and  {\it cut-1} and {\it cut-2b} for Scenario
C. The numbers for the signal significance show that events in
Scenario A have no chance of being observed. For Scenario B, only
monolepton signals generated via chargino pair seem to be promising
since they have signal significance greater than 10 events after the
{\it cut-2a} set. A better option is the trilepton signal with no
jets from Scenario C. The signal significance with $r=1$ is not
calculated since there are no other channels giving significant
trilepton signals with no jets.

\begin{table*}[htbp]
\begin{center}
\caption{\label{tab:cuts-sign}\sl\small The cross sections for
signal events and signal to background significance at the LHC with integrated
luminosity ${\cal L}=100\xfb^{-1}$ after the {\it cut-1} and {\it cut-2a(b)} for Scenarios A, B (C). See the text for the definition of
$\beta_\alpha^{ij}(r)$. $\beta_\alpha^{ij}(0)$ is the significance
with no contamination from others channels. }
\begin{tabular*}{1.01\textwidth}{@{\extracolsep{\fill}} ccccccccccc}
\hline\hline
 $\rm Signal $& Channel& Cut &$\rm S A[fb]$&$\rm S B[fb]$&$\rm S C[fb]$&$\beta_A^{ij}(0)$&$\beta_B^{ij}(0)$&$\beta_C^{ij}(0)$&$\beta_A^{ij}(1)$&$\beta_B^{ij}(1)$
  \\ \cline{1-4}\cline{1-4}
 \hline\hline
$1\ell 2 j \EmissT$&\\
&$\tilde\chi_{i}^{+}\tilde\chi_{j}^{-}$ & {\it cut-1}&3.4&51.9&$-$&0.2&3.3&$-$&0.2&3.3\\
&&{\it cut-2a}&0.3&12.5&$-$&0.25&10.4&$-$&0.25&10.4\\
&$\tilde\chi_{i}^0\tilde\chi_{j}^0$ &{\it cut-1}&$-$&0.3&$-$&$-$&0.02&$-$&$-$&0.02\\
&&{\it cut-2a}&$-$&0.03&$-$&$-$&0.02&$-$&$-$&0.02\\
 &$\tilde\chi_{i}^{0}\tilde\chi_{j}^{\pm}$ &{\it cut-1}&7.6&2.0&$-$&0.49&0.1&$-$&0.49&0.1\\
&&{\it cut-2a}&0.05&0.07&$-$&0.04&0.06&$-$&0.04&0.06\\

$2\ell 2j \EmissT$&\\
&$\tilde\chi_{i}^{+}\tilde\chi_{j}^{-}$ &{\it cut-1}&0.2&0.9&$-$&0.1&0.56&$-$&0.1&0.56\\
&&{\it cut-2a}&0.03&0.1&$-$&0.1&0.4&$-$&0.1&0.4\\
&$\tilde\chi_{i}^0\tilde\chi_{j}^0$ & {\it cut-1}&0.07&1.7&$-$&0.04&1.0&$-$&0.04&1.0\\
&&{\it cut-2a}&0.01&0.12&$-$&0.04&0.42&$-$&0.04&0.42\\
 &$\tilde\chi_{i}^{0}\tilde\chi_{j}^{\pm}$ &{\it  cut-1}&0.01&0.4&$-$&0.0&0.25&$-$&0.0&0.25\\
&&{\it  cut-2a}&0&0&$-$&0.0&0.0&$-$&0.0&0.0\\

$3\ell 0j \EmissT$\\
&$\tilde\chi_{i}^{0}\tilde\chi_{j}^{\pm}$ &{\it  cut-1}&$-$&$-$&55.0&$-$&$-$&260.0&$-$&$-$\\
&&{\it  cut-2b}&$-$&$-$&10.0&$-$&$-$&156.4&$-$&$-$\\
\hline\hline
\end{tabular*}
\end{center}
 \end{table*}

\begin{figure*}[htbp]
\begin{center}$
    \begin{array}{cc}
\hspace*{-1.6cm}
    \includegraphics[width=2.8in,height=2.2in]{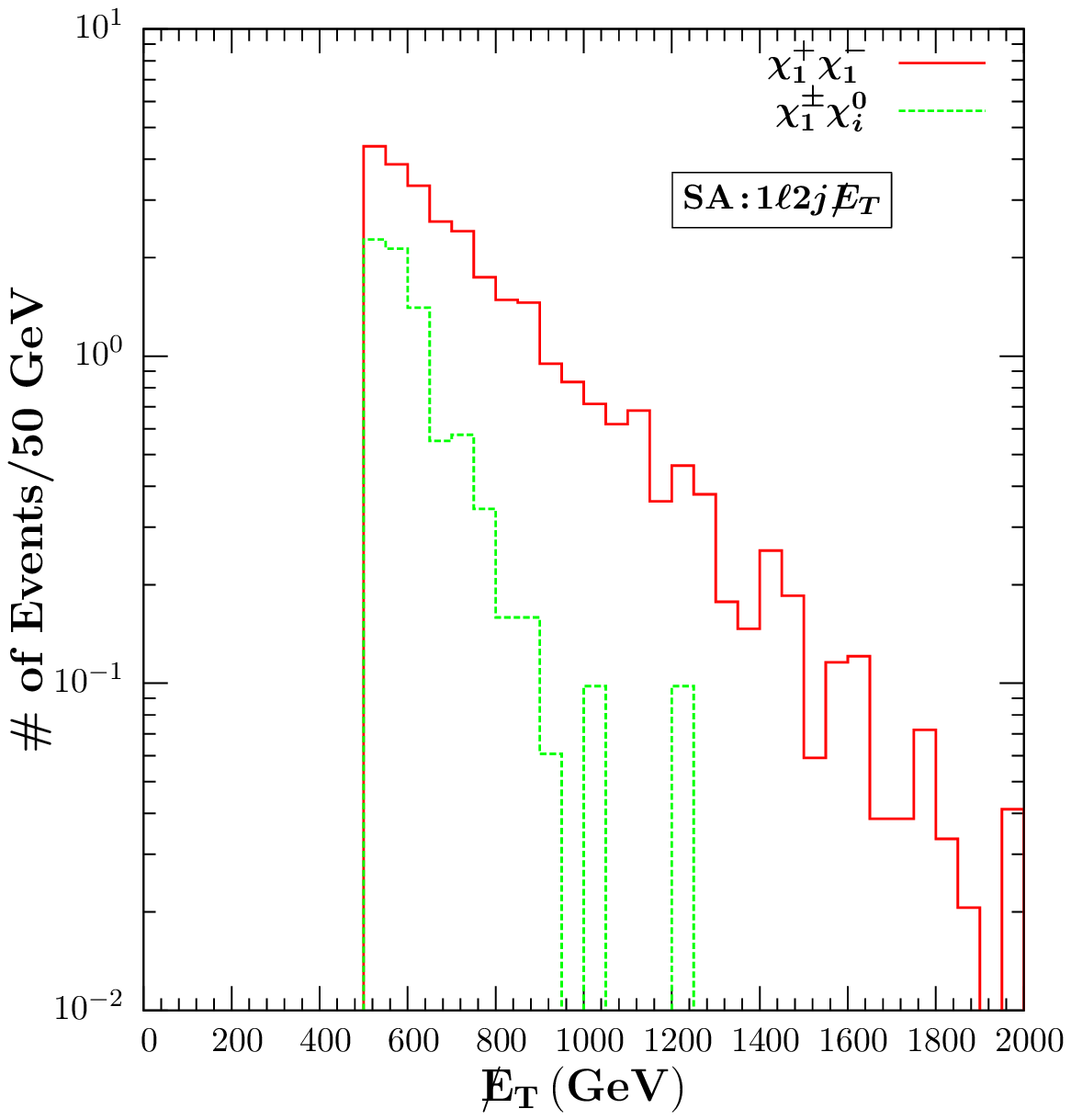}
&\hspace*{-1.1cm}
    \includegraphics[width=2.8in,height=2.2in]{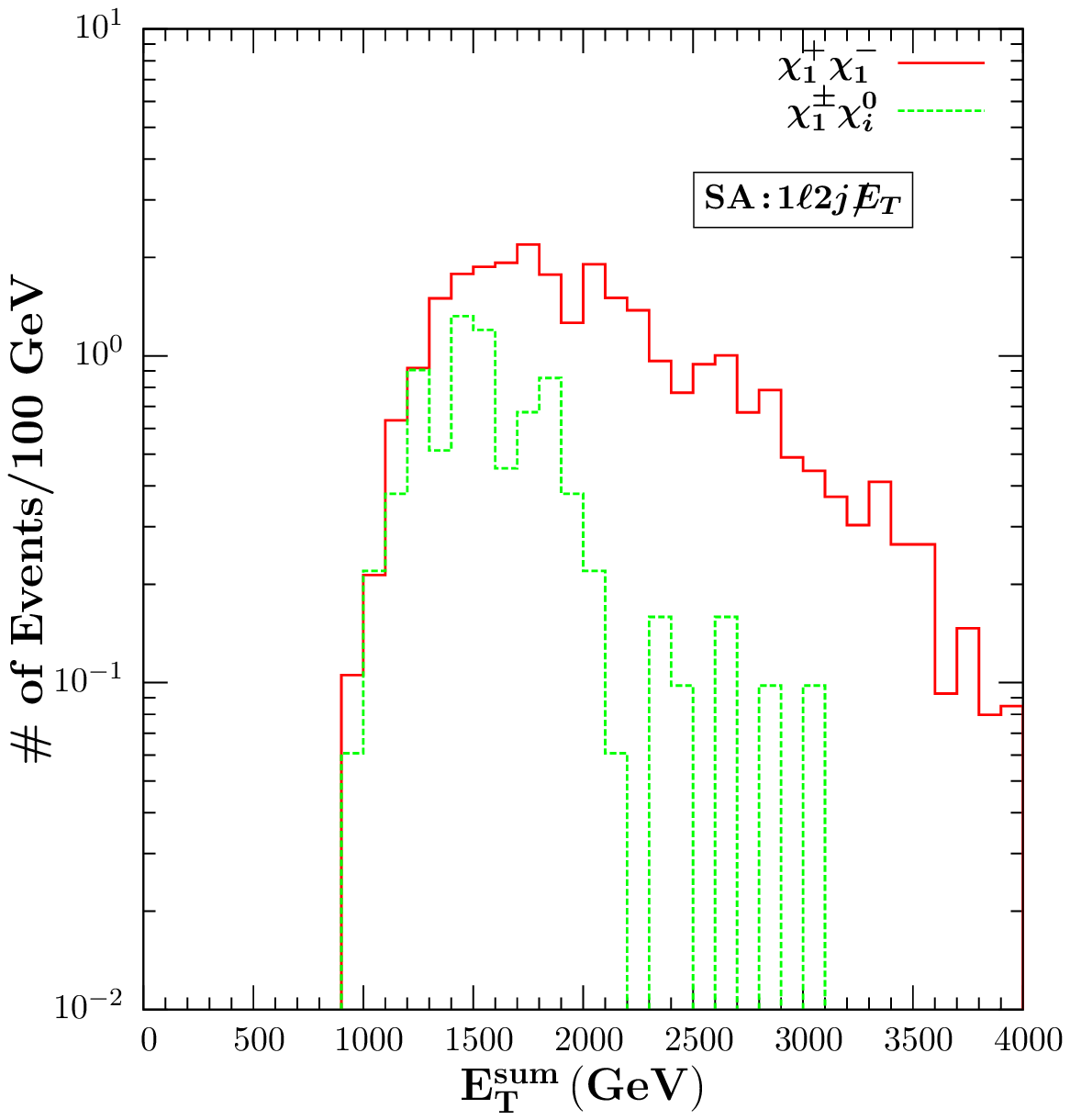}\\
\hspace*{-1.6cm}
      \includegraphics[width=2.8in,height=2.2in]{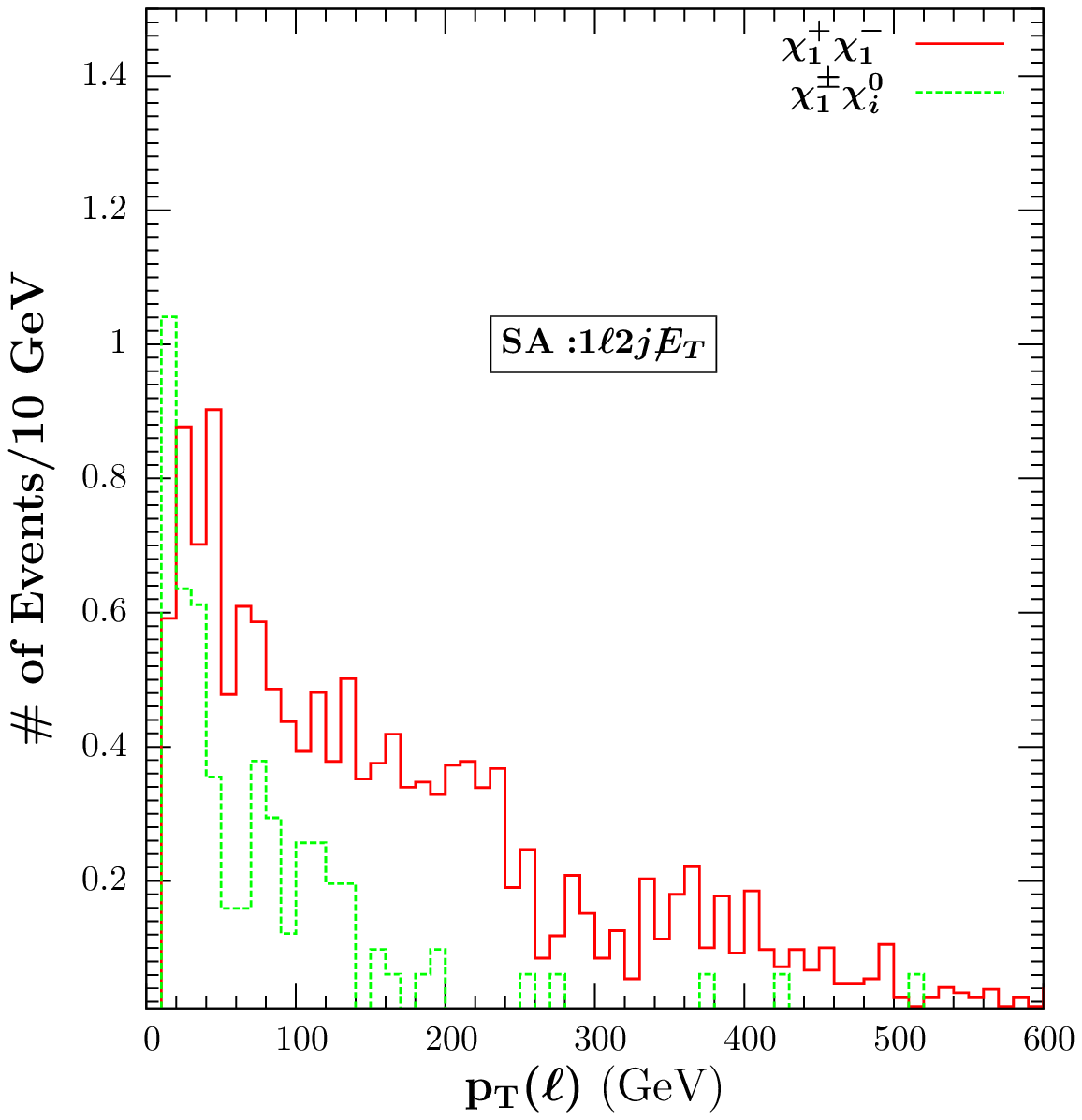}
&\hspace*{-1.0cm}
    \includegraphics[width=2.8in,height=2.2in]{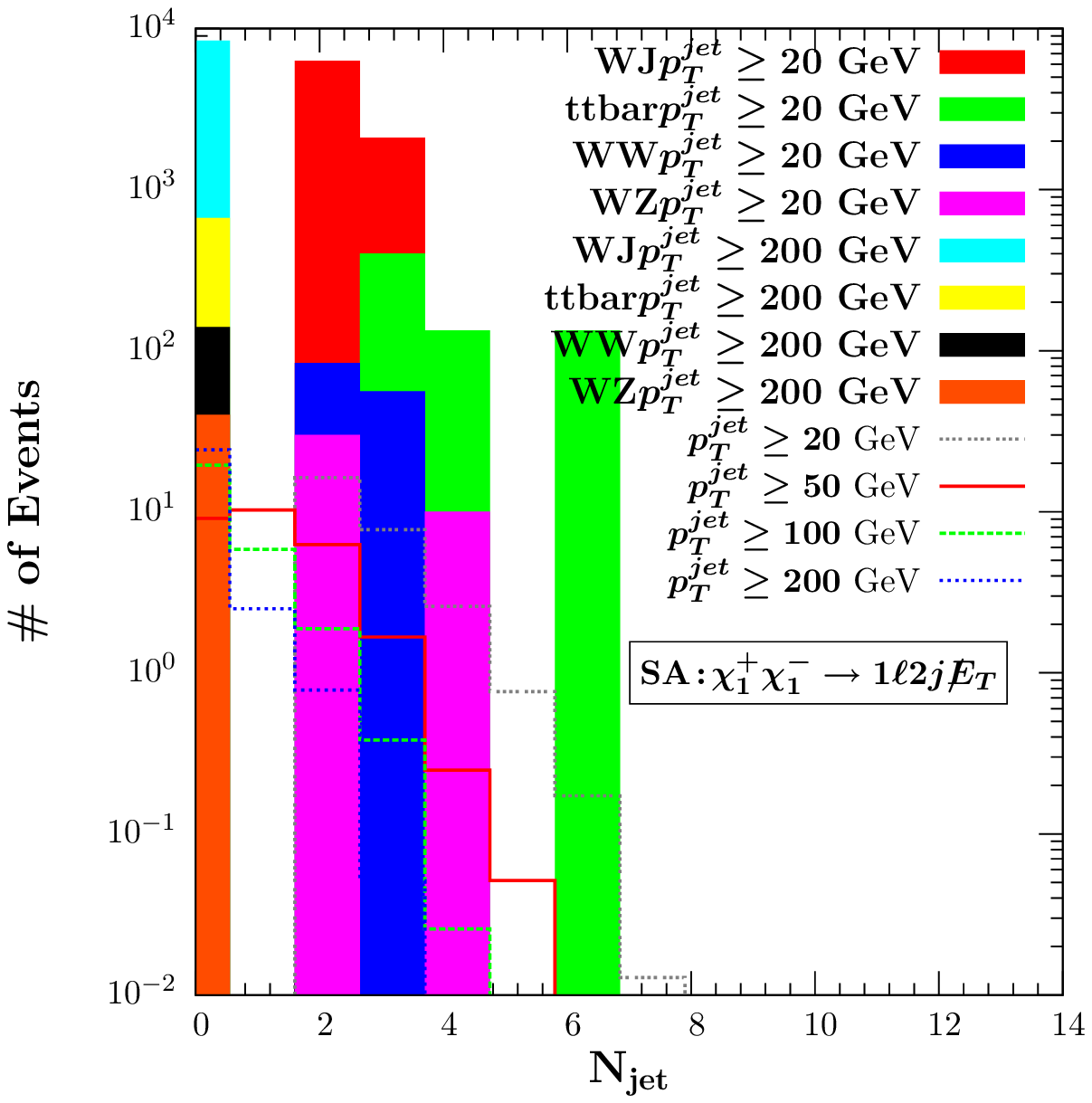}\\
\hspace*{-1.5cm}
    \includegraphics[width=2.9in,height=2.2in]{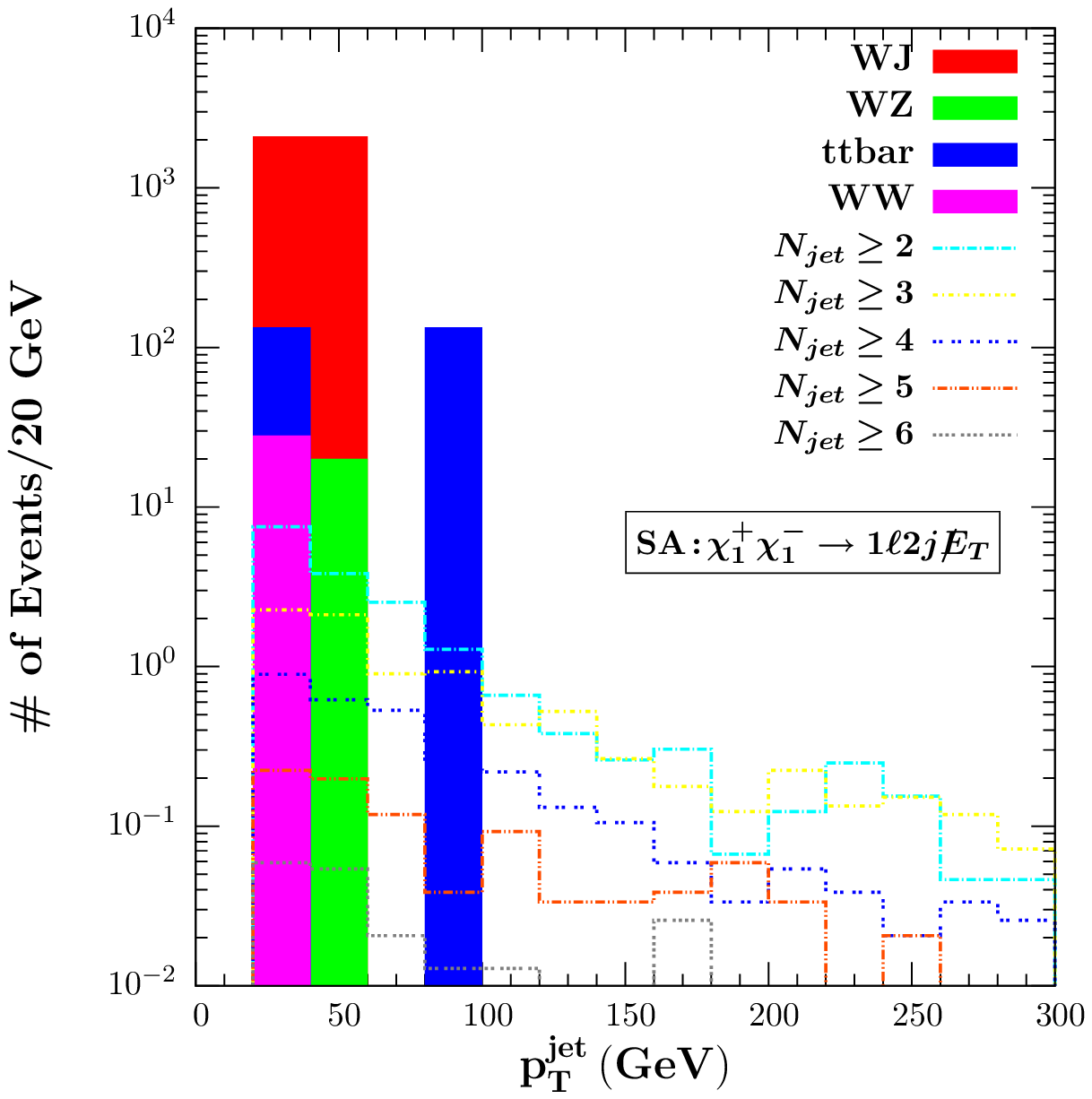}
    &
    \end{array}$
    \end{center}
\vskip -0.1in
      \caption{(color online). \sl\small The $\EmissT$,  $\ETsum$, $p_T(\ell)$, $N_{jet}$ and $p_T^{jet}$ distributions
of the $1\ell + 2j + \EmissT$ signal at $14 \tev$
with integrated luminosity ${\cal L}=100\xfb^{-1}$,
for Scenario A, after {\it cut-2a}.  We also include the backgrounds from WJ ($W+jet$), WZ, WW and $t{\bar t}$. } \label{fig:1lepA}
\end{figure*}

\begin{figure*}[htbp]
\begin{center}$
    \begin{array}{cc}
\hspace*{-1.6cm}
    \includegraphics[width=2.8in,height=2.2in]{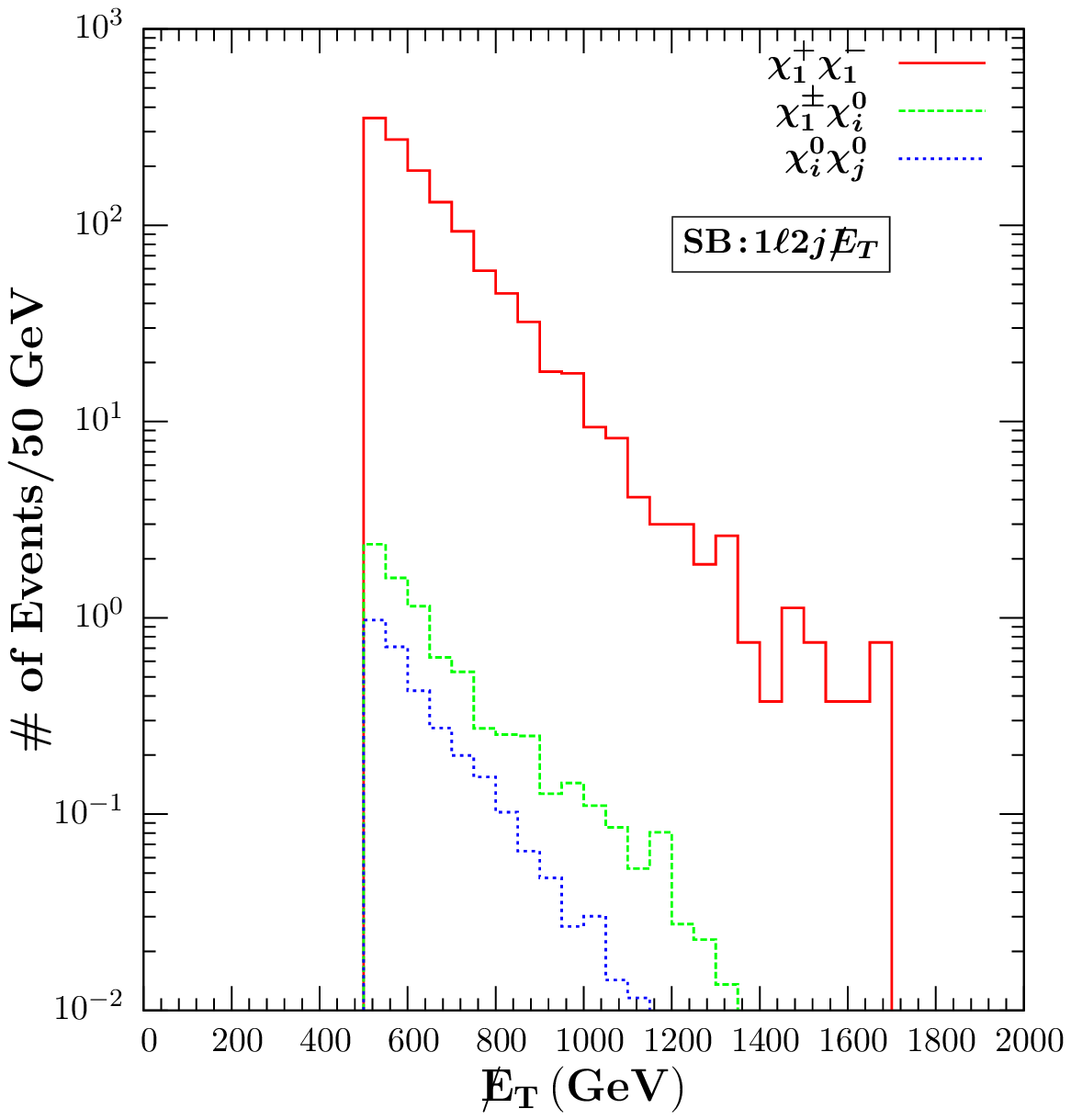}
&\hspace*{-1.5cm}
    \includegraphics[width=2.8in,height=2.2in]{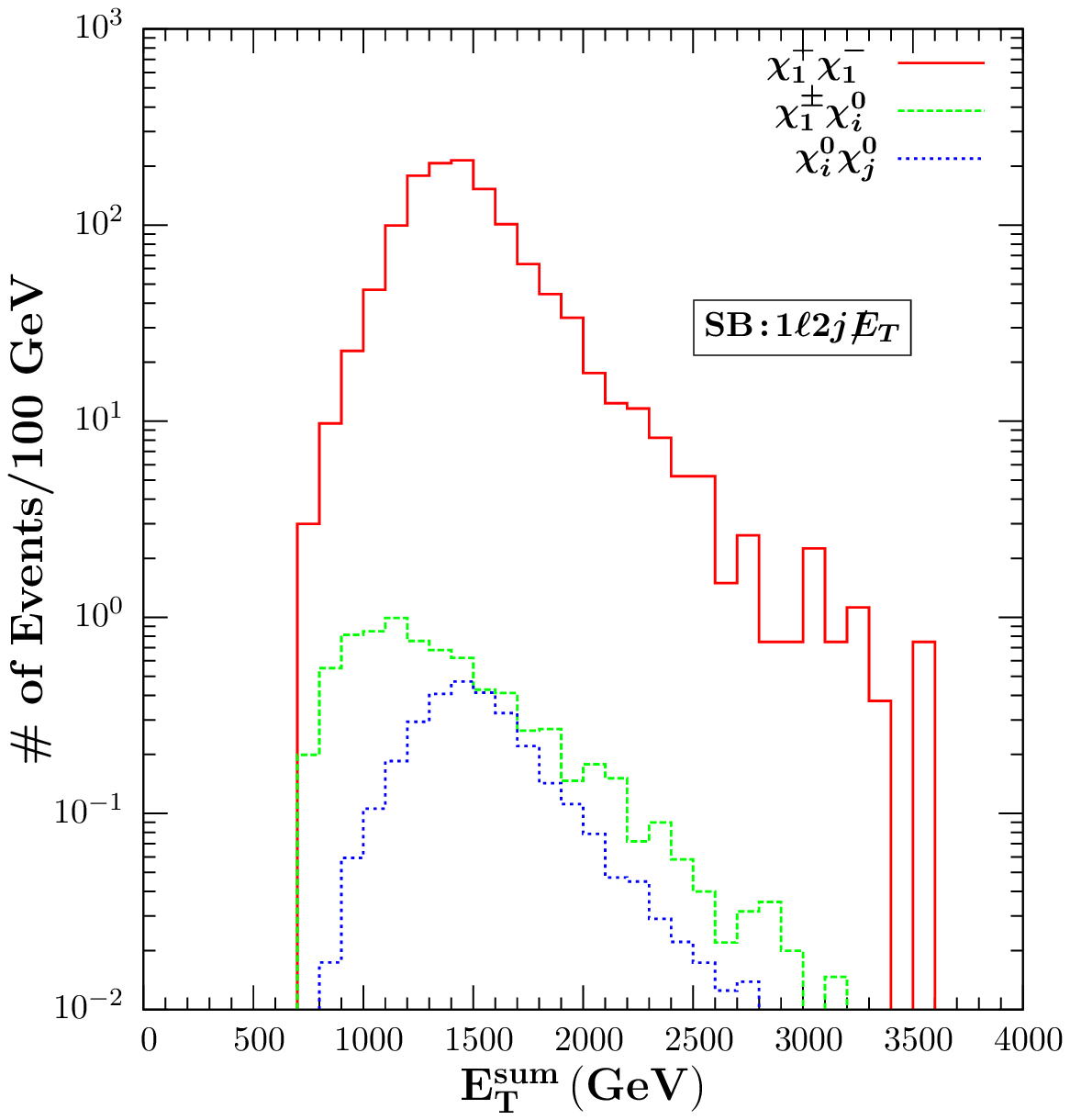}\\
\hspace*{-1.3cm}
      \includegraphics[width=2.8in,height=2.2in]{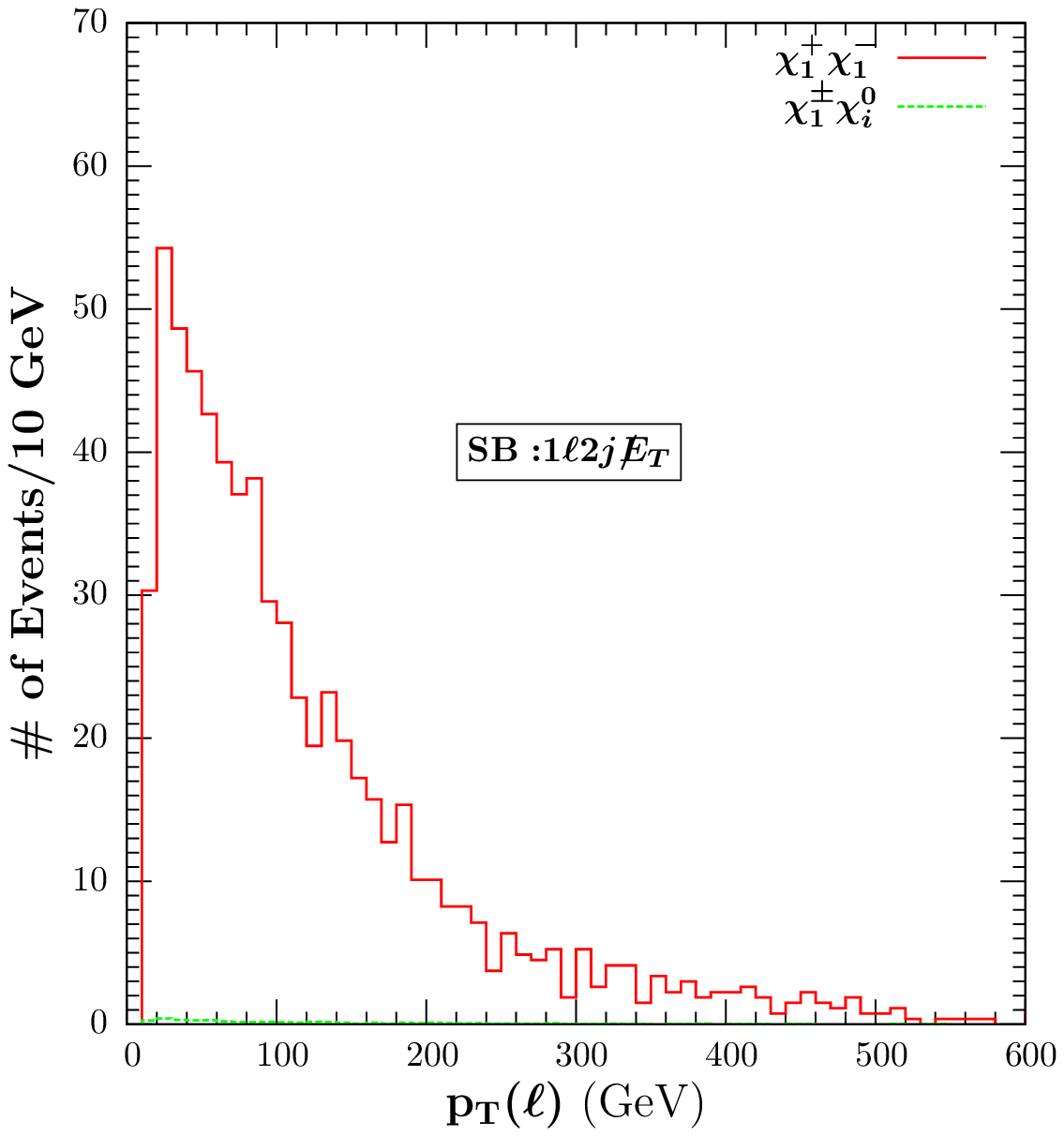}
&\hspace*{-1.5cm}
    \includegraphics[width=2.7in,height=2.2in]{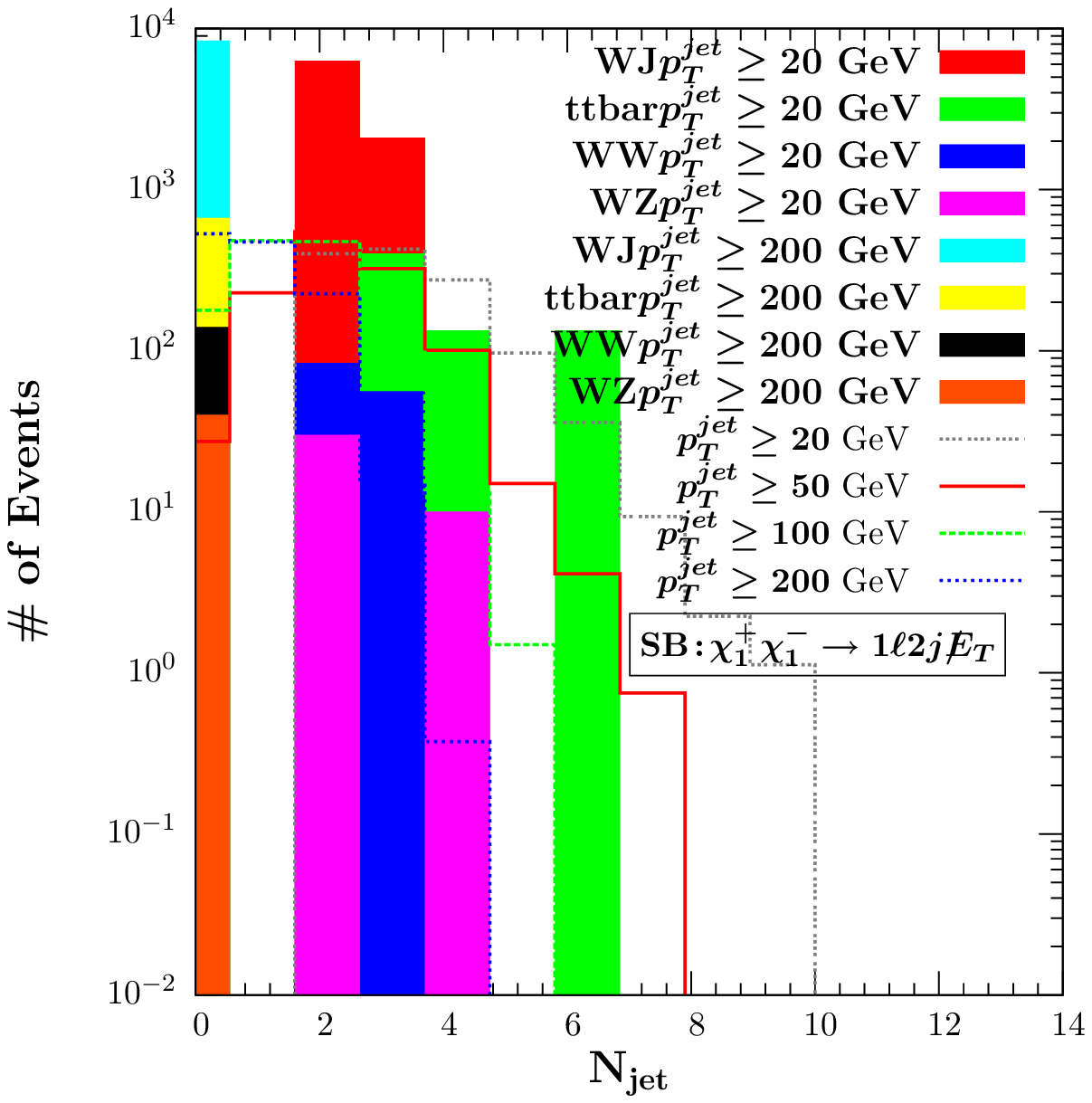}\\
\hspace*{-1.5cm}
\includegraphics[width=2.9in,height=2.2in]{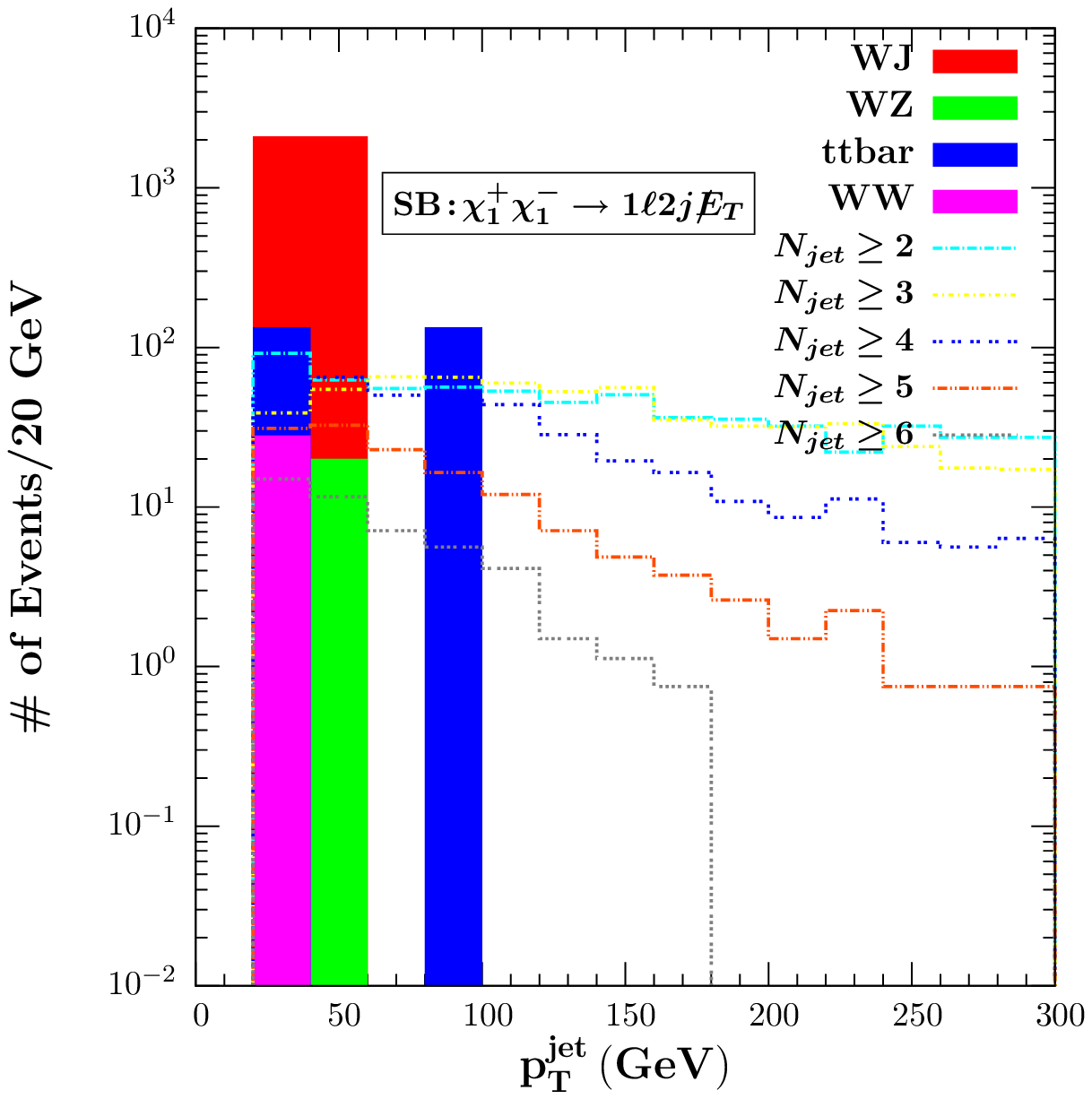}
&
\end{array}$
\end{center}
\vskip -0.1in
      \caption{(color online). \sl\small The $\EmissT$,  $\ETsum$, $p_T(\ell)$, $N_{jet}$ and $p_T^{jet}$ distributions
of the $1\ell + 2j + \EmissT$ signal at $14 \tev$
with integrated luminosity ${\cal L}=100\xfb^{-1}$,
for Scenario B, after {\it cut-2a}. We also include the backgrounds from WJ ($W+jet$), WZ, WW and $t{\bar t}$.} \label{fig:1lepB}
\end{figure*}
\begin{figure*}[htbp]
\begin{center}$
    \begin{array}{cc}
\hspace*{-1.5cm}
    \includegraphics[width=2.7in,height=2.2in]{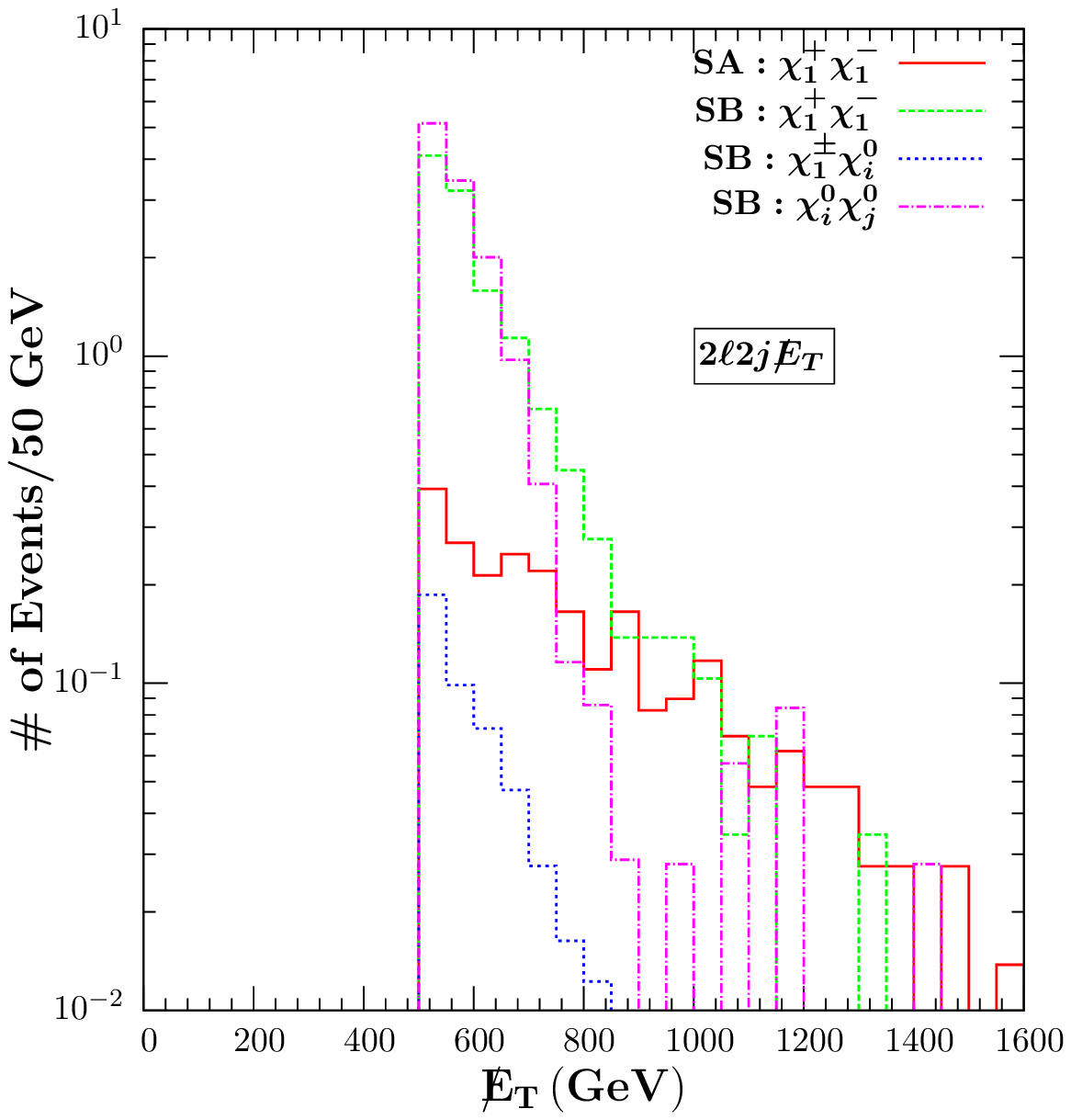}
&\hspace*{-1.5cm}
    \includegraphics[width=2.7in,height=2.2in]{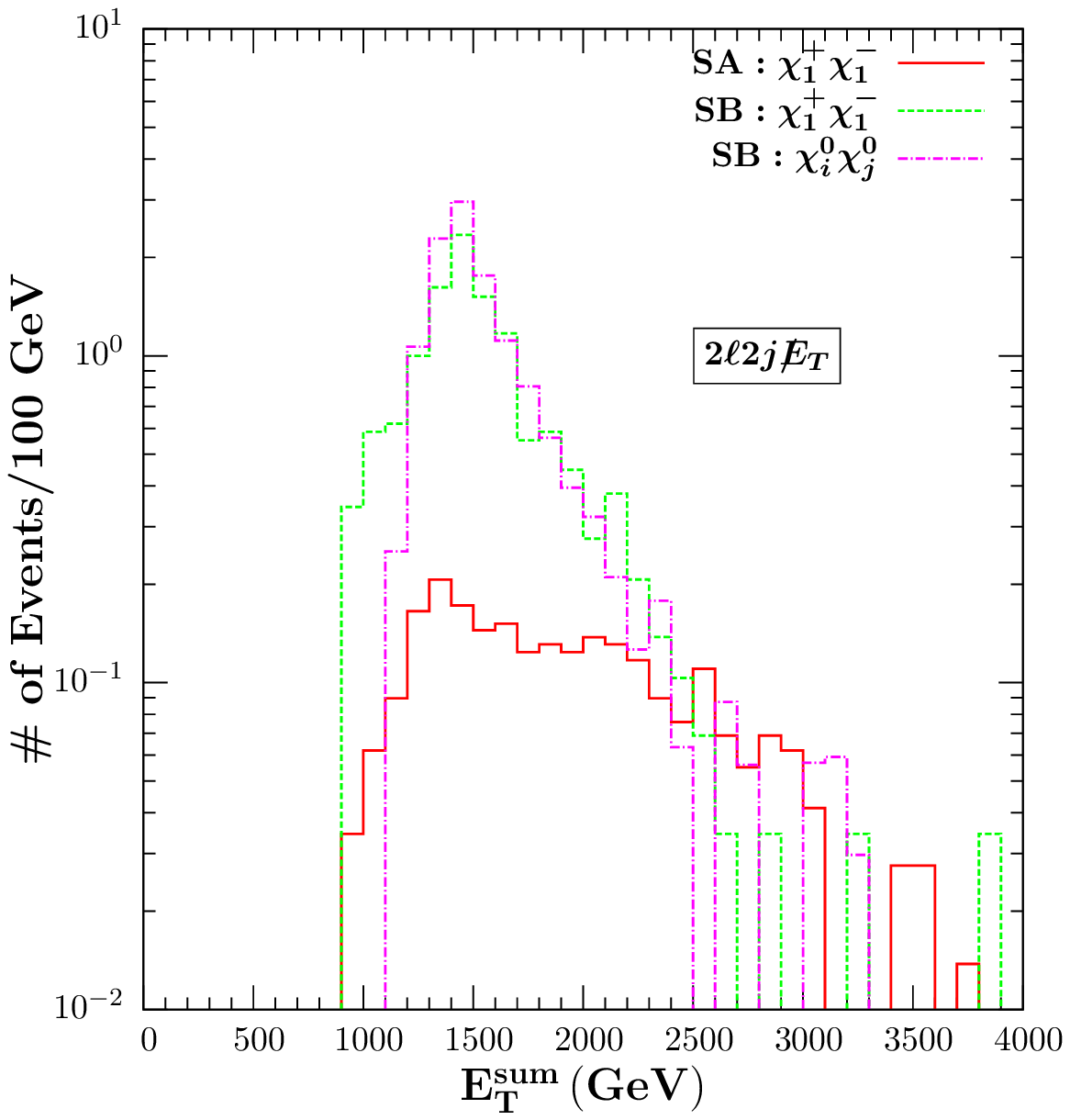}\\
 \hspace*{-1.0cm}
      \includegraphics[width=2.7in,height=2.2in]{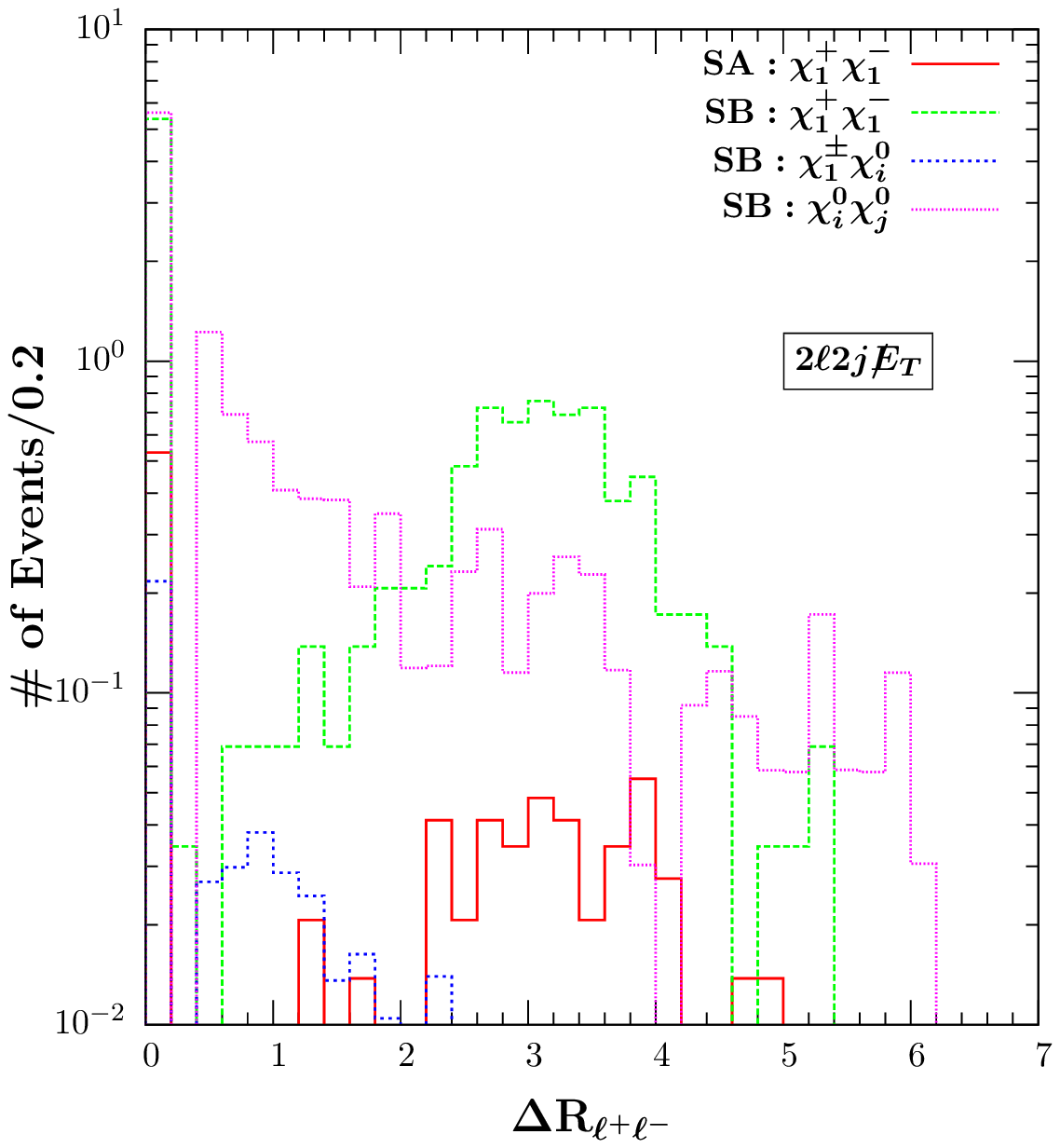}
    &  \hspace*{-1.5cm}
    \includegraphics[width=2.7in,height=2.2in]{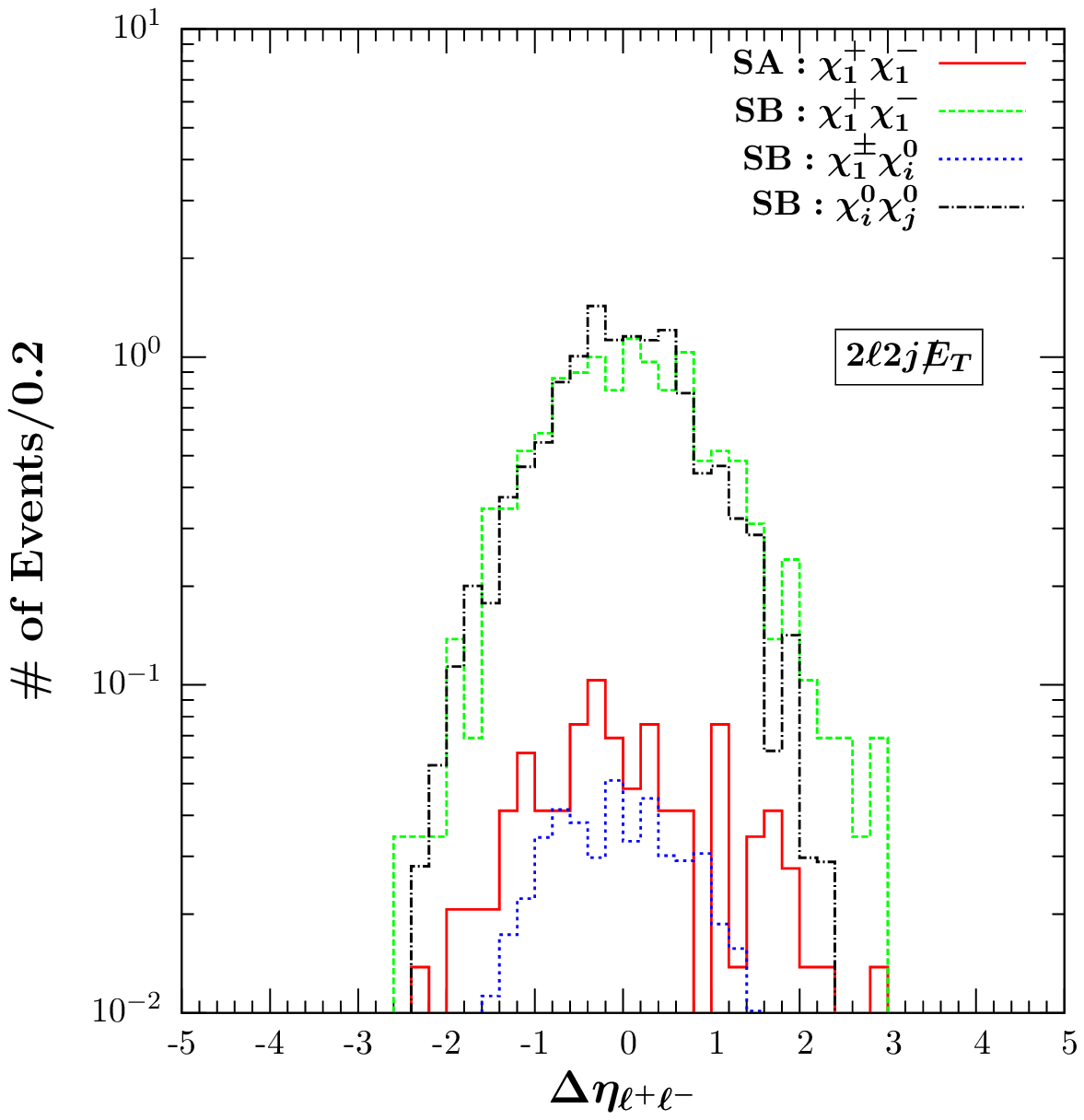}\\
\hspace*{-1.5cm}
    \includegraphics[width=2.7in,height=2.2in]{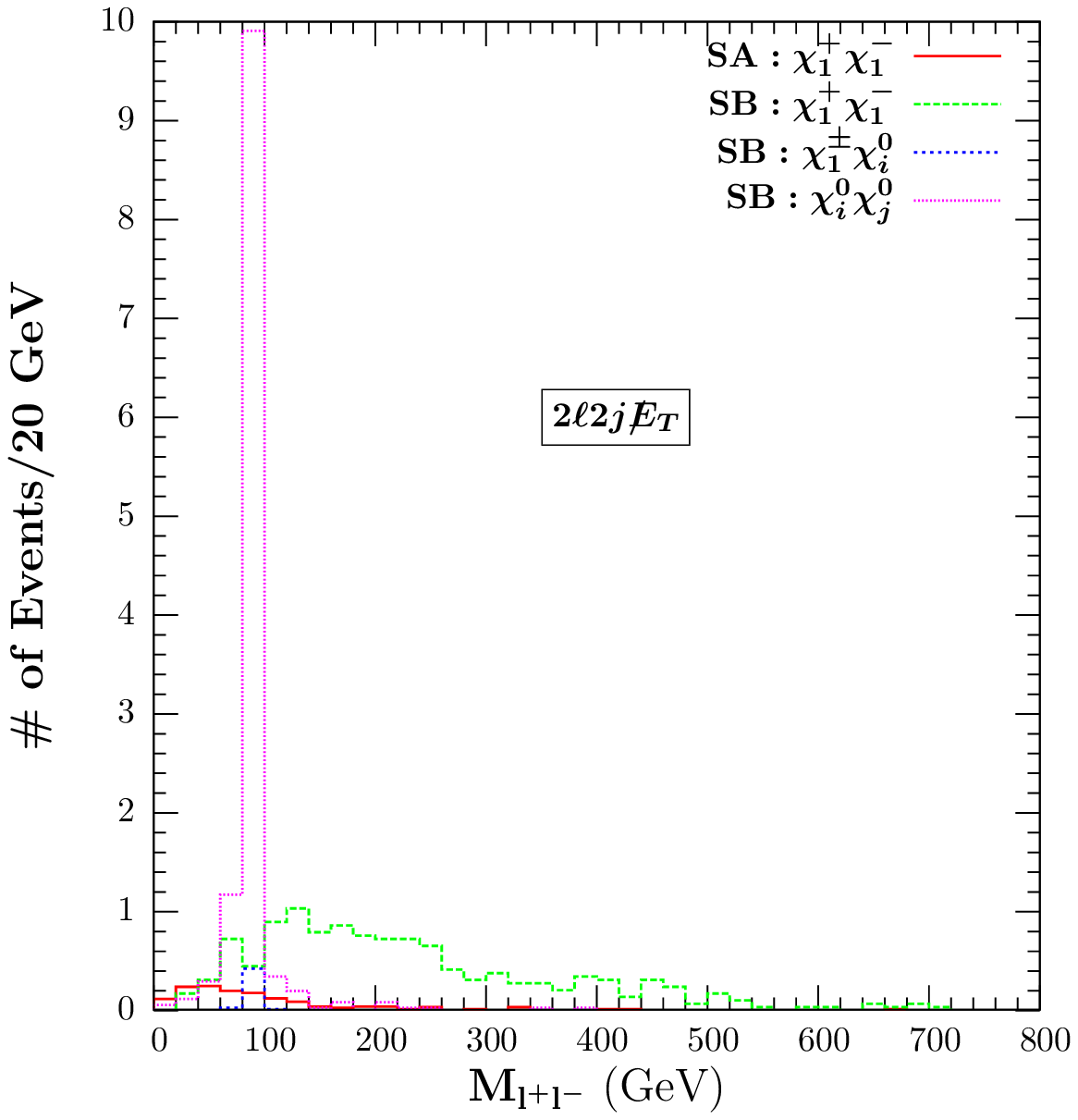}
    &  \end{array}$
\end{center}
\vskip -0.1in
 \caption{ (color online). \sl\small The $\EmissT$, $\ETsum$, $\Delta R_{\ell^+\ell^-}$ and $\Delta
\eta_{\ell^+\ell^-}$ and invariant mass distributions
of the $2\ell + 2j +\EmissT$ signal at $14 \tev$
with integrated luminosity ${\cal L}=100\xfb^{-1}$,
for  Scenario A and Scenario
B,  after {\it cut-2a}.} \label{fig:2lepEDelta}
\end{figure*}

\begin{figure*}[htbp]
\begin{center}$
    \begin{array}{cc}
\hspace*{-1.7cm}
    \includegraphics[width=2.7in,height=2.2in]{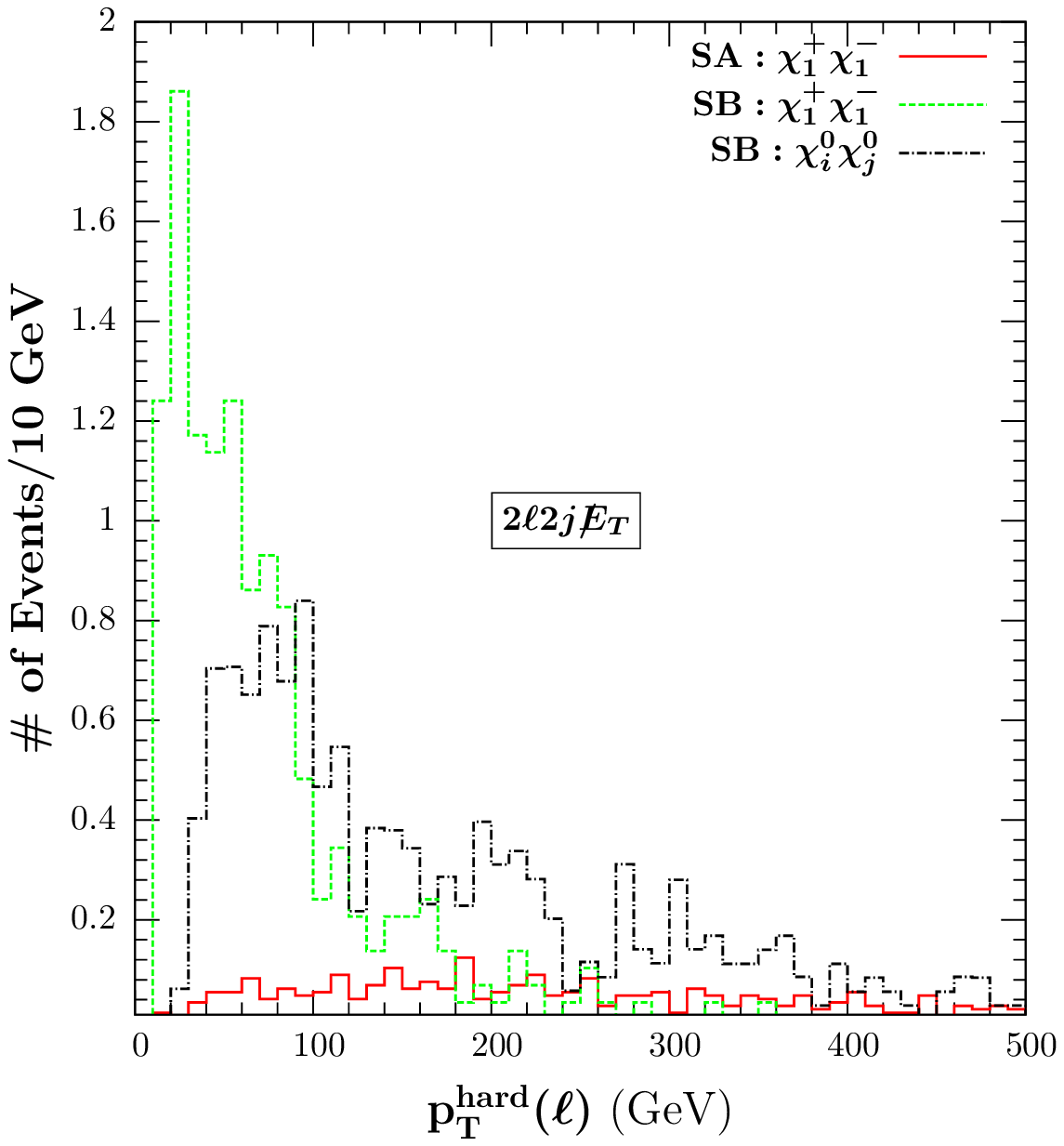}
&\hspace*{-1.5cm}
    \includegraphics[width=2.7in,height=2.2in]{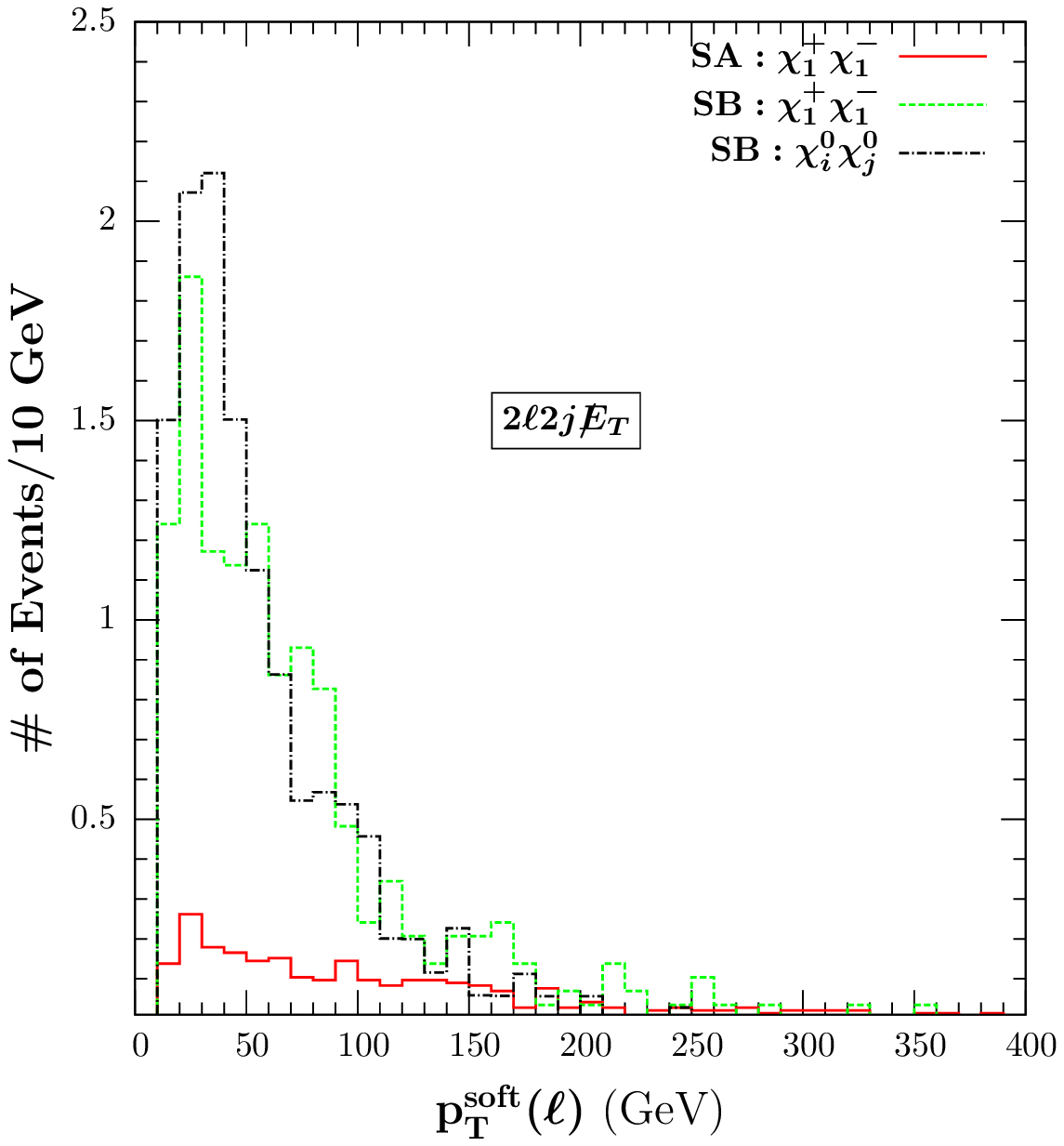}\\
    \hspace*{-1.7cm}
    \includegraphics[width=2.7in,height=2.2in]{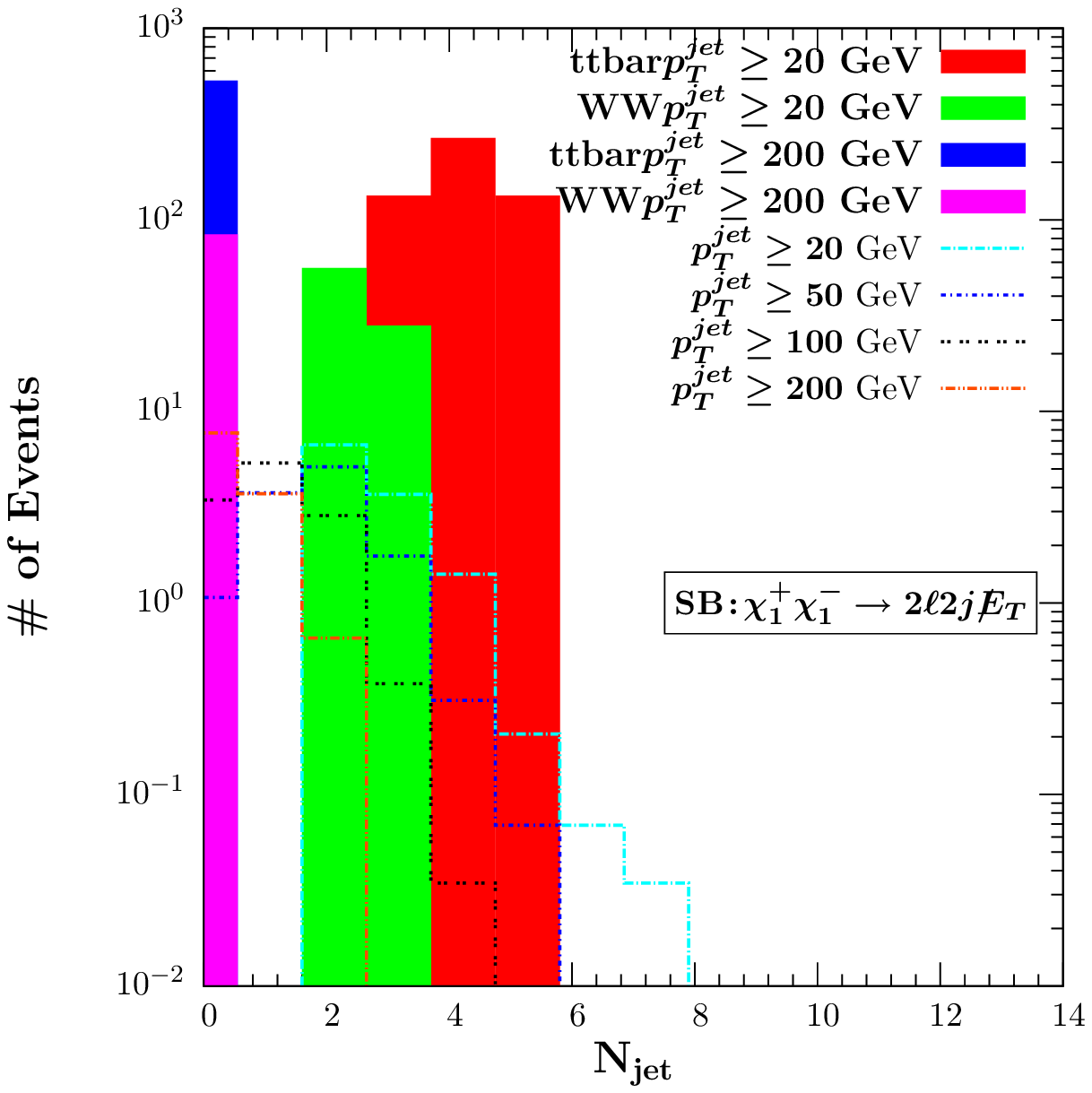}
&\hspace*{-1.5cm}
    \includegraphics[width=2.7in,height=2.2in]{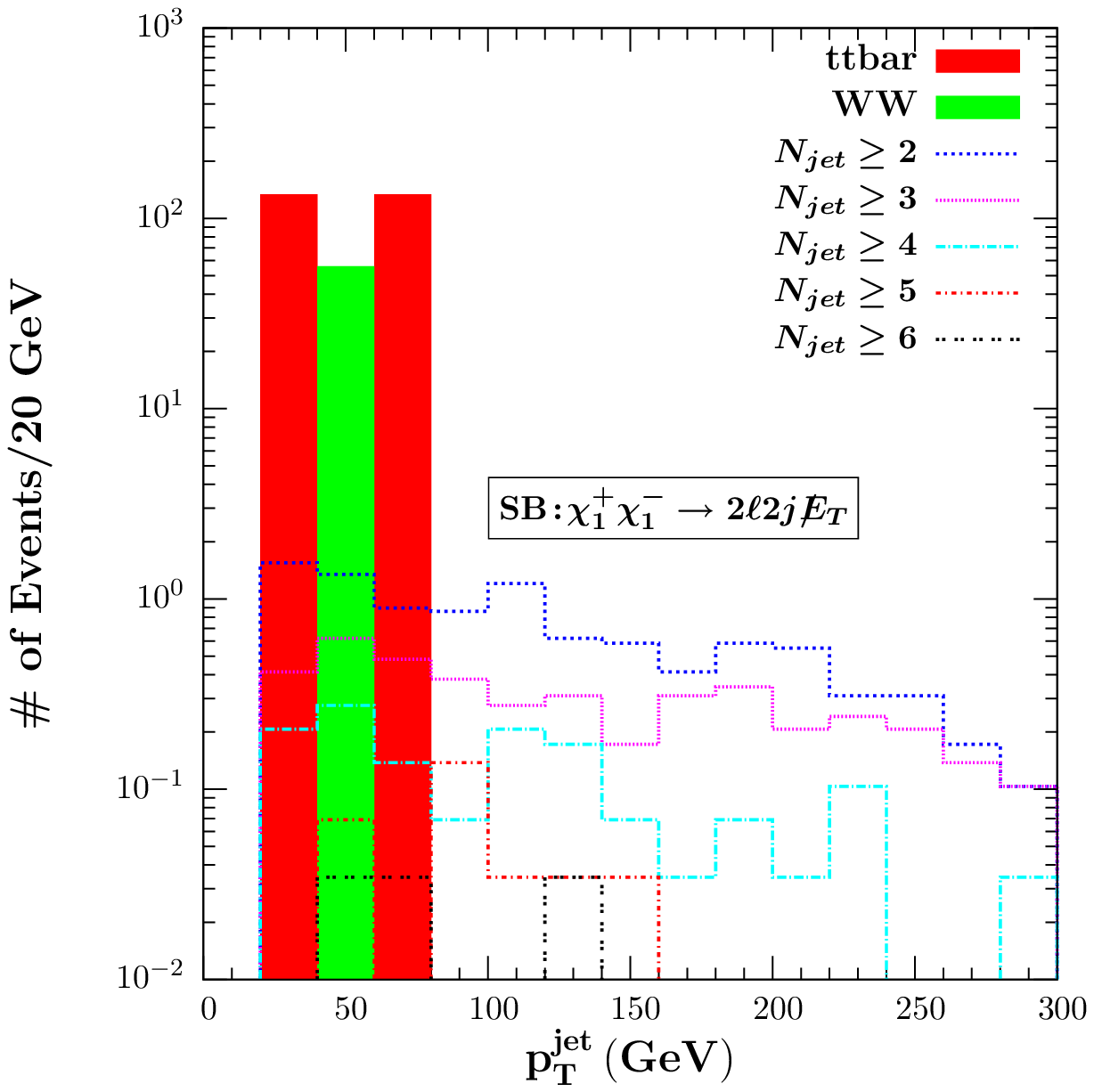}
    \end{array}$
\end{center}
\vskip -0.1in
      \caption{(color online). \sl\small The  $p_T$ distribution for Scenario A and B, and the $N_{jet}$ and $p_T^{jet}$ distributions
of the $2\ell + 2j + \EmissT$
$(\tilde\chi_{i}^{+}\tilde\chi_{j}^{-})$ signal at $14 \tev$
 with integrated luminosity ${\cal L}=100\xfb^{-1}$,
for Scenario B, after {\it cut-2a}.  We also include the backgrounds from WW and $t{\bar t}$. For Scenario A, the background is too large and completely obliterates the signal.}
\label{fig:2leppT}
\end{figure*}

\begin{figure*}[htb]
\begin{center}$
 \begin{array}{ccc}
         \hspace*{-2.2cm}
      \includegraphics[width=2.7in,height=2.2in]{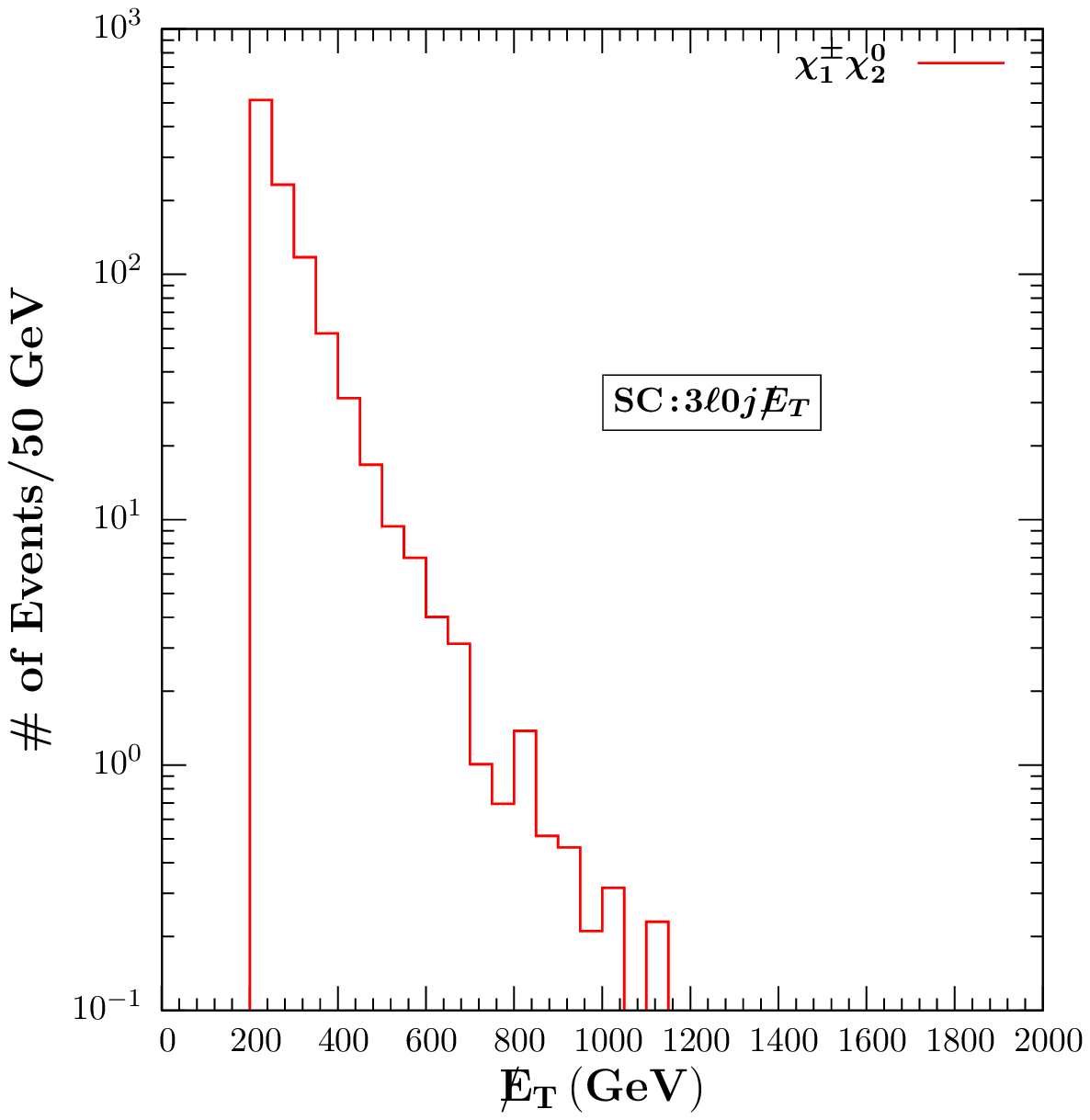}
&\hspace*{-1.2cm}
    \includegraphics[width=2.7in,height=2.2in]{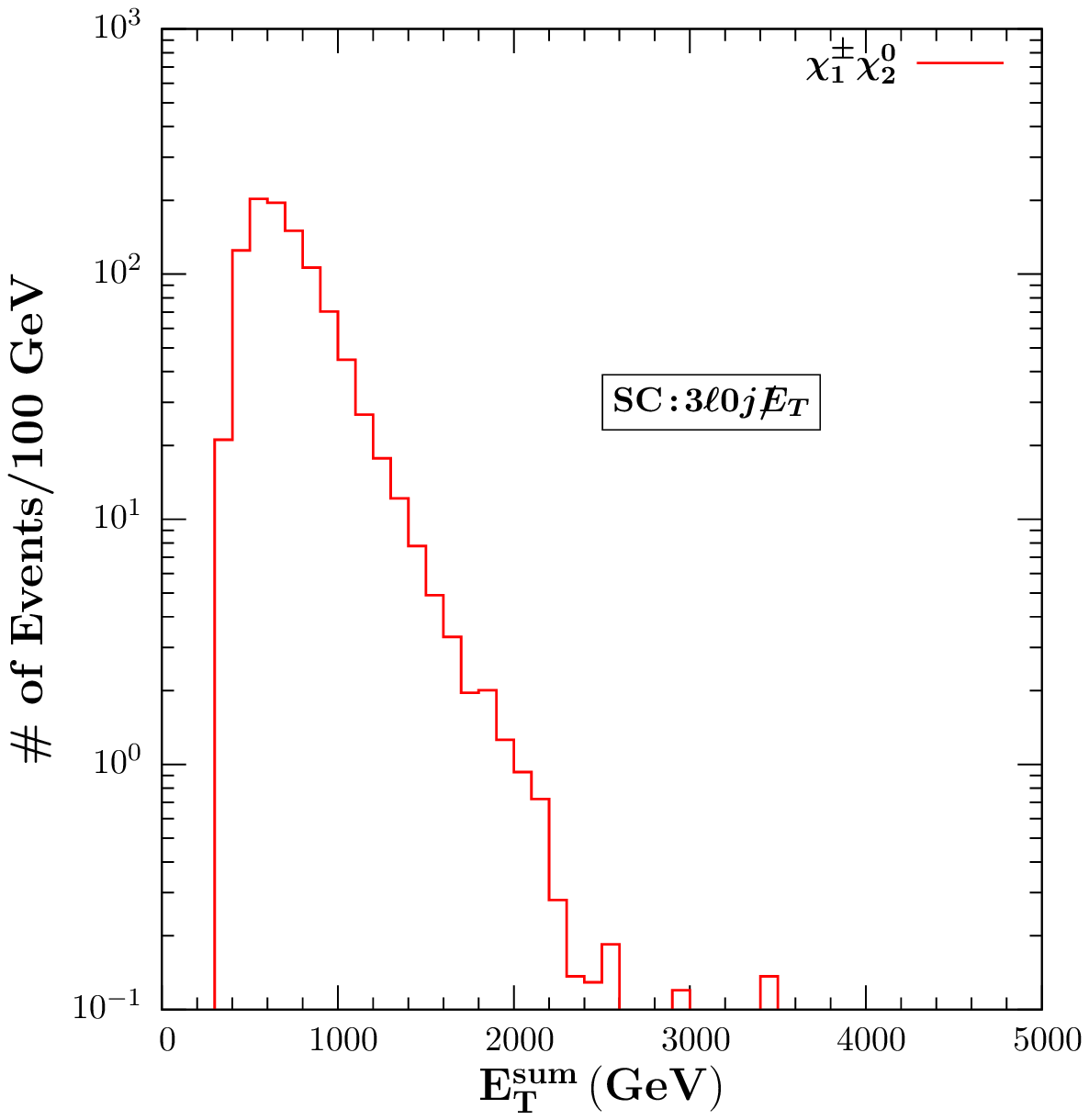}
&\hspace*{-1.2cm}
    \includegraphics[width=2.7in,height=2.2in]{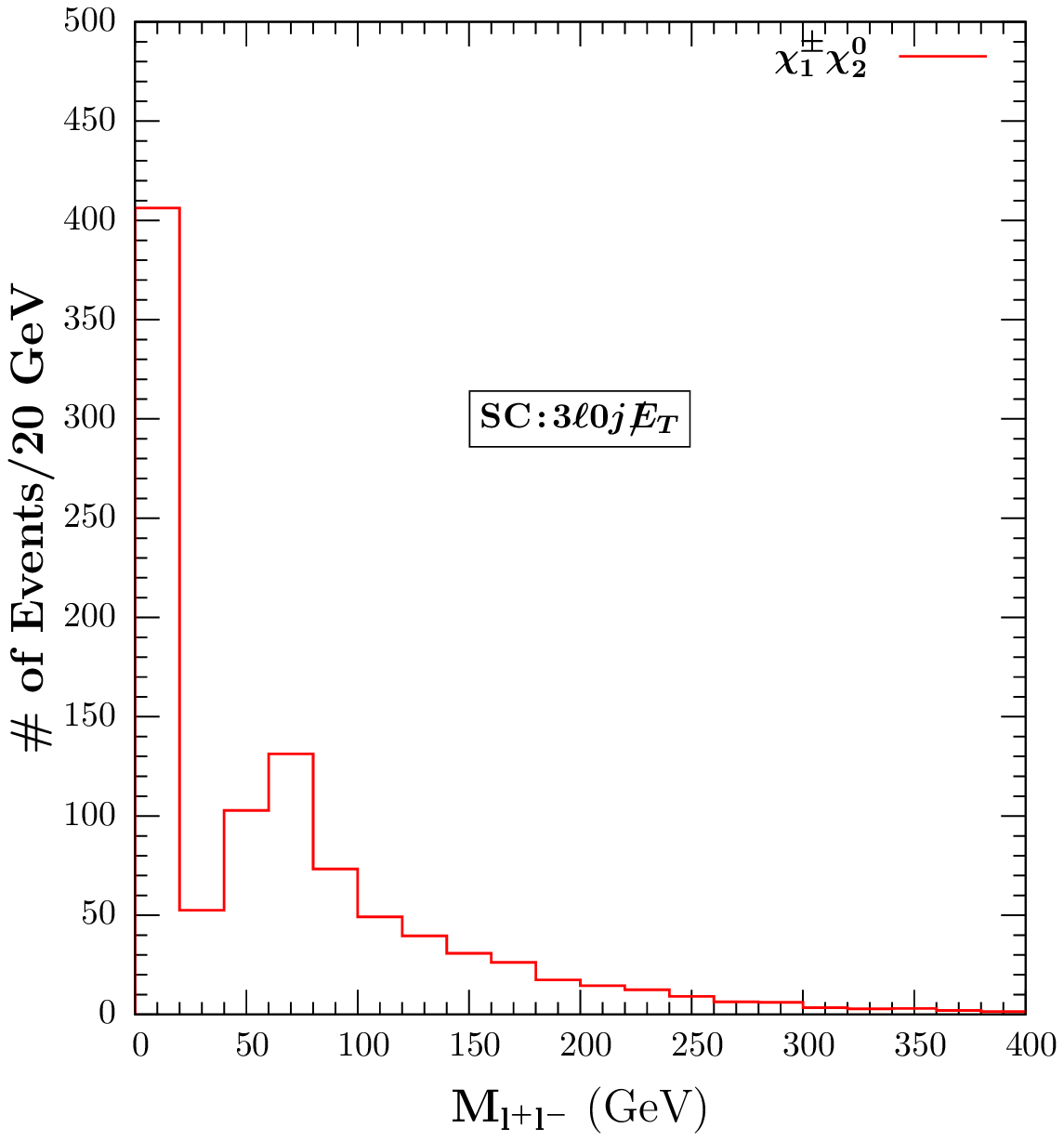} \\
\hspace*{-2.2cm}
    \includegraphics[width=2.7in,height=2.2in]{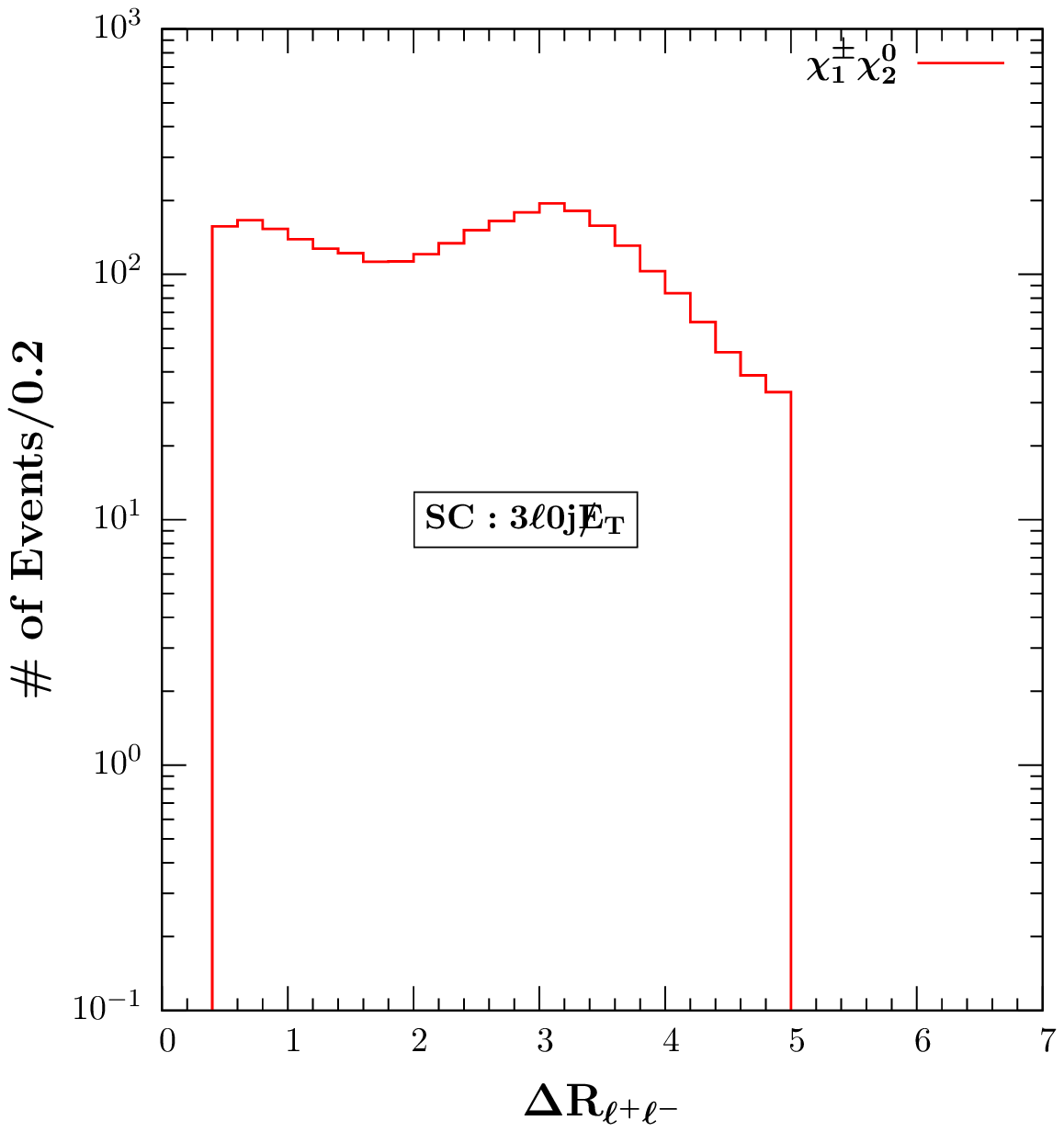}
&\hspace*{-1.0cm}
    \includegraphics[width=2.7in,height=2.2in]{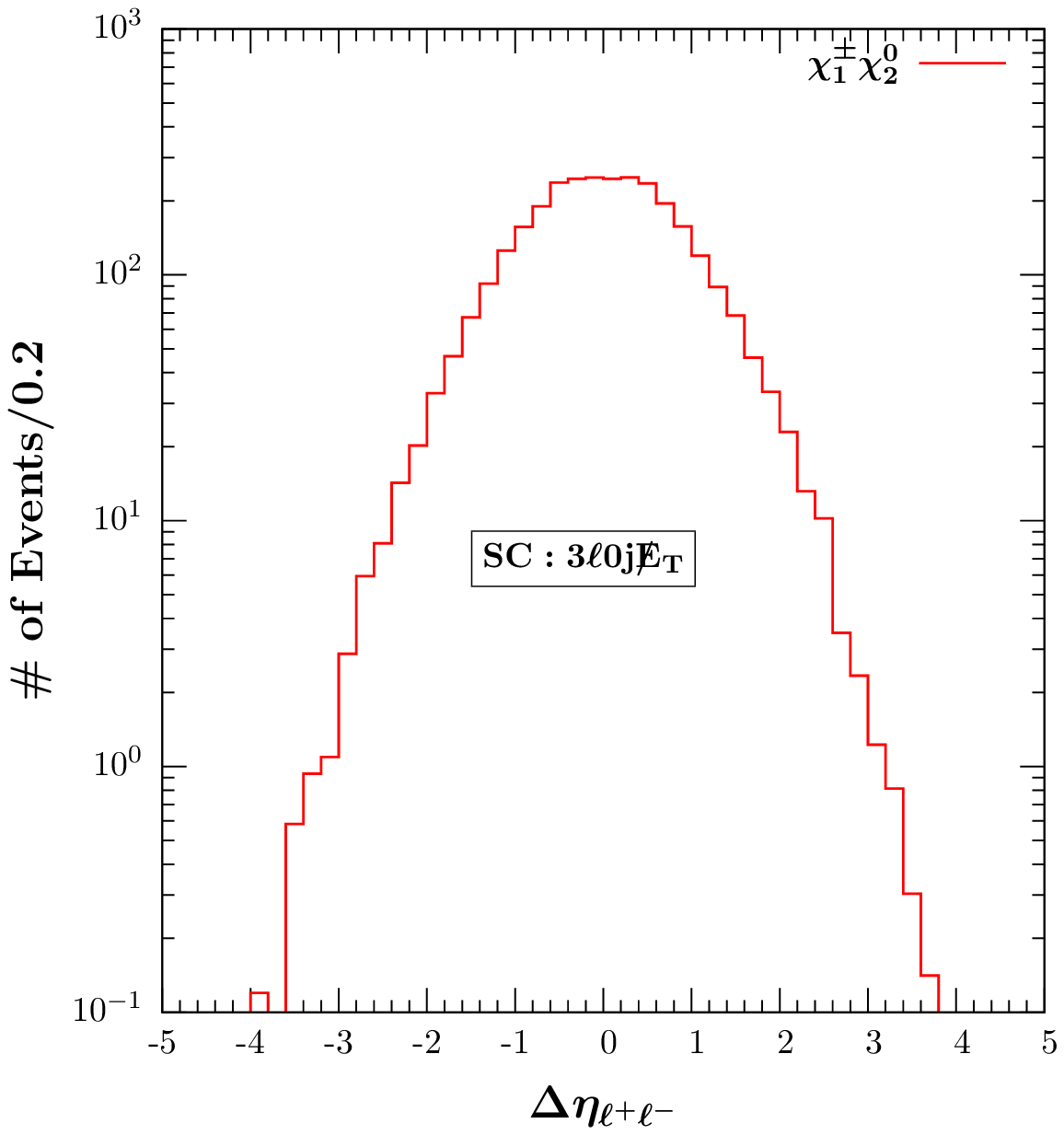}
    &\hspace*{-1.0cm}
    \includegraphics[width=2.7in,height=2.2in]{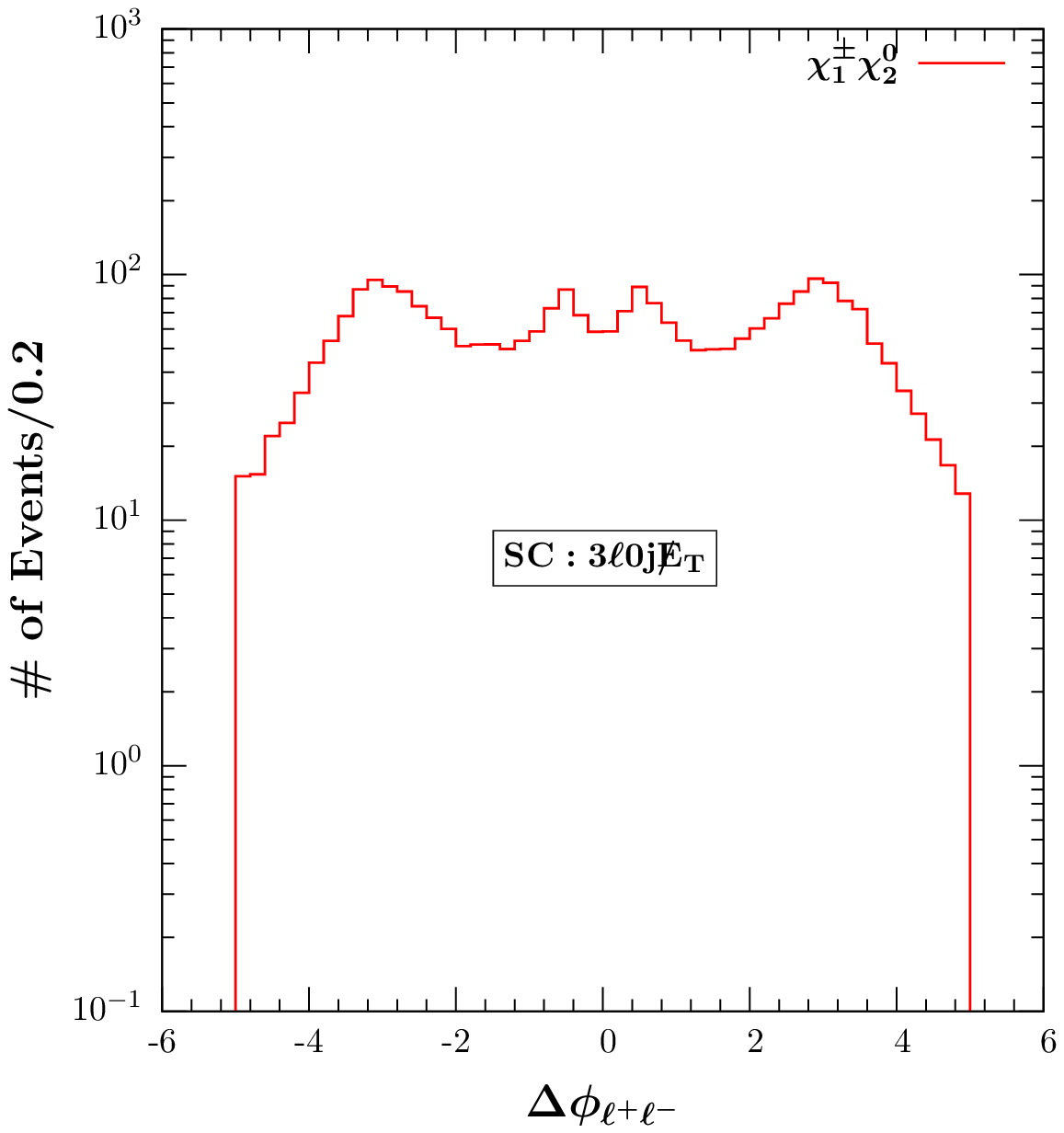}\\
 \hspace*{-2.2cm}
    \includegraphics[width=2.7in,height=2.2in]{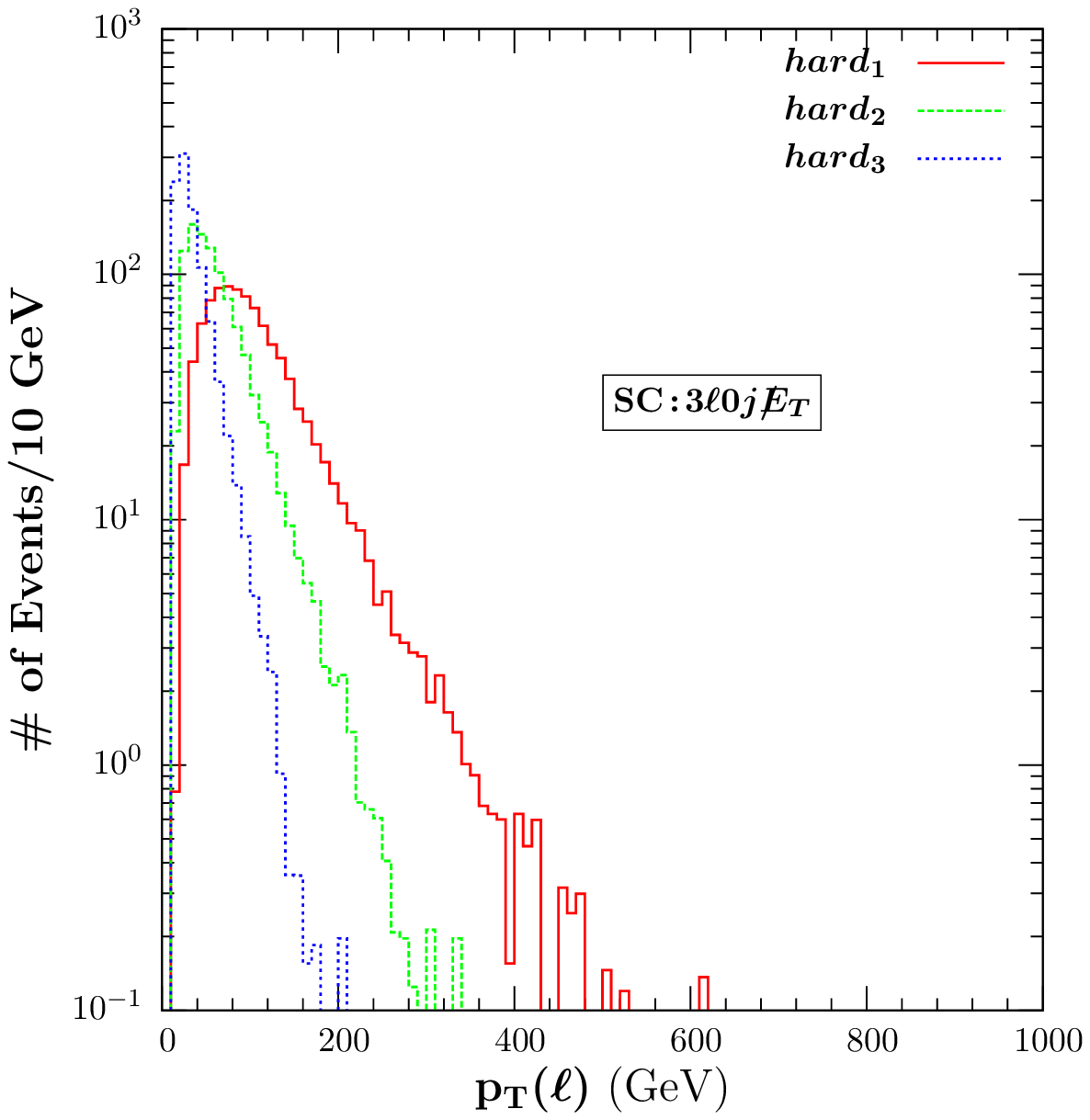} & &
    \end{array}$
\end{center}
\vskip -0.1in
      \caption{(color online). \sl\small The $\EmissT$,  $\ETsum$, invariant mass, $\Delta R_{\ell^+\ell^-}$,  $\Delta
\eta_{\ell^+\ell^-}$, $\Delta \phi_{\ell^+\ell^-}$ and $p_T$
distributions of the $2\ell + 2j + \EmissT$ signal at $14 \tev$
with integrated luminosity ${\cal L}=100\xfb^{-1}$,
for Scenario C, after {\it cut-2b}. }
\label{fig:3lep1_}
\end{figure*}


We are interested in signals with leptons in the final state, as
these would be clear to identify at the LHC. We analyzed the
signal  with missing energy only $\nolep$, but unfortunately, although strong, this
signal is completely overwhelmed by the background, mostly  QCD multijet production, $Z$+jets and Drell-Yan production
$pp \to ZZ$ and $pp \to WW$. Cuts for $p_T^{jet}>20$ GeV, for
$N_{jet}>2$, $E_T^{\rm sum}>1.2$ TeV,  and $\EmissT>1$ TeV may yield
some signal, but we found that this signal would still be difficult to isolate and
distinguish securely from background.

We thus concentrate our analysis on monolepton, dilepton, (accompanied by jets), and
trilepton final states. We note that in the rest of the signal simulations we used
the software {\tt PGS 4} \cite{pgs4} to include LHC detector effects.
{\tt PGS 4} uses a jet algorithm which assumes that jets are confined in a cone with
diameter $\Delta R_{jj}=0.5$, together with a hadronic calorimeter
energy resolution as $\sigma(E_T^{\rm jet})=0.8\sqrt{E_T^{\rm
jet}}$. For Scenarios A and B, we generated about $10^5$ events for the signal, and $3 \times 10^5$ for the background. We now summarize the results for each signal.


\subsubsection{The Monolepton Signal: $\onelep$}
We analyze first the case of a single charged lepton with at least two jets in the signal.  For this type of events, for looser selections,  $W$ production with additional QCD jets can dominate \cite{bkgd1l}, while $t{\bar t}$ production is expected to be the dominant background for tight selections of one lepton in SUSY.  Smaller background contributions come from $Z+$jets, di-bosons, single top and QCD processes\footnote{Requiring one lepton in the final state will significantly reduce the QCD multijet production.}.  For LHC energies, $Z+$jets cross section is at least one order of magnitude smaller than $W+$jets, and less resistant to our {\it cut-1}. We used $2 \times10^5$ events for generating $W+$jets, and $8 \times10^5$ events for generating $Z+$jets for {\it cut-2a}, and included Drell-Yan production through virtual $Z^*/\gamma^*$ which contributes with a similar topology to $Z+$jets.  We analyzed the $W+$jets and $Z+$jets processes  contributing to our signals. The hard cross sections are huge: about $1.3 \times 10^8$ fb for $W+$jets and $4 \times10^7$ fb for $Z+$jets. We then applied the basic cuts for one lepton signal. For the $W+$jets case, after the first set of cuts ({\it cut-1}), the cross section goes down to 21288 fb, which is still large. Using the second set of cuts ({\it cut-2a}), that is  increasing $\EmissT$ cut value from 100 GeV to 500 GeV, reduces significantly the events which passed {\it cut-1}.   The situation is similar for the $Z+$jets case, but in this case there are two leptons produced, one of which needs to be veto-ed based on our selection criteria, so that it effectively acts like one lepton at the detector. This happens if one of the leptons is  soft or not well separated from the other lepton, or too close to the beamline, etc.  Requiring large $\EmissT$ from $Z+$jets events  reduces this background further.

Di-boson production, with high $p_T$ leptons is also smaller, with $WW$ production dominating this type of background. While $WW$ and $WZ$ yield some contribution after {\it cut-1},  this is controlled effectively by {\it cut-2a}. The $ZZ$ background is smaller. Here,  we used $8 \times10^5$ events for generating $ZZ$ signals. Single top production where the top decays leptonically has a much smaller cross section than $t{\bar t}$ production.

In Fig. \ref{fig:1lepA} we plot the relevant distributions $\EmissT$,
$\ETsum$, $p_T(\ell)$, $N_{jet}$ and $p_T^{jet}$ distributions of
the $1\ell + 2j + \EmissT$ signal at $14 \tev$ with integrated
luminosity ${\cal L}=100\xfb^{-1}$, for Scenario A.  The figures represent the signal and background after {\it cut-2a}.
We define in general $\ETsum$ (often referred to as $m_{eff}$), in terms of missing transverse energy and transverse momentum for leptons and jets
\begin{equation}
\ETsum=\EmissT +\sum_{\ell} p_T(\ell) +\sum_{jets} p_T (j).
\label{eq:ETsum}
\end{equation}
In general, we expect the threshold for $\ETsum$ and $\EmissT$ to be different, as cascade decays of heavy particles, which often have  large $p_T^{jet}$ \cite{Aad:2009wy} and increase final state jet multiplicity. The dominant
signals  ${\tilde \chi}_1^+ {\tilde \chi}_1^- $  and ${\tilde
\chi}_1^\pm {\tilde \chi}_i^0$ show distinguishing distributions in
$\EmissT$,  $\ETsum$. The   decay into chargino pairs has a tail at large
$\EmissT$,  $\ETsum$, while the chargino-neutralino signal peaks at
low $\EmissT$,  $\ETsum$. In $p_T(\ell)$ the two-charginos signal is
the largest, while the chargino-neutralino signal peaks at low
$p_T(\ell) <100$ GeV. The number of events per bin-size as a
function of the number of jets is completely overwhelmed by
backgrounds ($t \bar t,~ WW,~W+$jets and $WZ$) for $N_{\rm jets} \le 6$, while as a function of
$p_T^{jet}$, secluded $U(1)'$ events are visible for $p_T^{jet}>100$
GeV. Additional cuts on $N_{\rm jets}$ would eliminate $W+$jets or $WW$ background, while $t{\bar t}$ would remain approximately constant.

The same analysis for Scenario B, shown in Fig. \ref{fig:1lepB},
yields non-negligible distributions for ${\tilde \chi}_1^+ {\tilde
\chi}_1^- ,~{\tilde \chi}_1^\pm {\tilde \chi}_i^0$ and ${\tilde
\chi}_i^0 {\tilde \chi}_j^0 $ (the last being the smallest). The
chargino pair decay is again dominant  but its tail at large $\EmissT$,
$\ETsum$  falls more abruptly than that in Scenario A, while the
chargino-neutralino signal peaks at low $\EmissT$, $\ETsum$. As
in Scenario A, in the  $p_T(\ell)$ distribution the two chargino
signal is  dominant, and significant for  $p_T(\ell) \le 150-200$ GeV,
while the other two are not visible. The number of events as a function of the number of jets is, as in Scenario A,
completely overwhelmed by backgrounds for $N_{\rm jets}\le 6$,
while as a function of $p_T^{jet}$, secluded $U(1)'$ events are
visible for $p_T^{jet}>100$ GeV,  and there the number of events per bin
size exceeds those in Scenario A by an order of magnitude. Given the
abundance of the neutralinos in the signal, we expect some
enhancement in the total signal (cross section) with respect to the
MSSM.


\subsubsection{The Dilepton Signal: $\twolep$}
We analyze the $\twolep$ (two {\it same-flavor opposite sign leptons})   in a similar fashion to the $\onelep$
presented in the previous subsection.  The  largest SM background for the dilepton signal comes from top quark pair production \cite{Mukhopadhyay:2009qa}, and from Drell-Yan via $Z^*/\gamma^*$ (showing us as $Z+$jets).  {\it Cut-1} is designed to mostly eliminate the Drell-Yan background, but the $t{\bar t}$ signal survives. We also include the di-boson signal, subdominant for the background. Of di-bosons, only WW gives any significant contribution, with $ZZ$ and $WZ$ signals being much smaller.
In particular,  $WZ$ background for two leptons requires a) that $W \to l \nu$,  but the lepton is not detected (this has a low probability and is further reduced by the pseudorapidity cut); or b) that $ Z\to  \tau^+ \tau^-$, and one $\tau$ decay hadronically, and the other leptonically, which has a very small probability again. The background for the $WZ$ channel  cross section, which is around $30460$ fb before applying the cuts,  is reduced to $2.1$ fb after the {\it cut-1}, and no event  survives after the {\it  cut-2a}. We have used $9\times 10^5$ events to generate this background after  {\it cut-2a}.  Our results are confirmed in  \cite{Davoudiasl:2004aj}. For the two lepton contribution from the $ZZ$ channel to have large enough $\EmissT$, the main contribution must come from the case where one $Z$ decays to neutrinos and the other leptonically, but out of $8\times 10^5$ events  the cross section was reduced to $4.3 $ fb after  {\it cut-1},  and was not visible after {\it cut-2a}.

The main results are shown in
Fig. \ref{fig:2lepEDelta}, where we plot the $\EmissT$, $\ETsum$,
$\Delta R_{\ell^+\ell^-}$ and $\Delta \eta_{\ell^+\ell^-}$ and the
invariant mass distributions of the $2\ell + 2j + \EmissT$ signal at
$14 \tev$ with integrated luminosity ${\cal L}=100\xfb^{-1}$, for
 Scenario A and Scenario B; and in Fig. \ref{fig:2leppT}
 where we give the  $p_T$ distribution for Scenario A and B, and the $N_{jet}$ and $p_T^{jet}$ distributions
of the dominant
$\tilde\chi_{i}^{+}\tilde\chi_{j}^{-} \to 2\ell + 2j + \EmissT$ signal at $14 \tev$ with
integrated luminosity ${\cal L}=100\xfb^{-1}$, for Scenario B, where the signal survives background cuts. The figures depict the signal and background after {\it cut-2a}. For
Scenario A, the background is too large and completely obliterates
the signal.

For the  the $\EmissT$ and $\ETsum$ graphs, the dominant signals are
${\tilde \chi}_1^+ {\tilde \chi}_1^-$ in Scenario A, and  ${\tilde
\chi}_1^+ {\tilde \chi}_1^-$ and ${\tilde \chi}_i^0 {\tilde
\chi}_j^0$ in Scenario B (where the last two give an enhanced number
of events over the signal in Scenario A). Note however that the
number of events per bin decreases by roughly an order of magnitude
with respect to the $\onelep$ signal. The signal peaks around 600
GeV for $\EmissT$ and 1500 GeV for $\ETsum$. Looking  at angular
variables for the two-lepton final signal, such as the cone size
between two charged leptons, the pseudorapidity and the two-lepton
invariant mass distinguish between the signals. The chargino pair production
 in Scenario B gives the largest signal in cone size and
pseudorapidity distributions, peaked respectively around 3 and 0;
while in the two-lepton invariant mass distribution $M_{l^+l^-}$ the
signal  ${\tilde \chi}_i^0 {\tilde \chi}_j^0$ in Scenario B peaks
sharply around 80 GeV, and is negligible elsewhere.  For
Scenario B, in ${\tilde \chi}_1^+ {\tilde \chi}_1^-$, the number of
events per bin size is smaller by a factor of about 10, but the
signal is visible round $M_{l^+l^-}\sim 50-200$ GeV. Looking at the
distributions of both $p_T^{\rm hard}$ and $p_T^{\rm soft}$ in Fig.
\ref{fig:2leppT}, the visible signals are the chargino-pair production ${\tilde
\chi}_1^+ {\tilde \chi}_1^-$ and the neutralino production ${\tilde
\chi}_i^0 {\tilde \chi}_j^0$ in Scenario B, which peak for both distributions
around 50 GeV. As for the $\onelep$ case, the number of events  as a function of the number of jets $N_{\rm jets}$ falls under
the background for $N_{\rm jets} \le 6$, while as a function
of $p_T^{jet}$, secluded $U(1)'$ events are visible for
$p_T^{jet}>80$ GeV, and events with $N_{\rm jets}\ge 2$ dominate.
In the  $\EmissT$ and  $\ETsum$ graphs the signal is dominated by ${\tilde
\chi}_i^0 {\tilde \chi}_j^0$ neutralinos, thus the
dilepton distribution would show more deviation from the MSSM than the
monolepton (as there are many more neutralino processes here).


\subsubsection{The Trilepton Signal: $\threelep$}
An excess of trilepton events, or
isolated dileptons with $\EmissT$, exhibiting a characteristic signature in the
$l^+l^-$ invariant mass distribution, could be the first manifestation of
production of supersymmetric particles. The LHC sensitivity to this channel reaches 320 GeV (720) GeV is the NLSP decays through intermediate gauge bosons(light sleptons) in the extreme case where the LSP is massless \cite{atlascharginos,CMScharginos}. Neither Scenario A nor Scenario B give any significant signals for
the trilepton signal, considered to be a signature for direct ${\tilde \chi}^\pm_1 {\tilde \chi}^0_2$ Drell-Yan production, and theoretically most reliable. For these final states, one expects events containing three hard isolated
leptons (including {\it two same-flavour opposite-sign leptons}, $\mu$ or $e$ and a third ``tagging'' lepton) and $\EmissT$, with no jets, and small SM backgrounds.  To
highlight this signal, we have set up another alternative, Scenario
C, where the dominant signal is ${\tilde \chi}_1^\pm {\tilde
\chi}_2^0$, yielding $\ell_i^\pm \ell_j^+ \ell^-_j +\EmissT$. We generated  about $2.7 \times 10^5$ signal events, to enhance the event to background ratio.

We present our results in Fig. \ref{fig:3lep1_}, together with the
dominant SM background coming from $WZ$. For  the modified cut $\EmissT >200$ GeV ({\it cut-2b}),
 this signal is almost background-free and will be distinguished by measurement of both
$\EmissT$ and $\ETsum$, and of the
angular correlations, as the cone size between two charged leptons,
the pseudorapidity and the two-lepton azimuthal angle. For
$\EmissT$  and $\ETsum$,  the signal is strong
till 800 and 2000 GeV, respectively. In the two-lepton invariant
mass (plotted here for the correlated {\it two same-flavor opposite-sign leptons}, distinguished by their separation), the signal is strong for low $M_{l^+l^-}=0-20$
GeV,  and shows a wide peak in the $40-100$ GeV region. The specific shape of the $M_{l^+l^-}$ distribution reveals details about $\tilde \chi_2^0$ production and decay \cite{Abdullin:1998pm}. In Scenario C, the decay of the NLSP proceeds through a two-body channel as $\tilde \chi_2^0 \to (\tilde l) l \to (l \tilde \chi_1^0) l$. We expect
$$M_{l^+l^-}^{\rm max} =\frac {\sqrt{ \left ( M^2_{{\tilde \chi}^0_2}-M^2_{\tilde l} \right ) \left ( M^2_{\tilde l}-M^2_{{\tilde \chi}^0_1}  \right) } } {M_{\tilde l}}=85.5 ~{\rm GeV}.$$
 The peak at low $M_{l^+l^-}$ indicates that many  events produced with large  $\EmissT> 200$ GeV, as we imposed in {\it cut-2b}, while the peak in the $50-80$ GeV region is near $M_{l^+l^-}^{\rm max} $.
 For angular variables, the production in terms of $\Delta R_{\ell \ell}$ has a sharp edge due to {\it cut-1} ($\Delta R_{\ell \ell}>0.4$), and is peaked around pseudorapidity $ \Delta \eta_{\ell \ell}=0$\footnote{ Note that the graphs in the frames 4-6 are interconnected, as $\Delta R_{\ell \ell}=(\Delta \eta_{\ell \ell}^2+ \Delta \phi_{\ell \ell}^2)^{1/2}$.}. These angular distributions indicate the region most likely to detect the $\threelep$ signal. We also show the $p_T(\ell)$ for the three hard leptons in the last panel, distinguished by their increasingly broader peaks. This
signal could be detected at the $14 \tev$ LHC, as more than 100
events per year would be observed. Though promising, this signal
looks  at first similar to the $\threelep$ signal in
MSSM, where the trilepton signal is also dominated by  ${\tilde
\chi}_1^\pm {\tilde \chi}_2^0$ production, and where similar cross
sections are expected \cite{Baer:1994nr}. However, while the production of the chargino-neutralino pair in this scenario is MSSM like, the decay of the chargino (which is wino-like)  is into a singlino-like LSP, whereas the NLSP in this model is bino-like and resembles the LSP in MSSM.  Thus the dark matter candidate and decays of the chargino-neutralino pair into the LSP  differ  from the MSSM.  This exclusive channel is expected to play a central role in the precise determination of ${\tilde \chi}^0_1$, the dark matter candidate \cite{Abdullin:1998pm}. The decays patterns are shown in Fig. \ref{fig:decayC} in the Appendix.



\section{Summary and Conclusion
\label{sec:conc}}
We studied the production of neutralinos and charginos at the LHC
in the context of the secluded $U(1)'$, in which  singlet fields are
added to supersymmetric models with extra $U(1)$'s  to
stabilize the $Z-Z'$ mass splitting. The model has five additional
neutralinos (in addition to the four in MSSM), which could enhance
the signals observed at the LHC. In fact, as the additional Higgs singlets are expected to be heavy, analyzing the neutralino sector would be a promising test of the secluded sector of the model. We perform the analysis for LHC
operating at $14 \tev$ with integrated luminosity ${\cal
L}=100\xfb^{-1}$.

As discovery of
supersymmetry at the LHC is expected
to occur through the observation of large excesses of events in missing $\EmissT$
plus jets, or with one or more isolated leptons,
we classify and analyze the final signals based on
the number of leptons emitted, and look at final states with
$\onelep$, $\twolep$ and $\threelep$. There are very few events
generated with more than three leptons in the final states in this
model, and, though spectacular,  thus these are not likely to be seen at the LHC, even at
$14 \tev$. For each signal, we study a parameter space where the
signals could be enhanced. In two of the Scenarios, A and B, the
largest cross section is obtained for the production of the lightest
chargino pair, or the lightest chargino with neutralinos. Both seem most promising
to be observed in $\EmissT$ and $\ETsum$ plots, or in $p_T^{jet}$,
with a cut   $p_T^{jet}>80-100$ GeV, to enhance the signal to
background response. Increasing the number of leptons in the final
state produces fewer events, however with a reduction in the
background as well. We found that for the $\twolep$ in Scenario A, the background completely overwhelms the signal, while in Scenario B the signal remains promising.  Events with large $\EmissT$ and $\ETsum$, or large $N_{\rm jets}$ are more likely to be signal than background. For highlighting the $\threelep$ scenario, we
analyzed another region of the parameter space, Scenario C, where
the dominant cross section is to the lightest chargino and second
lightest neutralino, resulting in a $\threelep$ final state.
We find that plots for events  yield observable results, and with judicious cuts ($\EmissT>200$ GeV) they are almost
background-free, and could yield ${\cal O}(10^2)$
events per energy bin at the LHC.

Our benchmark scenarios and signals are qualitatively  and quantitatively different from the MSSM.  We list below some of the sources for  the expected differences.
\begin{itemize}
\item The three benchmark parameters scenarios, all distinct and all chosen to highlight some of the features of the $U(1)^\prime$ model. All scenarios satisfy conditions for the stability of the vacuum, predict a SM-like Higgs at the LHC at 125 GeV, and fulfil relic density constraints. All scenarios have light singlinos, to distinguish them from MSSM, and several more light neutralinos than in MSSM, to enhance their production.
\item   The composition of the dark matter candidate, the LSP, and of the NLSP, have been chosen to be non-MSSM-like. For instance, in Scenario A, the LSP is mostly bino $\tilde B$, while the NLSP is mostly singlino.  In Scenario B the LSP is mostly singlino  while  the NLSP is  bino, with singlino admixtures. In Scenario C the LSP is  singlino, while the NLSP is mostly wino.  Thus in all the scenarios chosen, one or more of the singlinos are light, resulting in differences in the production and decay patterns for neutralinos and charginos between this model and the MSSM.
\item   In particular, we  highlight the decay pattern of the NLPS into the LSP. Even for the three lepton decay in Scenario C, which at first seems to be very similar to that in  MSSM, the decay pattern of the wino NLSP into a singlino LSP, which does not couple directly to fermions would yield a distinguishable signature of the model from the decay of the NLSP into a bino LSP in MSSM.
\item   A clear difference between our model and MSSM is the fact that we need to have $\tan \beta$ very close to 1 to satisfy vacuum stability bounds in the Higgs potential, while yielding a Higgs mass $\sim 125$ GeV, in contrast the  MSSM  prefers a medium to large value for $\tan \beta$. This fact makes satisfying constraints from $B_{d,s}$ decays in $U(1)^\prime$ natural. The topic is beyond our study, but  we showed that the bounds are satisfied.
\item In MSSM, there are 21 different reactions for direct chargino-neutralino pair production: 3 for ${\tilde \chi}^\pm_i {\tilde \chi}^\mp_j$, 8 for ${\tilde \chi}^\pm_i {\tilde \chi}^0_j$ and 10 for ${\tilde \chi}^0_i {\tilde \chi}^0_j$, while in $U(1)^\prime$ there are 56 possibilities: 3 for ${\tilde \chi}^\pm_i {\tilde \chi}^\mp_j$, 18 for ${\tilde \chi}^\pm_i {\tilde \chi}^0_j$ and 35 for $ {\tilde \chi}^0_i {\tilde \chi}^0_j$. Thus the model has significant quantitative differences from the MSSM.
\end{itemize}
  MSSM, like our model, has a large parameter space, and it could be possible that in one corner of that space, some signals would overlap with ours. What would be highly unlikely is that, in any of the MSSM parameter space, {\it all the signals} would be the same as those resulting from $U(1)^\prime$.
  Given our choice of light neutralino states and dark matter candidate, we would have a next-to-impossible task to reproduce a similar scenario for MSSM.  Thus most of our comparisons are with existing LHC data.

 As the pressure put on the constrained and phenomenological MSSM
 by present measurements at the LHC mounts, the analysis
presented here provides a map of possible signals in neutralino production of physics beyond MSSM, which
should be easily confirmed or ruled out at $14 \tev$. If the secluded $U(1)^\prime$ is
the correct supersymmetric scenario, the difference with MSSM  should manifest itself
in the $\onelep$ and $\twolep$ signal, where the cross sections (and
the correlated number of events) should be above what one expects in
the minimal model, and in the $\threelep$ scenario, where the decays of the chargino-neutralino pair are into a singlino-like LSP, shedding light on the nature of dark matter.

\section{Acknowledgments}
M.F. would like to thank Benjamin Fuks  for many
discussions on the topic of chargino and neutralino production. The
work of  M.F.  is supported in part by NSERC under grant number
SAP105354. The research of L. S. is  supported in part by The
Council of Higher Education of Turkey (YOK).

\appendix


 \section{Feynman diagrams for decays channels}
 \label{app}

\begin{figure*}[htbp]
\begin{center}$
    \begin{array}{c}
      \includegraphics[width=4.8in,height=2.0in]{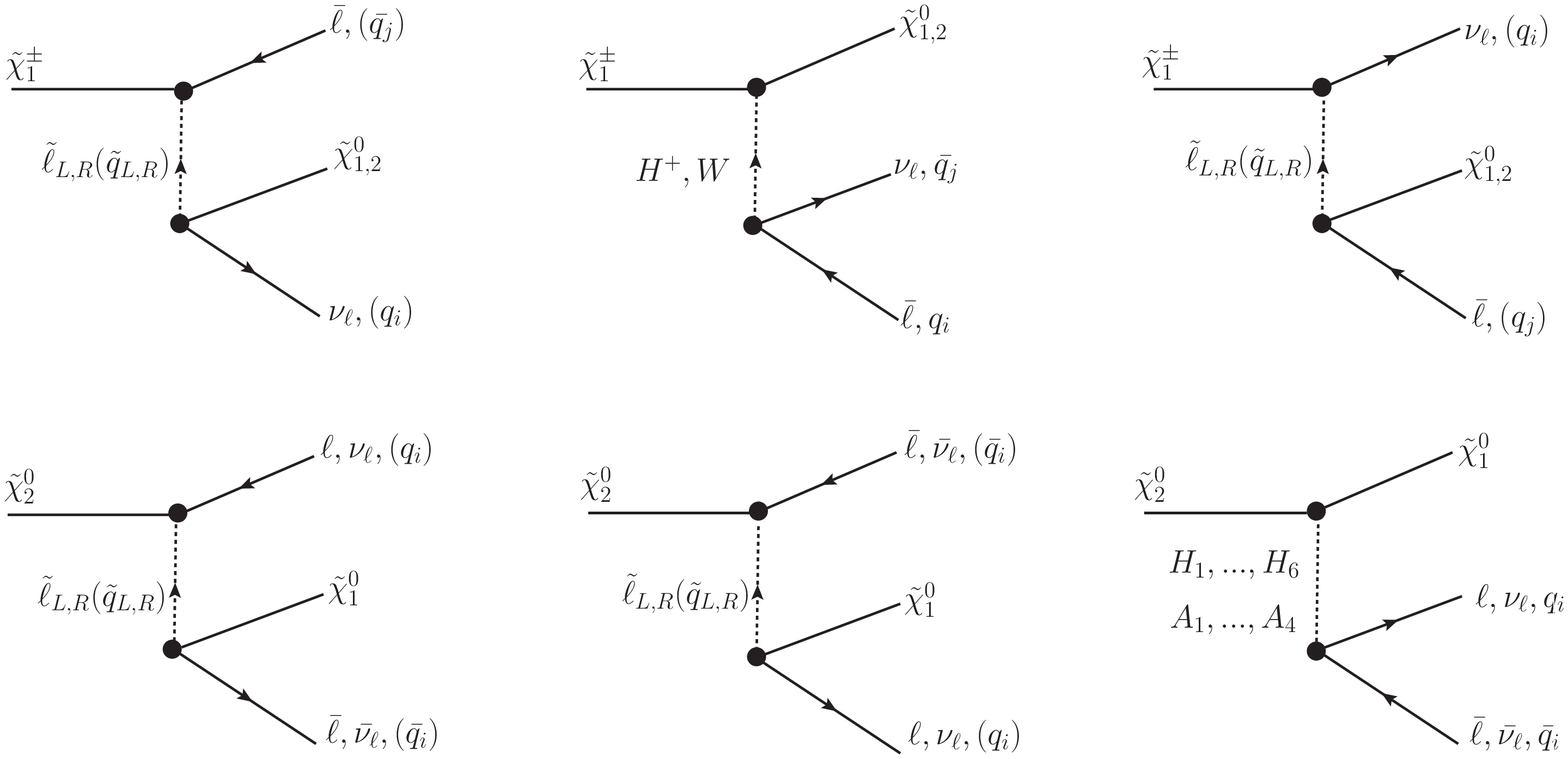}\\
    \includegraphics[width=4.8in,height=0.8in]{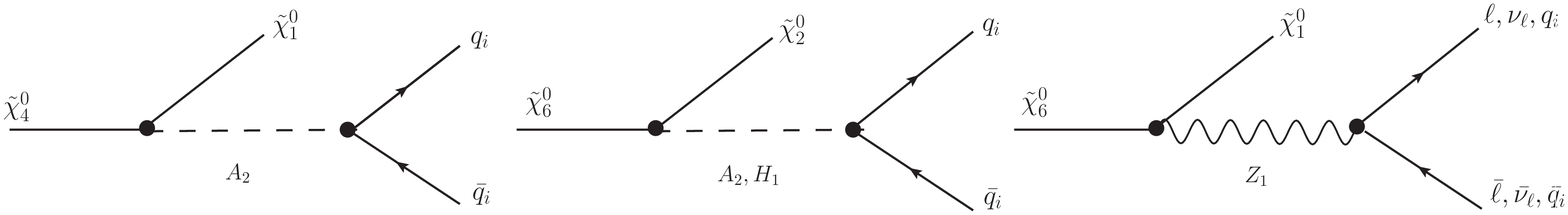}\\
    \end{array}$
\end{center}
\vskip -0.1in
      \caption{Generic Feynman diagrams for the decays of the
chargino ${\tilde \chi}_1^\pm$ and neutralinos ${\tilde \chi}_2^0,~{\tilde \chi}_4^0$, and ${\tilde \chi}_6^0$ in Scenario A of the secluded $U(1)^\prime$ model. Here $\tilde l$ are scalar leptons, $H_i,~A_j$ are scalar and pseudoscalar Higgs bosons, and $W$ and $Z$ are gauge bosons. }
\label{fig:decayA}
\end{figure*}

\begin{figure*}[htbp]
\begin{center}$
    \begin{array}{c}
      \includegraphics[width=4.8in,height=2.0in]{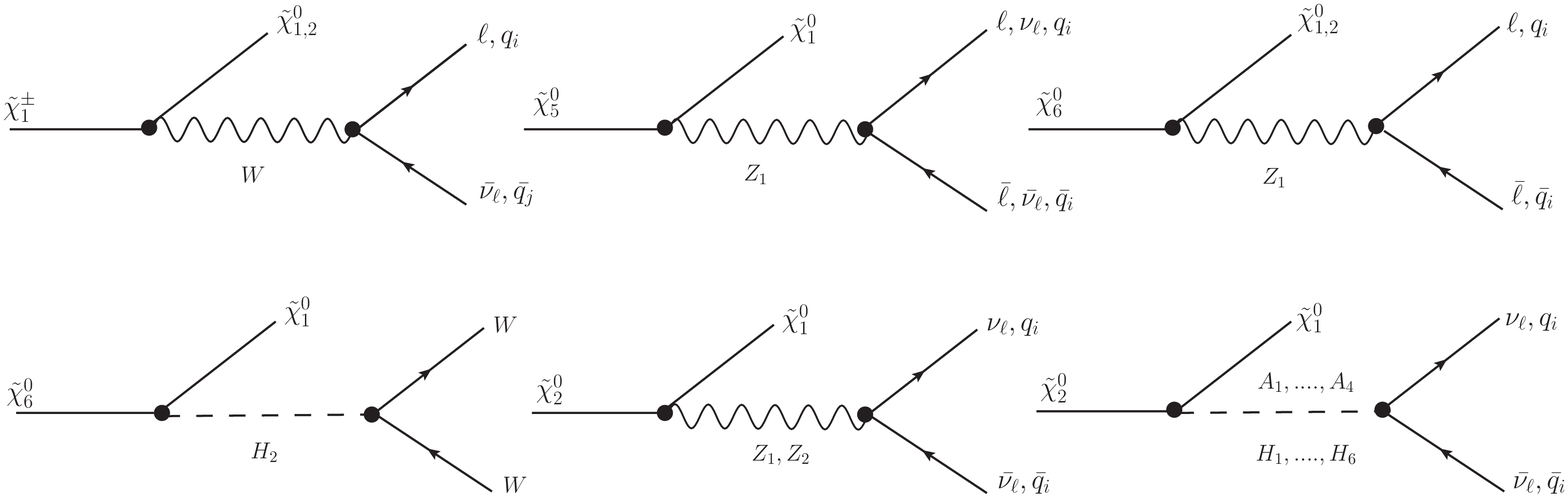}\\
    \includegraphics[width=4.8in,height=0.9in]{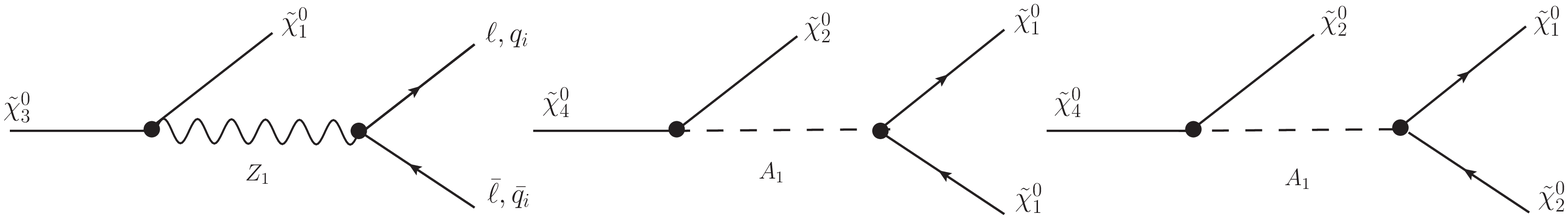}\\
    \end{array}$
\end{center}
\vskip -0.1in
      \caption{Generic Feynman diagrams for the decays of the
chargino ${\tilde \chi}_1^\pm$ and neutralinos ${\tilde \chi}_i^0,~i=2, \ldots 6$ in Scenario B of the secluded $U(1)^\prime$ model. Intermediate particle notation is the same as in Fig. \ref{fig:decayA}. }
\label{fig:decayB}
\end{figure*}

\begin{figure*}[htbp]
\begin{center}$
    \begin{array}{c}
      \includegraphics[width=4.8in,height=1.0in]{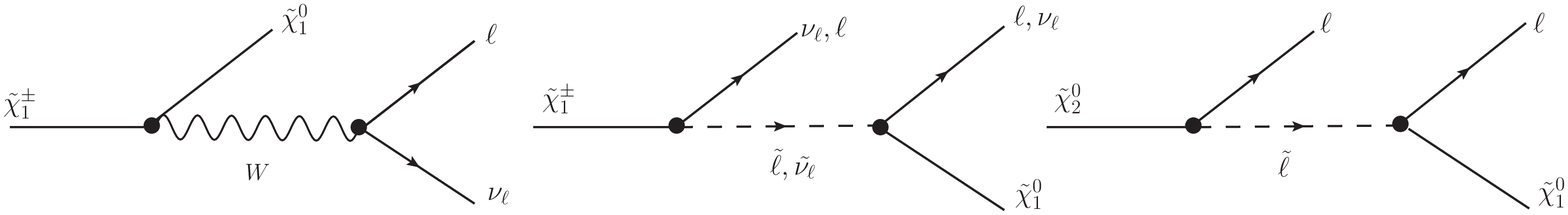}\\
    \end{array}$
\end{center}
\vskip -0.1in
      \caption{Generic Feynman diagrams for the decays of the
chargino ${\tilde \chi}_1^\pm$ and neutralino ${\tilde \chi}_2^0$ in Scenario C of the secluded $U(1)^\prime$ model. Intermediate particle notation is the same as in Fig. \ref{fig:decayA}, and $\tilde \nu_l$ is the scalar neutrino.}
\label{fig:decayC}
\end{figure*}

 We list the main decay channels of chargino and neutralinos in Scenario A, (Fig.  \ref{fig:decayA}), Scenario B, (Fig.  \ref{fig:decayB}) and Scenario C (Fig. \ref{fig:decayC}).

\clearpage

\end{document}